\documentclass[a4paper,11pt]{article}
\pdfoutput=1
\usepackage{float}
\usepackage{tabulary}
\usepackage{verbatim}
\usepackage[dvipsnames]{xcolor}
\usepackage{textcomp}
\usepackage{gensymb}
\usepackage{url}

\usepackage{jcappub} 
\title{\boldmath Dark matter implications of the WMAP-Planck Haze}

\author[a]{Andrey E.~Egorov,}
\author[b,c]{Jennifer M.~Gaskins,}
\author[a]{Elena Pierpaoli,}
\author[d,e,f] {and Davide Pietrobon}

\affiliation[a]{University of Southern California, \\ 3620 McClintock Ave., SGM 408, Los Angeles, CA 90089, USA}
\affiliation[b]{California Institute of Technology, \\ 1200 E.~California Blvd., Pasadena, CA 91125}
\affiliation[c]{GRAPPA, University of Amsterdam, \\ Science Park 904, 1098 XH Amsterdam, Netherlands}
\affiliation[d]{University of California, Berkeley, Space Sciences Laboratory, \\ 7 Gauss Rd, Berkeley CA 94720}
\affiliation[e]{Lawrence Berkeley National Laboratory, 1 Cyclotron Road, Berkeley, CA 94720}
\affiliation[f]{HERE, 2075 Allston Way, Berkeley CA 94704}
\emailAdd{egorov@usc.edu}
\emailAdd{jgaskins@uva.nl}
\emailAdd{pierpaol@usc.edu}
\emailAdd{daddeptr@gmail.com}

\abstract{
Gamma rays and microwave observations of the Galactic Center and surrounding areas indicate the presence of anomalous emission, whose origin remains ambiguous.  The possibility of dark matter annihilation explaining both signals through prompt emission at gamma rays and secondary emission at microwave frequencies from interactions of high-energy electrons produced in annihilation with the Galactic magnetic fields has attracted much interest in recent years. We investigate the dark matter interpretation of the Galactic Center gamma-ray excess by searching for the associated synchrotron emission in the WMAP and Planck microwave data. Considering various magnetic field and cosmic-ray propagation models, we predict the synchrotron emission due to dark matter annihilation in our Galaxy, and compare it with the WMAP and Planck data at 23-70 GHz.   In addition to standard microwave foregrounds, we separately model the microwave counterpart to the Fermi Bubbles and the signal due to dark matter annihilation, and use component separation techniques to extract the signal associated with each template from the total emission.
We confirm the presence of the Haze at the level of $\approx$7\% of the total sky intensity at 23 GHz in our chosen region of interest, with a harder spectrum ($I \sim \nu^{-0.8}$) than the synchrotron from regular cosmic-ray electrons. The data do not show a strong preference towards fitting the Haze by either the Bubbles or dark matter emission only. Inclusion of both components provides a better fit with a dark matter contribution to the Haze emission of $\approx$20\% at 23 GHz, however, due to significant uncertainties in foreground modeling, we do not consider this a clear detection of a dark matter signal.  We set robust upper limits on the annihilation cross section by ignoring foregrounds, and also report best-fit dark matter annihilation parameters obtained from a complete template analysis.  We conclude that the WMAP and Planck data are consistent with a dark matter interpretation of the gamma-ray excess.   }

\keywords{dark matter theory, cosmic ray theory, absorption and radiation processes}
\arxivnumber{1509.05135}

\begin{document}
\maketitle
\flushbottom

\section{Introduction}
\label{sec:intro}

The existence of dark matter (DM) has been firmly established by a variety of observations \cite{Bertone-book}, however its nature remains a compelling mystery.  DM is often assumed to be in the form of Weakly Interacting Massive Particles (WIMPs), which may pair annihilate and/or decay to produce Standard Model particles.
Indirect searches for the detectable products of DM annihilation and decay, including photons, charged cosmic rays (CRs), and neutrinos, offer the possibility of detecting this elusive particle and constraining its properties, as well as determining its spatial distribution.  For typical WIMP DM models, the Standard Model particles immediately resulting from annihilation or decay are in the MeV to TeV energy range, making gamma rays a particularly promising channel for DM detection.

With the great leap in capabilities for detection of GeV gamma rays provided by the Fermi Large Area Telescope (LAT) aboard the Fermi Gamma-ray Space Telescope \cite{Atwood:2009ez}, much progress has been made in indirect searches for gamma rays from DM annihilation and decay.  Analyses of Fermi LAT observations of possible sources of annihilation signals have provided strong constraints on the WIMP mass and annihilation cross section \cite{Hooper:2012sr, Ackermann:2015zua, Buckley:2015doa}.   These studies probed annihilation cross sections below the canonical value for a thermal relic WIMP DM particle $\langle\sigma v\rangle = 3 \cdot 10^{-26}$ cm$^3$/s, for WIMP masses below a few tens of GeV for some Standard Model final states. 

The Galactic Center (GC) is expected to be the brightest source in the sky of gamma rays from DM annihilation due to its large overdensity of DM and its close proximity. In recent years, numerous studies have firmly revealed the presence of an excess at GeV energies over the expected astrophysical gamma-ray emission from the GC and Inner Galaxy that is consistent with expectations for a DM annihilation signal (e.g., \cite{Goodenough:2009gk, Hooper:2010mq, Abazajian:2012pn, Gordon:2013vta, Daylan:2014rsa, Abazajian:2014fta, Calore:2014xka}). In this work we will use GC to refer to the dynamical center of the Galaxy, and Inner Galaxy to refer to the region extending as much as tens of degrees away from the GC\@.  In addition to the DM interpretation, other explanations considered for this excess include a population of unresolved gamma-ray millisecond pulsars (MSPs) \cite{Gordon:2013vta, Hooper:2013nhl, Cholis:2014lta, Yuan:2014rca, Petrovic:2014xra, Abazajian:2014fta, Brandt:2015ula} and gamma rays from CRs injected in bursts near the GC~\cite{Petrovic:2014uda, Carlson:2014cwa, Cholis:2015dea}. Recent results suggest that the majority of the excess emission arises from a population of unresolved sources such as MSPs \cite{Bartels:2015aea, tracey15}, however those sources have not been identified in multi-wavelength studies (see, e.g., \cite{Linden:2015qha}), and there remains some debate in the community, leaving open the exciting possibility that the excess represents a detection of DM annihilation.

Multi-wavelength observations can provide critical tests of DM interpretations of high-energy signals, particularly in a complex environment like the GC\@. Along with prompt gamma-ray emission, WIMP annihilation can also inject other energetic particles into the interstellar medium, including electrons and protons and their antiparticles, which give rise to lower-energy photons. Protons propagate and participate in hadronic interactions with gas, yielding neutral pions which produce gamma-ray photons when they decay. Electrons and positrons (referred as electrons in our paper, since they behave equivalently for our purposes in producing secondary emission), on the other hand, generate substantial secondary emission via inverse Compton scattering (ICS) on the interstellar radiation fields, bremsstrahlung on ionized gas, and synchrotron emission in magnetic fields (MF).  DM annihilation in the Inner Galaxy thus generates associated radiation from radio to gamma rays (with a rather broad spectrum from each mechanism mentioned above). Thus, to test the DM interpretation for the GC gamma-ray excess, one may look for the associated lower-frequency emission. In particular, one can look for the synchrotron emission at microwave frequencies from the propagation of injected high-energy electrons in Galactic MF\@.

A similarly intriguing excess over conventional astrophysical models at microwave frequencies was identified several years ago: the WMAP Haze~\cite{Finkbeiner:2003im, Hooper:2007kb}.  The WMAP Haze refers to excess emission in the Inner Galaxy identified in the WMAP data, observed to be approximately symmetric around the GC, extending tens of degrees away from the GC, and exhibiting a harder spectral index than expected for synchrotron associated with the Galactic CR population from astrophysical sources.  The spectrum of the WMAP Haze suggests that it originates from a distinct population of high-energy electrons. Initially, secondary emission from WIMP annihilation was proposed as a possible explanation for the Haze, and an associated gamma-ray signal was predicted~\cite{Hooper:2007kb}.  

More recently, a new feature in the gamma-ray sky has been identified: the Fermi Bubbles~\cite{Su:2010qj, Fermi-LAT:2014sfa}.  The Bubbles are a bipolar structure centered on the GC and extending above and below it out to  $|b| \approx 50\degree$, and they are clearly visible  in the LAT gamma-ray data.  Similar structures had previously been identified in X-ray and 408 MHz radio data~\cite{Sofue:1999zf} as well as in IR maps~\cite{BlandHawthorn:2002ij}, and they had been suggested to result from starburst-driven winds in the GC\@.  Refined observations in microwave with Planck confirmed the presence of the excess  emission previously identified as the Haze in WMAP data, and revealed a correlated morphology between the Haze and Fermi Bubbles~\cite{:2012rta}. This indicates a potentially common origin of these two phenomena. In addition, recent observations of polarized radio emission with the S-PASS survey at 2.3 GHz revealed that the radio features are also closely aligned with those observed in gamma rays~\cite{Carretti:2013sc}, again pointing to a consistent interpretation (e.g., \cite{Crocker:2014fla}).  

The Bubble-like structures observed over a broad range of energies are now presumed to be multi-wavelength views of the same phenomenon, and compelling non-exotic explanations have been put forth (e.g. reverse shocks in the GC's giant outflows \cite{Crocker:2014fla}).  Moreover, due to the unique morphology of these structures, it is quite unlikely that they are associated with DM annihilation. Thus, a search for a DM signal in the WMAP-Planck data requires an understanding of not only ``standard'' astrophysical foregrounds but also the microwave emission associated with the Fermi Bubbles. Despite the presence of the Bubbles in multi-frequency observations, and its likely presence in the microwave band,  it is still possible that some part of the observed Haze is due to DM annihilation.  Throughout this paper, the term  ``Haze''  will refer to  the microwave emission not accounted for by common foregrounds (i.e. Bubbles and/or DM annihilation),  and ``Bubbles''  will refer  to the microwave counterpart of the Fermi Bubbles.

In this work, we revisit the problem of multi-wavelength DM searches in the Galaxy using the WMAP-9 and Planck 2013 data sets, focusing on DM models that could explain the observed GC gamma-ray excess.  We calculate the microwave emission associated with WIMP annihilation in the Galaxy, and test the consistency of that signal with the observations by performing a complete model-fitting analysis of the WMAP-Planck data. The predicted microwave emission due to WIMP annihilation is subject to uncertainties in the modeling of the Galactic environment.  In particular, uncertainties in the MF distribution in the Galaxy and  in the propagation parameters for electrons in the interstellar medium can lead to significant variations in the expected emission.  We address these sources of uncertainty by varying the MF and propagation models in order to understand their impact on our results. 
We then compare our models with the data in order to assess the potential contribution of DM to the total signal.

We  adopt  two different approaches for model-data comparison: first we assume that the data only contain noise, CMB fluctuations and DM signal (but no other foregrounds or Bubbles). This leads to an overestimate of the potential DM contribution  which translates to a conservative but robust  upper limit on the DM annihilation cross section.
Next, we perform a component separation procedure based on template fitting, this time also considering all relevant foregrounds and the Bubbles.
In this case, we present the best-fit DM parameters obtained.
With these analyses we aim to address the following issues: {\it i)} the nature of the  Haze -- whether the data favor (and at what significance) a DM component in the Haze emission; and {\it ii)} the compatibility of WIMP models needed to fit the GC gamma-ray excess with the microwave data.

We present the DM models and the computation of the microwave emission from DM annihilation in \S\ref{sec:Computation}. In \S\ref{sec:roi} we define the two regions of interest (ROIs) we consider, and in \S\ref{sec:dmspectra} we show the predicted DM spectra in the Haze region.  \S\ref{sec:cmb_constraints} is dedicated to conservative WIMP model constraints obtained from the data when considering CMB and DM emission as the only contributions, neglecting all other foregrounds.  A comprehensive component separation procedure, which includes models of expected foregrounds, is described along with its results in \S\ref{sec:Fitting}. We discuss the implications of these findings for DM models and a DM interpretation of the GC gamma-ray excess in \S\ref{sec:Discussion}.

\section{Microwave emission from DM annihilation}
\label{sec:Computation}

The propagation of high-energy electrons in the Galactic MF gives rise to synchrotron emission at microwave frequencies. The emission associated with DM annihilation depends on the spatial distribution of the injection sites, the injected spectrum of these particles, as well as the Galactic environment.  

\subsection{DM distribution}
\label{sec:DMDensityDistr}

The exact DM density distribution in our Galaxy is a matter of active debate. A wide range of DM density profiles from cored to very cuspy are in agreement with numerical simulations and observational data (e.g., \cite{Iocco:2011jz, Nesti:2013uwa, Calore:2015oya, Pato:2015dua, Iocco:2015xga}). In this work we restrict our analysis to profile shapes that have been shown to provide a good fit to the GeV excess in the Inner Galaxy \cite{Goodenough:2009gk, Hooper:2010mq, Abazajian:2012pn, Gordon:2013vta, Daylan:2014rsa, Abazajian:2014fta, Calore:2014nla}.
In particular, we adopt for the DM distribution a generalized form of the Navarro-Frenk-White (NFW) density profile \cite{NFW}
\begin{equation}\label{rho_DM}
	\rho_{\rm DM}(r) = \frac{\rho_{\rm s}}{\left(\frac{r}{r_{\rm s}}\right)^{\gamma}\left(1+\left(\frac{r}{r_{\rm s}}\right)^{\alpha}\right)^{(\beta-\gamma)/\alpha}},
\end{equation}
where $\rho_{\rm DM}(r)$ is the DM density at radius $r$ from the center of the halo, $\rho_{\rm s}$ is a scale density, $r_{\rm s}$ is a scale radius, and $\alpha$, $\beta$, and $\gamma$ are shape parameters. We adopt the common convention to fix $\alpha=1$ and $\beta=3$, and allow only the inner profile slope, parametrized by $\gamma$, to vary.
Following the results of \cite{Goodenough:2009gk, Hooper:2010mq, Abazajian:2012pn, Gordon:2013vta, Daylan:2014rsa, Abazajian:2014fta, Calore:2014nla}, 
we consider values of $\gamma$ ranging between 1.1 and 1.3 (considering three discrete values 1.1, 1.2 and 1.3). We adopt $r_{\rm s} = 23$~kpc and fix the local density at the solar position to $\rho_{\rm DM}(r_{\odot} = 8.3\mbox{ kpc}) = 0.34$ GeV/cm$^3$ (which are similar to the values adopted in \cite{Goodenough:2009gk, Hooper:2010mq, Abazajian:2012pn, Gordon:2013vta, Daylan:2014rsa, Abazajian:2014fta, Calore:2014nla}). 

The density distribution is indeed highly uncertain in a very close vicinity of the center. For this reason, authors typically do not extrapolate profiles like Eq.~\ref{rho_DM} down to zero radius and instead truncate it at some distance $r_{\rm core}$ leaving the density constant inside $r_{\rm core}$. A natural lower limit on $r_{\rm core}$ is about the Schwarzschild radius of the central black hole $4.2 \cdot 10^{-7}$ pc. An upper limit allowed by the fit of the gamma excess was found to be $\sim 10$ pc \cite{Daylan:2014rsa}. We computed that the DM emission intensities in our relevant ROIs only vary by about $\sim 1\%$ when we vary $r_{\rm core}$ over the mentioned allowed range $4.2 \cdot 10^{-7} \mbox{ pc} \lesssim r_{\rm core} \lesssim 10$ pc. Hence, although this parameter is highly uncertain, its specific choice does not matter in our case.
 
Substructures are also known to play an important role  in boosting the luminosity of a DM halo (see e.g. \cite{Kamionkowski:2010mi}).  We modeled the substructure contribution with a local boost factor to the annihilation rate everywhere in the Galaxy $\xi_{\rm DM}(r)$, taking its radial distance dependence from Fig.~4 of \cite{Kamionkowski:2010mi}. We found that substructure increases the signal in our chosen ROI (discussed in \S\ref{sec:roi}) only by $\approx (30-35)\%$. This is because the majority of our signal comes from areas close to the GC where, due to tidal disruption, we do not expect a large amount of substructure.

\subsection{Electron injection from DM annihilation}
\label{sec:dmspect}

The spectrum of electrons produced by DM annihilation depends on the DM mass and the branching ratios to different Standard Model final states.  We consider DM masses from 7.0 to 52 GeV, broadly motivated by the range of WIMP masses found in prior work to be able to provide, for some choice of final state(s), a good fit to the spectrum of the observed gamma-ray excess around the GC\@.  

Also in line with much prior work on the gamma-ray excess, in this work we consider two benchmark annihilation channels: $b\bar{b}$ and $\tau^{+}\tau^{-}$.  The former represents a typical hadronic channel which yields a soft spectrum of injected electrons, while the latter is a leptonic channel that produces more electrons with a harder spectrum.  We take the yields of electrons from WIMP annihilation from the PPPC4DMID package~\cite{Cirelli:2010xx, PPPC} (including electroweak corrections; see, e.g.~\cite{Ciafaloni:2010ti}).

\subsection{Magnetic field distribution}
\label{sec:MF}

Synchrotron emission depends crucially on the MF distribution in the Galaxy.  Although sophisticated MF configurations are possible (see, e.g., \cite{Dobler:2011mk}), for the sake of simplicity  we model the MF  with an axisymmetric distribution, as also adopted in earlier works, e.g. \cite{Malyshev:2010xc, Mambrini:2012ue}:
\begin{equation}\label{MF}
	B(R,z)=B_0 \exp(-(R-R_{\odot})/R_B-|z|/z_B),
\end{equation}
where $R$ and $z$ represent radial-in-plane and plane-orthogonal distances in the Galaxy, respectively.

The parameter $B_0$ in Eq.~\ref{MF} represents the total MF value at the Solar location and is taken to be 6 $\mu$G following \cite{Fornengo:2011iq}. The radial scale $R_B$  in Eq.~\ref{MF} naturally sets the central field value and is more uncertain. On this point,  we rely on the lower limit $B(0,0) = 50~\mu$G derived in \cite{Crocker:2010xc} as our preferred choice. This sets the respective $R_B = 4.0$ kpc, which we consider to be the most realistic value. However, we also compute our results for $B(0,0) = 100~\mu$G as a  limiting case which would produce higher DM emission.  Such high central MF values are encountered in the literature \cite{Crocker:2010rd}. However, the field would reach these values only very close to the GC (within $\sim 0.2$ kpc), and in that case the field radial dependence in Eq.~\ref{MF} might not be steep enough to produce the proper field attenuation at larger radii. For this reason, we consider the $B(0,0) = 100~\mu$G (with respective $R_B = 3.0$ kpc) configuration to be generally less realistic.

The $z_B$ parameter in Eq.~\ref{MF} sets the vertical characteristic scale of the MF, which is also somewhat uncertain. Various studies of the MF distribution in the Milky Way and other similar galaxies (e.g. \cite{2008A&A...480...45S, Haverkorn:2011jg, Haverkorn:2014jka}) estimate $z_B$ to be robustly constrained in the range $\approx (1-10)$ kpc with a characteristic ``average'' value of 4-6 kpc. At the same time, $z_B$ is naturally linked to the CR propagation parameters (discussed in \S\ref{sec:diffusion}), as the MF is the driving reason for CR diffusion, and where the MF is small or null CR can freely propagate. We therefore expect the MF spatial extent  $z_B$ to be comparable to the spatial extent of the diffusion region, parametrized by $h$ (the vertical half-height of the diffusion zone).
Proportionality between these two parameters was employed in multiple earlier works to estimate $z_B$ (e.g. \cite{Fornengo:2011iq, Mambrini:2012ue}). We follow this approach and, similarly to \cite{Fornengo:2011iq}, we compute all our results for two values of $z_B$: $z_B = \delta \cdot h$ and $z_B = 0.5 \cdot h$. The parameter $\delta$ denotes the spectral slope of the spatial diffusion coefficient  (described below).

\subsection{CR propagation using GALPROP}
\label{sec:galprop}

We computed all-sky intensity maps of the synchrotron emission from DM annihilation using the GALPROP package \cite{GP}. GALPROP is a sophisticated code designed to compute CR propagation in the Milky Way. It solves the transport equation in its most general version for a given distribution of CR sources in 2 or 3 spatial + time + particle momenta dimensions. It also incorporates detailed knowledge about the Galaxy such as the neutral and ionized gas distributions, interstellar radiation field, and astrophysical CR sources. Moreover, GALPROP is able to compute all sky maps of secondary emission from CRs resulting from a variety of processes, including synchrotron emission.

We employed GALPROP v54r2423 to model the propagation of electrons from DM annihilation in the Galaxy. Specifically, GALPROP was used to numerically solve the following transport equation:
\begin{equation}\label{diff.eq.}
	\frac{\partial n}{\partial t} = q(r,p) + D_{xx}(p) \nabla^2 n + \frac{\partial}{\partial p} p^2 D_{pp}(p) \frac{\partial}{\partial p} \frac{1}{p^2} n - \frac{\partial}{\partial p} (\dot{p} n), 
\end{equation}
where $p$ is the electron momentum and $n \equiv n(\vec{r},p,t)$ is the electron concentration per unit momentum range everywhere in the Galaxy.  GALPROP solves Eq.~\ref{diff.eq.} in time until a stationary solution $n(\vec{r},p,t) \rightarrow n(\vec{r},p)$ is achieved.

The various terms in Eq.~\ref{diff.eq.} are defined as follows:

\begin{itemize}

\item $q(r,p)$ represents the spherically symmetric source term, which defines the electron production rate per unit volume per unit momentum range.  The current public version of GALPROP does not offer the option to include a DM annihilation source.
For this reason, we modified the source code to introduce a DM source of CRs, by injecting CRs according to the following equation:
\begin{equation}
	q(r,p) = \frac{1}{2} \langle \sigma v \rangle \left(\frac{\rho_{\rm DM}(r)}{m_{\chi}}\right)^2 \xi_{\rm DM}(r) \frac{dN_e}{dp}(p),
	\label{eq:q}
\end{equation}
where $\left\langle \sigma v \right\rangle$ denotes the WIMP annihilation cross section, $\rho_{\rm DM}(r)$ is the DM density distribution defined by Eq.~\ref{rho_DM}, $m_{\chi}$ is the WIMP mass, $\xi_{\rm DM}(r)$ is the DM annihilation rate boost factor due to substructures taken from \cite{Kamionkowski:2010mi}, and $\frac{dN_e}{dp}(p)$ represents the energy spectrum of electrons per annihilation (from the PPPC4DMID package \cite{PPPC}).  
As the standard astrophysical foregrounds are accounted for in our analysis using existing templates for microwave data analysis described in \S\ref{sec:Fitting}, we do not include any astrophysical (non-DM) sources of CRs in the GALPROP model.

\item $D_{xx}(p)$ is the spatial diffusion coefficient (assumed to be isotropic and spatially-independent):
\begin{equation}\label{D_xx}
	D_{xx}(p) = D_0 \left(\frac{p}{p_0}\right)^{\delta},
\end{equation}
where $p_0$ is set to the constant value 1 GeV/c (e.g., \cite{Fornengo:2011iq}). The choice of $D_0$ and $\delta$ constants is explained below. 

\item $D_{pp}(p)$ is the electron reacceleration coefficient. Particle reacceleration in the interstellar medium happens due to interactions with MHD waves. $D_{pp}(p)$ is defined here through the spatial diffusion coefficient (Eq.~\ref{D_xx}), see \cite{Strong:1998pw} for further details:
\begin{equation}\label{D_pp}
	D_{pp}(p) D_{xx}(p) = \frac{4 p^2 v_a^2}{3 \delta (4-\delta^2) (4-\delta) w},
\end{equation}
where $v_a$ is the Alfven speed in the media and $w$ is the ratio of MHD waves energy density to MF energy density, assumed to be 1 in GALPROP\@.

\item $\dot{p} \equiv dp/dt$ is the electron momentum loss rate due to various cooling processes: synchrotron radiation (computed using the MF distribution Eq.~\ref{MF}), emission from ICS on the interstellar radiation field (composed by starlight, CMB, and dust emission photons), bremsstrahlung radiation on ionized gas, and others.  We refer to reader to \cite{GP} for further details.

\end{itemize}

We run the version of GALPROP that is implemented in 3 spatial dimensions.  For the spatial dimensions we choose a resolution of 200 pc, which is sufficient for our purposes as our region of interest is somewhat far from the GC and therefore large variations in emissivity are not expected within this box size.  Reducing the box size to 100 pc varies the emission by $\approx (1-10)\%$, at the cost of highly increasing computing time.  Our implementation of the DM source in GALPROP was validated by comparison with similar calculations in the literature that were obtained by a variety of methods~\cite{Fornengo:2011iq, Hooper:2010im, Malyshev:2010xc}. Our GALPROP version with DM can be downloaded for a public use at \cite{GP_DM}.

\subsection{Diffusion model}
\label{sec:diffusion}

For the choice of the propagation parameters in Eqs.~\ref{diff.eq.}--\ref{D_pp}, to begin we followed the  canonical MIN/MED/MAX paradigm which has been used extensively in earlier works (see, e.g., \cite{Cirelli:2010xx, Fornengo:2011iq, Mambrini:2012ue}). These three diffusion parameter configurations were derived from various CR data sets.  MED is an average ``best-fit'' configuration, while MIN and MAX are intended to represent possible extremes. Because recent CR data seem to reliably exclude the MIN model \cite{Lavalle:2014kca}, we discarded this case and considered instead a newly proposed model derived in \cite{Trotta:2010mx}, which includes CR reacceleration effects in the fit. 
Specifically, we used the posterior mean parameter values from Table 2 of \cite{Trotta:2010mx}. 

Recalling that we consider two values for the MF vertical scale height $z_B$ (referred to as $z_{B1}$ and $z_{B2}$), which is related to $h$, we now define combined MF and propagation models.  In principle we would consider both values of $z_B$ with the three propagation models described above, however we discovered empirically that the intensity for the two $z_B$ values is very similar for the MAX propagation configuration, and so we use only the $z_{B1}$ case with the MAX model.  This results in five $z_B$/propagation models.  Table~\ref{tab:MF/prop} summarizes the parameter values for these five configurations.

For each of these five models, we consider the two central MF values mentioned previously, for a total of ten MF/propagation models.  These are meant to represent current systematic uncertainties on this aspect of the analysis.  For each of these models, we calculate the emission from each of the DM models described in \S\ref{sec:DMDensityDistr}
and \S\ref{sec:dmspect}.
In \S\ref{sec:DMTheoResults} we show the dependence of the DM emission intensity on the MF and propagation configuration.

\begin{table}[tbp]
\caption{Diffusion parameter values for the electron propagation (which enter Eqs.~\ref{diff.eq.}--\ref{D_pp}) and MF vertical scale heights (which enter Eq.~\ref{MF}) used in our work.}
\label{tab:MF/prop} 
\centering
\begin{tabulary}{1\textwidth}{|C|C|C|C|}
\hline
Parameter & MED & MAX & Reacc \\
\hline
Half-height of the diffusion box $h$, kpc & 4.0 & 15 & 5.4 \\
Diffusion coefficient normalization $D_0$, cm$^2$/s & $3.4\cdot 10^{27}$ & $2.3\cdot 10^{28}$ & $5.4\cdot 10^{28}$ \\
Diffusion coefficient energy dependence power $\delta$ & 0.70 & 0.46 & 0.31 \\
Alfven speed in the intragalactic media $v_a$, km/s & 0 & 0 & 38 \\
MF vertical scale height, version 1 $z_{B1} = \delta \cdot h$, kpc & 2.8 & 6.9 & 1.7 \\
MF vertical scale height, version 2 $z_{B2} = 0.5 \cdot h$, kpc & 2.0 & - & 2.7 \\
\hline
\end{tabulary}
\end{table}

 \S\ref{sec:MF} described the vertical sizes of the rectangular 3D diffusion box; for the horizontal size, we set the box edges to be $+/-$15 kpc along both $x$- and $y$-axes to approximately enclose the Galactic disc (see e.g. \cite{Fornengo:2011iq}).  Boundary conditions were set to the free escape mode.  For the energy boundaries, we take 0.01 to 200 GeV.  The lower boundary is mainly defined by the fact that the electrons with smaller energies have no contribution to the synchrotron radiation at the frequencies of interest. The upper boundary has to contain the most energetic electrons produced by our DM source. One might naively assume that they would approximately correspond to the largest DM particle rest mass involved ($\sim$~50 GeV),  however particle reacceleration (as in our Reacc model) can in principle raise the electron energies much beyond $m_{\chi} c^2$.  We empirically found that 200 GeV is an effective upper limit on the energies the electrons attain in our cases. 

Further details on all parameters we adopted in the GALPROP runs can be requested from the authors.

\subsection{DM emission maps}
\label{sec:DMTheoResults}

We implemented the modeling described in the previous sections in GALPROP and produced multi-frequency maps for all DM models needed to cover the parameter space. In total, we computed 300 parameter configurations, which are constituted by 5 different WIMP masses over the range of interest (we interpolated our results between the computed points), 2 annihilation channels ($b\bar{b},~\tau^+ \tau^-$), 3 profile slopes (see \S\ref{sec:DMDensityDistr}), and 10 MF/propagation models (see \S\ref{sec:MF} and \S\ref{sec:diffusion}). The maps have a resolution of $\approx 0.5\degree$ (HEALPix \cite{Healpix} resolution $N_{side}=128$) and are produced at  the WMAP and Planck frequencies of 23, 28, 33, 41, 44, 61, and 70 GHz.

\begin{figure}[t]
\begin{center}
\includegraphics[width=0.495\textwidth]{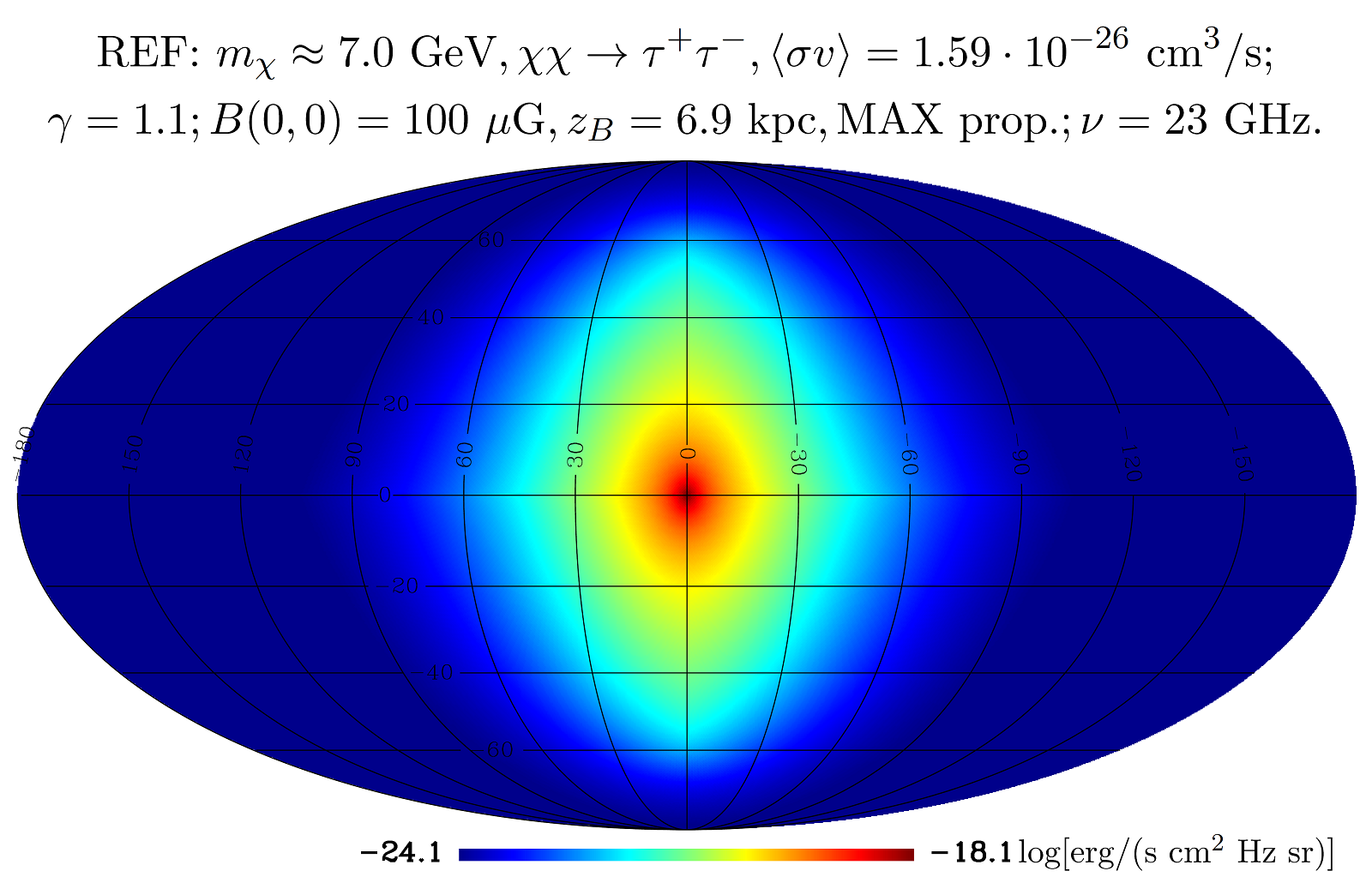}
\includegraphics[width=0.495\textwidth]{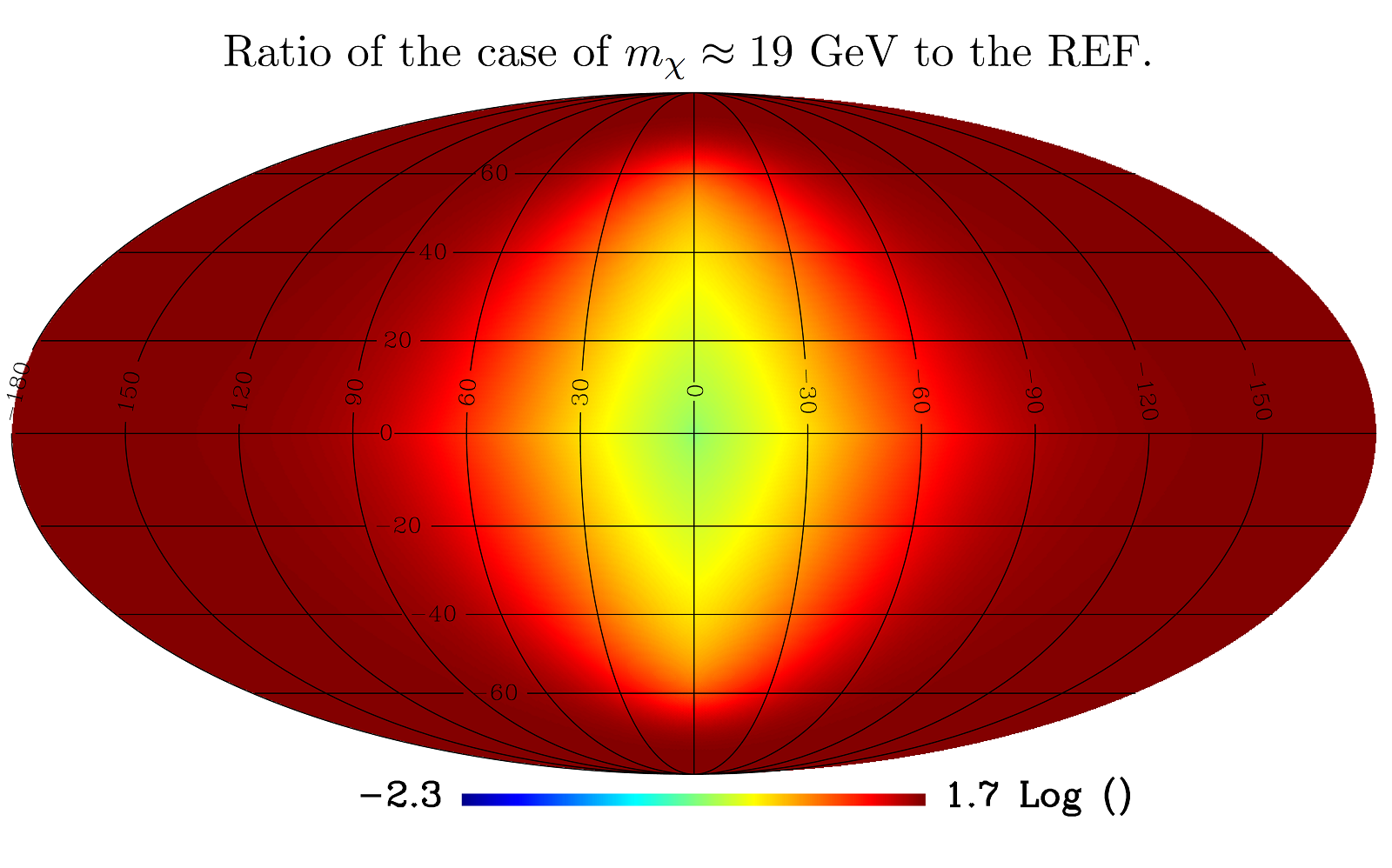}
\includegraphics[width=0.495\textwidth]{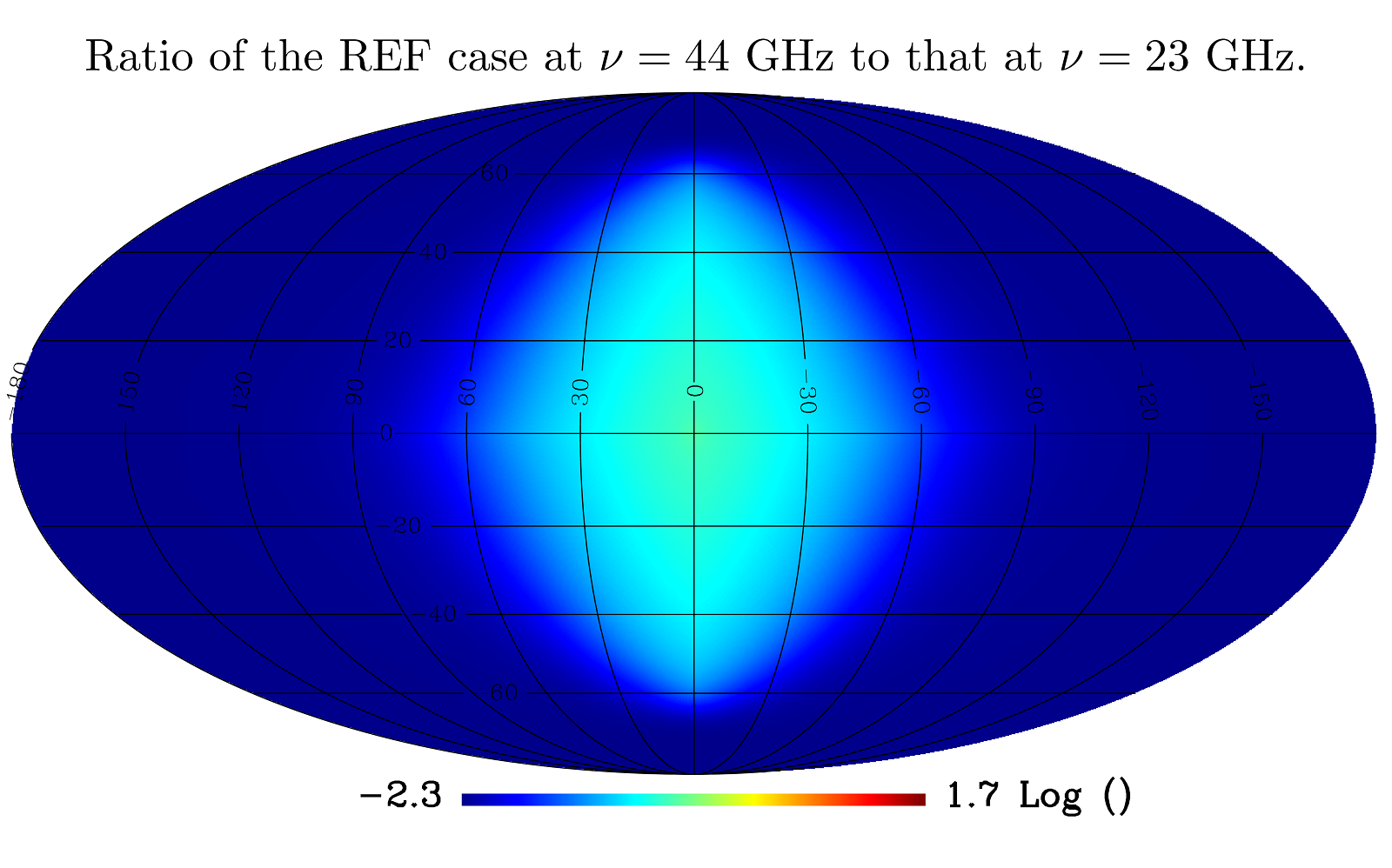}
\includegraphics[width=0.495\textwidth]{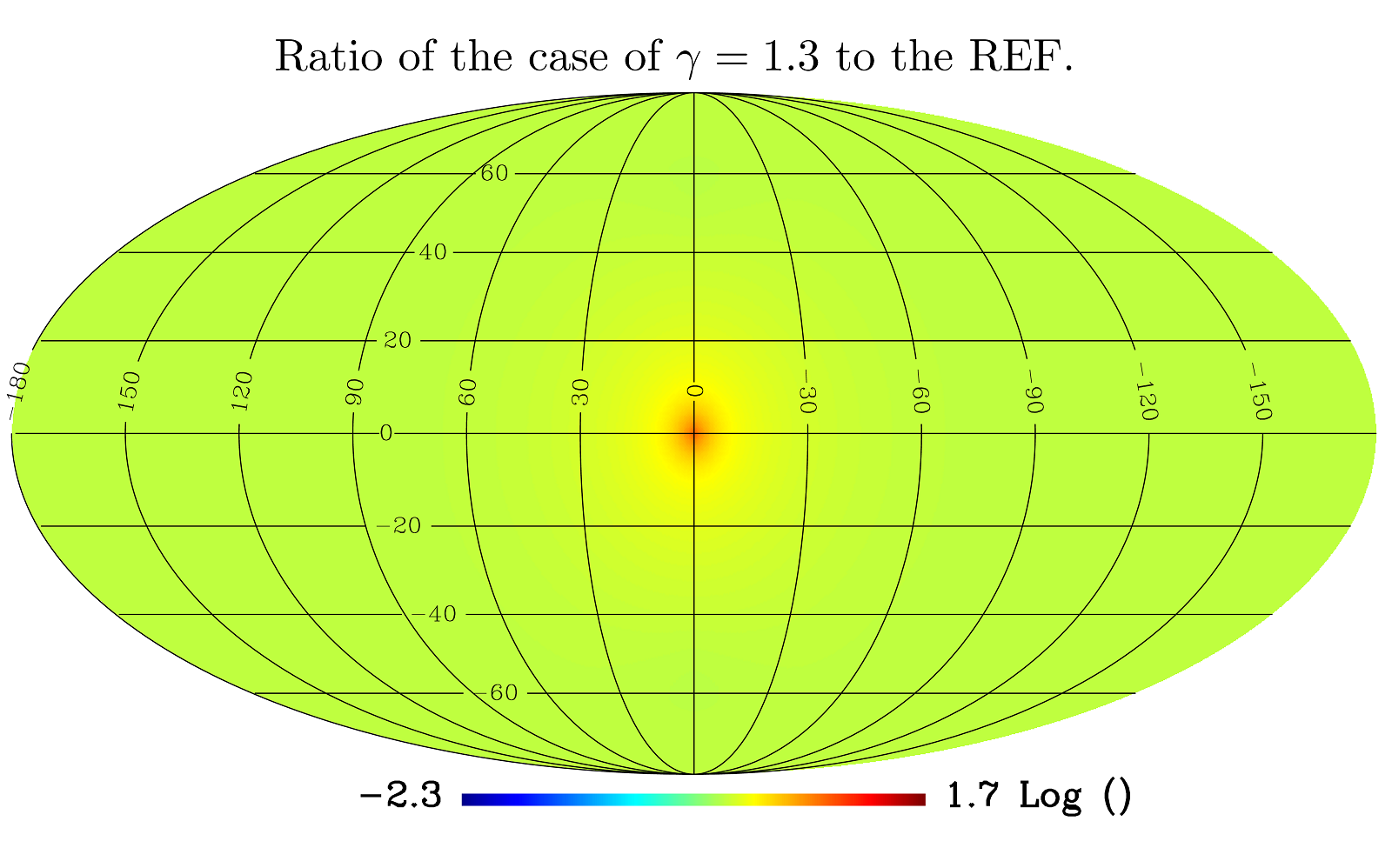}
\includegraphics[width=0.495\textwidth]{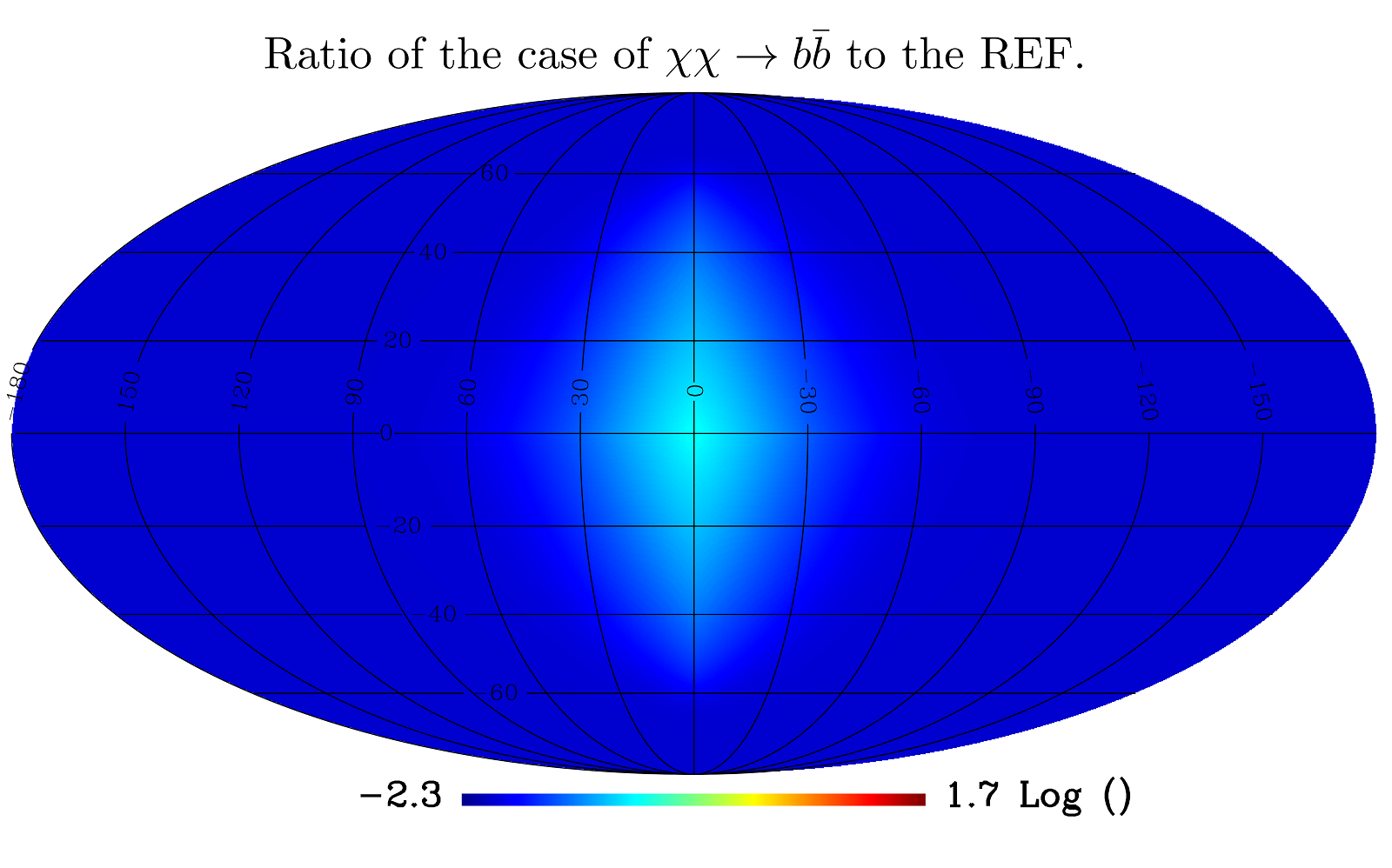}
\includegraphics[width=0.495\textwidth]{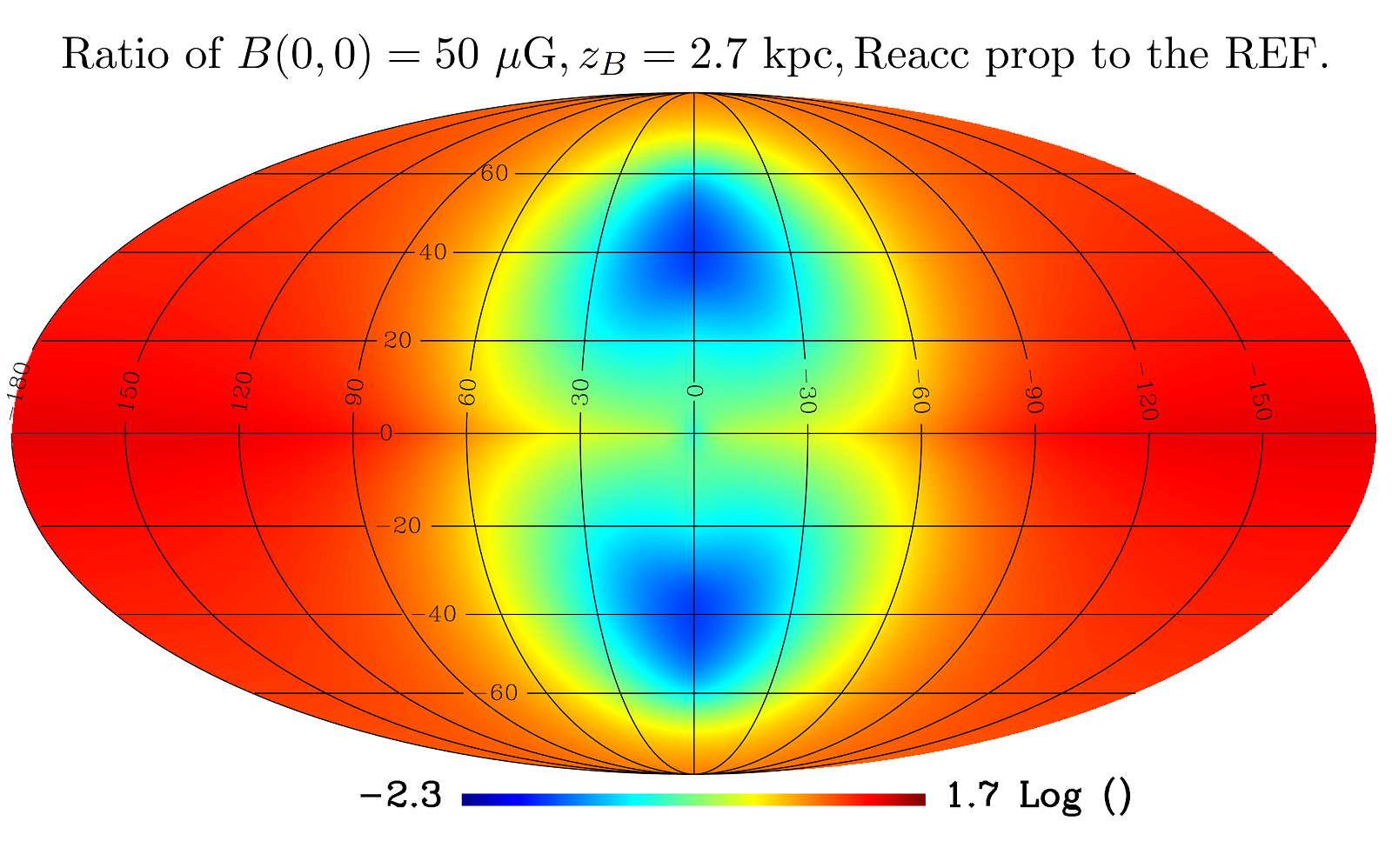}
\caption{(COLOR ONLINE FOR ALL FIGURES) Examples of DM emission intensity maps in Galactic coordinates, at the frequencies indicated. The top left map represents a reference DM parameter configuration (``REF'').  Specific parameters of the configuration are given above the map. Other maps show the difference with respect to the REF map when the model parameters are changed as noted. The difference is shown as the ratio $I_x(b,l)/I_{\rm REF}(b,l)$, where $I_x$ is the intensity of the map with one of parameters being changed and $I_{\rm REF}$ is the intensity of the REF map. To illustrate dependence on MF/propagation we changed the relevant parameters. For more details see \S\ref{sec:DMTheoResults}.}
\label{fig:dm_maps}
\end{center}
\end{figure}

Fig.~\ref{fig:dm_maps}, top left panel, shows the DM intensity map for the reference model of the data (``REF''); this is also the best-fit configuration found in \S\ref{sec:Fitting}. The figure shows an intensity distribution around the GC that is highly concentrated, spanning six orders of magnitude between the center and anti-center.  The other panels of Fig.~\ref{fig:dm_maps} show examples of the dependence of the emission on various parameters in terms of the ratio to the REF case.

We comment now on some general features.  The difference between the maps typically grows with distance from the GC\@.  This is seen when changing the DM particle mass (top right), and also interestingly, for the frequency dependence (middle row, left). However the steeper DM density profile map ($\gamma=1.3$ vs $\gamma=1.1$ for the REF model) differs noticeably only very close to the GC (middle row, right); this indicates that the profile slope does not have an important impact on the emission at higher Galactic latitudes. The $b\bar{b}$ annihilation channel produces significantly smaller intensities than $\tau^+ \tau^-$ one,  everywhere on the sky, as expected (bottom left).  A more conservative MF/propagation model (bottom right) differs non-trivially from the REF model: the intensity is much smaller around the center and much larger around the anti-center. This can be primarily attributed to an overall smaller MF,  which generates weaker synchrotron around the GC and  cools electrons less efficiently, allowing them to diffuse further while still emitting.

\section{Regions of interest on the sky}
\label{sec:roi}

For each of the two approaches to testing a DM contribution to the data, we selected a ROI in the sky on which to perform the analysis.  For the first approach, in which we derive conservative limits by assuming the total data is composed of only the CMB and a DM component, we chose a ROI inside which we compute the average intensity at every frequency, to be compared to the analogous quantity as derived from DM maps.

\begin{figure}[t]
\begin{center}
\includegraphics[width=0.495\textwidth]{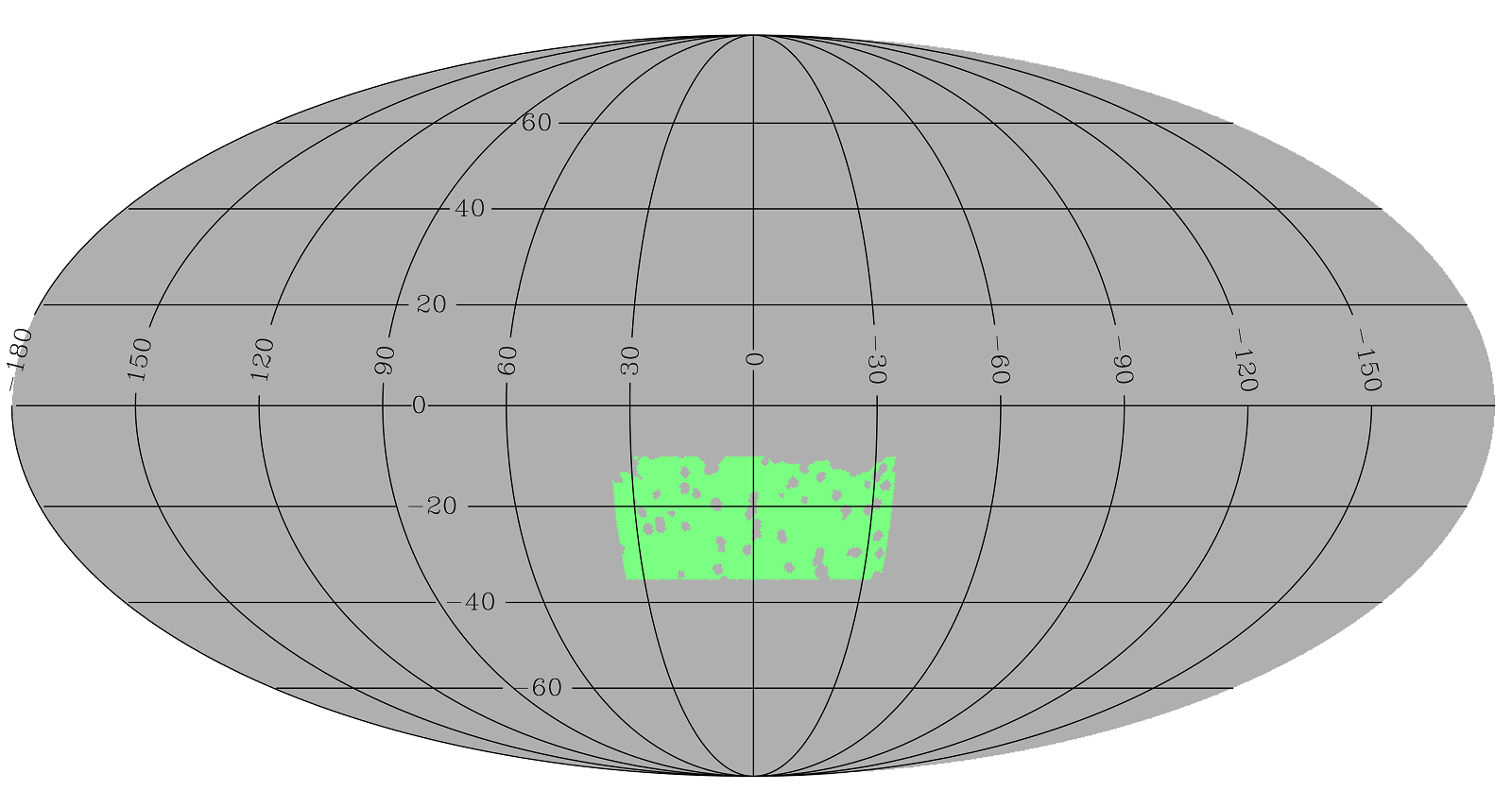}
\includegraphics[width=0.495\textwidth]{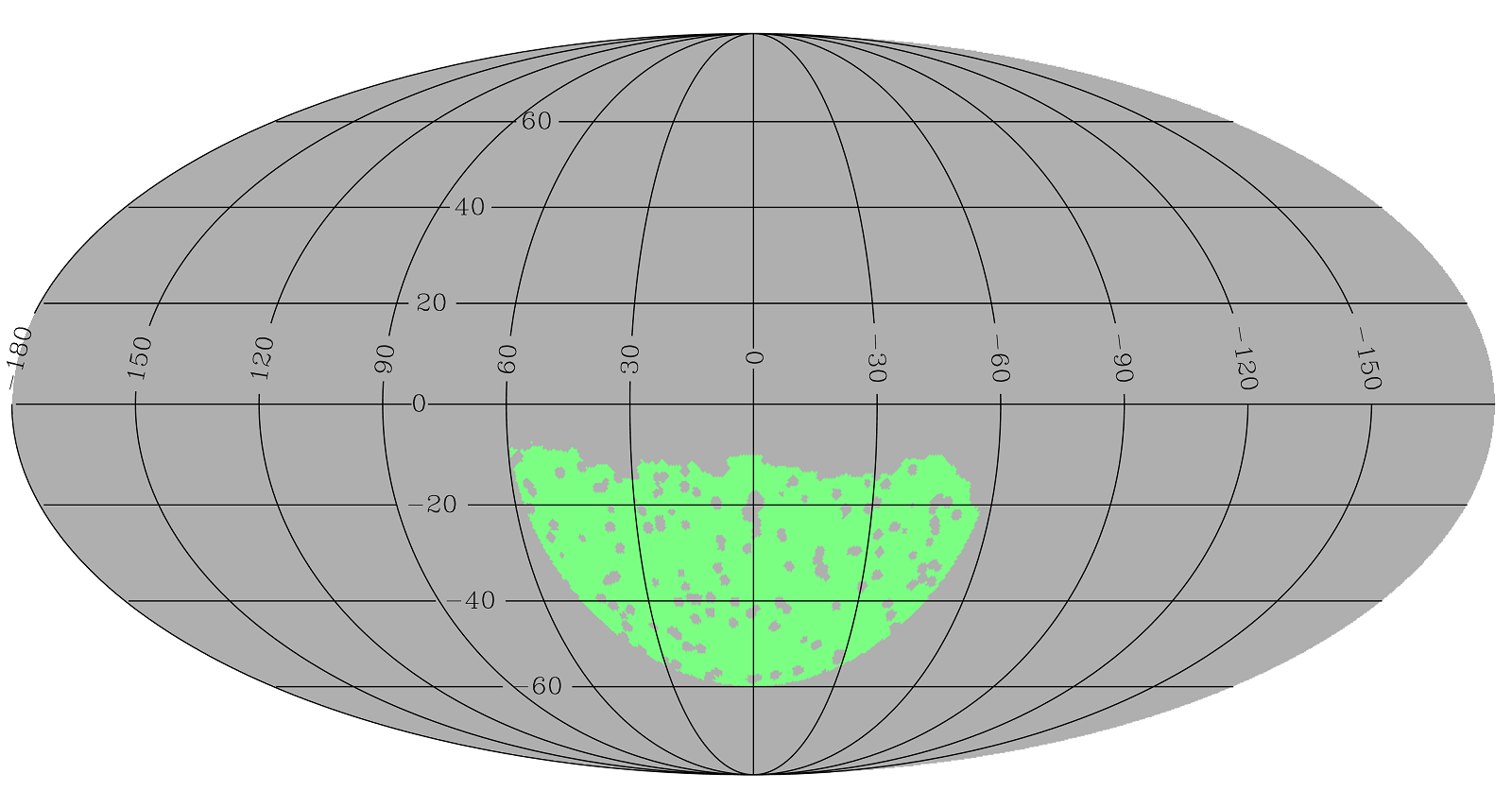}
\caption{\textit{Left:} Our ROI with masked compact sources chosen to produce the conservative DM constraints with CMB subtraction only. It was constructed as the rectangle $-35\degree \leq b \leq -10\degree$, $|l| \leq 35\degree$.  \textit{Right:} Our ROI with masked compact sources chosen for the regional fitting in the component separation procedure (see more in \S\ref{sec:Fitting}). Before point source masking it was constructed as a half-disc of $60\degree$ radius around GC, with very bright areas around the Galactic plane also masked. This ROI is also used in the presentation of example DM emission spectra in \S\ref{sec:dmspectra}.}
\label{fig:masks}
\end{center}
\end{figure}

The ROI choice was based on the Planck collaboration's study of the Haze~\cite{:2012rta}, 
and consists of a  rectangular area in  Galactic coordinates $-35\degree \leq b \leq -10\degree$, $|l| \leq 35\degree$. This choice is well-motivated for a DM search, since that ROI is sufficiently far from the bright foreground emission in the Galactic plane but still in a region where the DM emission is relatively strong. 
We also masked bright compact sources inside the ROI using the product of the WMAP and Planck source masks for each channel, smoothed to 1$\degree$~\cite{maskwmap,maskplanck}. Our final ROI for the conservative analysis is shown in Fig.~\ref{fig:masks} (left panel).

For the component separation analysis in which we determine the contribution to the data from DM by modeling Galactic foregrounds and the Bubbles, we adopt a slightly different ROI\@.
The  ROI  for this analysis is shown in right panel of Fig.~\ref{fig:masks}. It is a 60$\degree$ semi-disc centered on the GC, with bright compact sources masked as in the first ROI, and also strong Galactic plane emission masked~\cite{maskwmap}. The choice of 60$\degree$ addresses additional challenges for this approach, namely avoiding strong foreground degeneracies due to using too small an area while still selecting an area in which each foreground has a spatially-uniform frequency dependence and including the relevant Haze emission.

\section{DM microwave spectra in the Haze region}
\label{sec:dmspectra}

Considering the ROIs we have defined, it is useful to examine the  frequency dependence of the DM component from our simulated maps.  For this purpose, we consider the larger ROI used later in the component fitting analysis in \S\ref{sec:Fitting}, shown in the right panel of Fig.~\ref{fig:masks}.  Inside this ROI we average the DM intensity and plot it versus frequency of emission in
Fig. \ref{fig:dm_sp}.  In general a simple power law is not sufficient to describe the spectra, as a mild spectral index evolution is observed within the frequency range considered.  Intensities for the $\tau^+ \tau^-$ channel are typically five to ten times larger than those for $b\bar{b}$, for the same WIMP mass, density profile and MF/propagation model.

In addition to the amplitude of the emission, the slope of the spectrum strongly depends on the DM particle mass.  Approximating  the spectra by a simple power law $I \sim \nu^{\beta}$, we found that for the $b\bar{b}$ case $\beta$ grows with mass over the range $-3.3 \lesssim \beta \lesssim -1.6$. For $\tau^+ \tau^-$ the variation is even slightly larger: $-3.2 \lesssim \beta \lesssim -1.2$.  Essentially no dependence of the intensity or slope on the inner profile slope $\gamma$ is observed, as expected from inspecting the relevant maps in Fig.~\ref{fig:dm_maps}.

\begin{figure}[H]
\begin{center}
\includegraphics[width=0.495\textwidth]{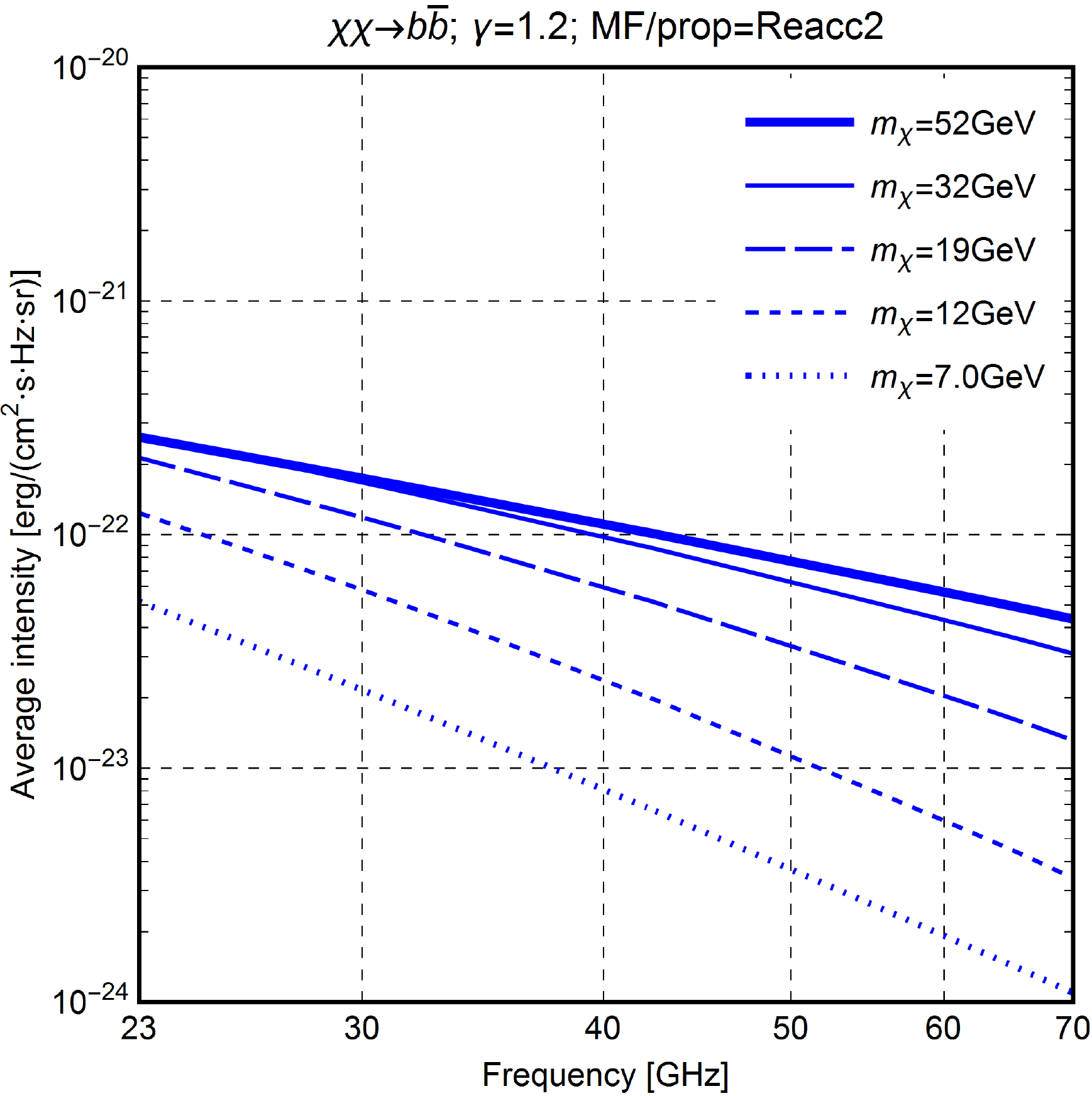}
\includegraphics[width=0.495\textwidth]{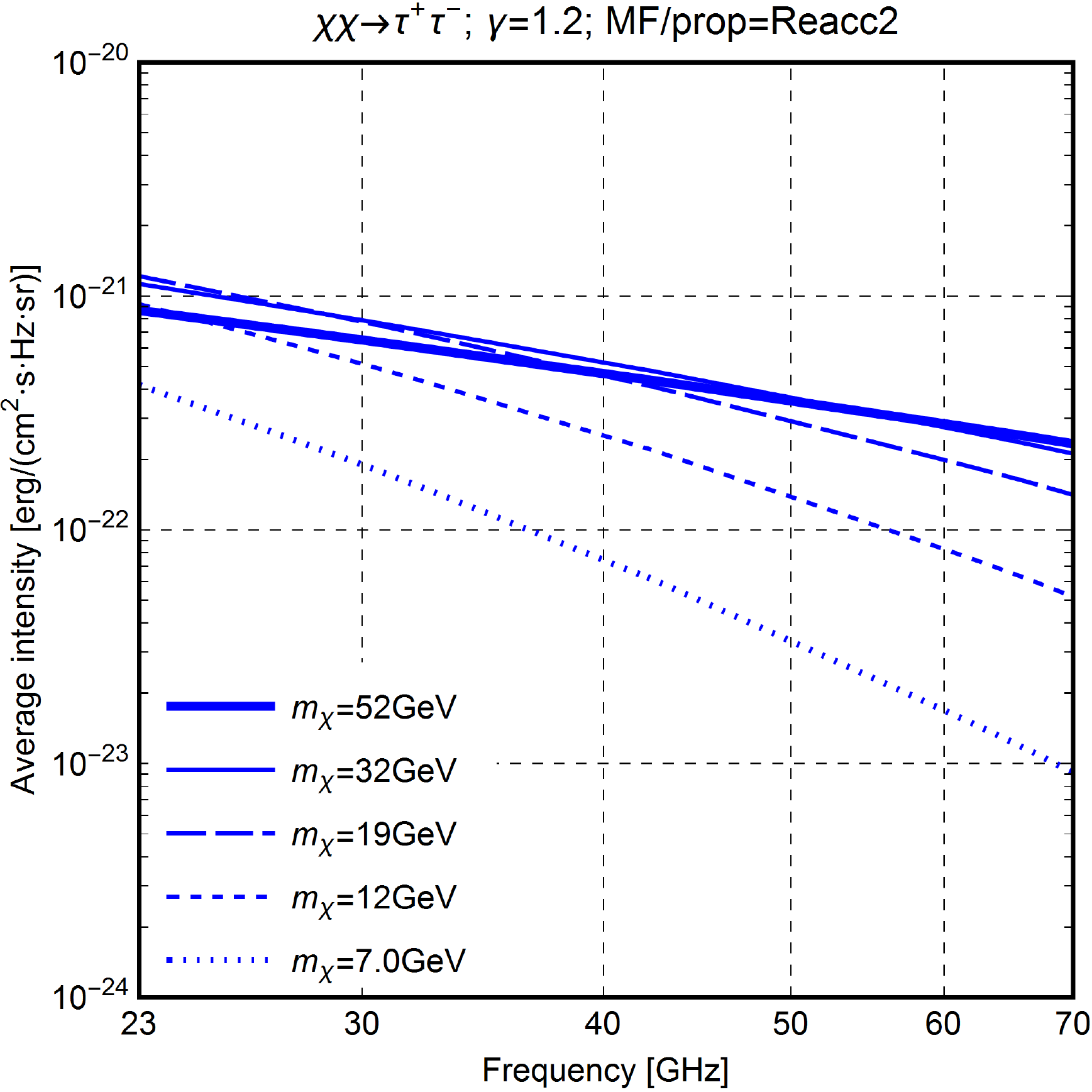}

\vspace*{0.25cm}

\includegraphics[width=0.495\textwidth]{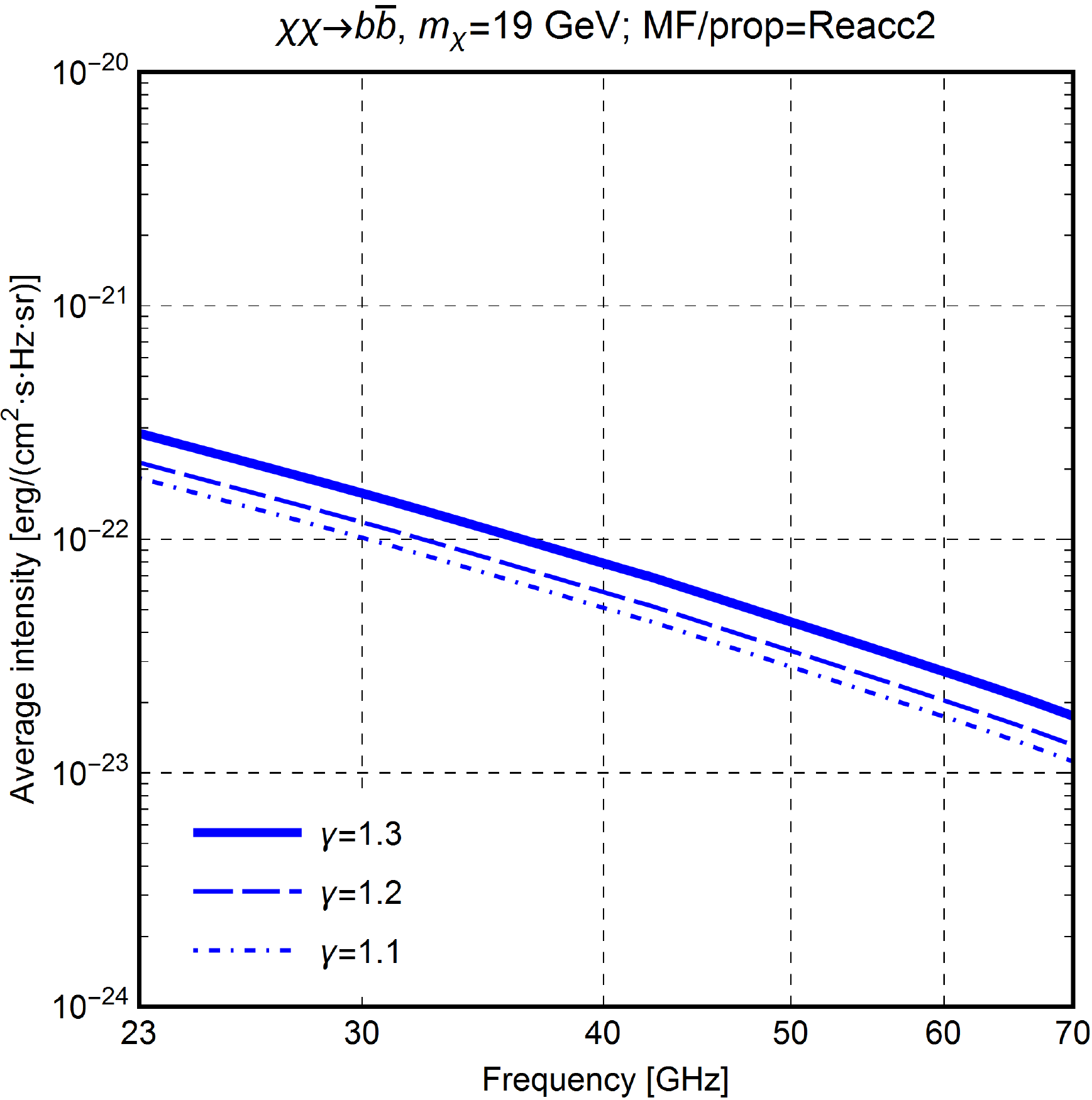}
\includegraphics[width=0.495\textwidth]{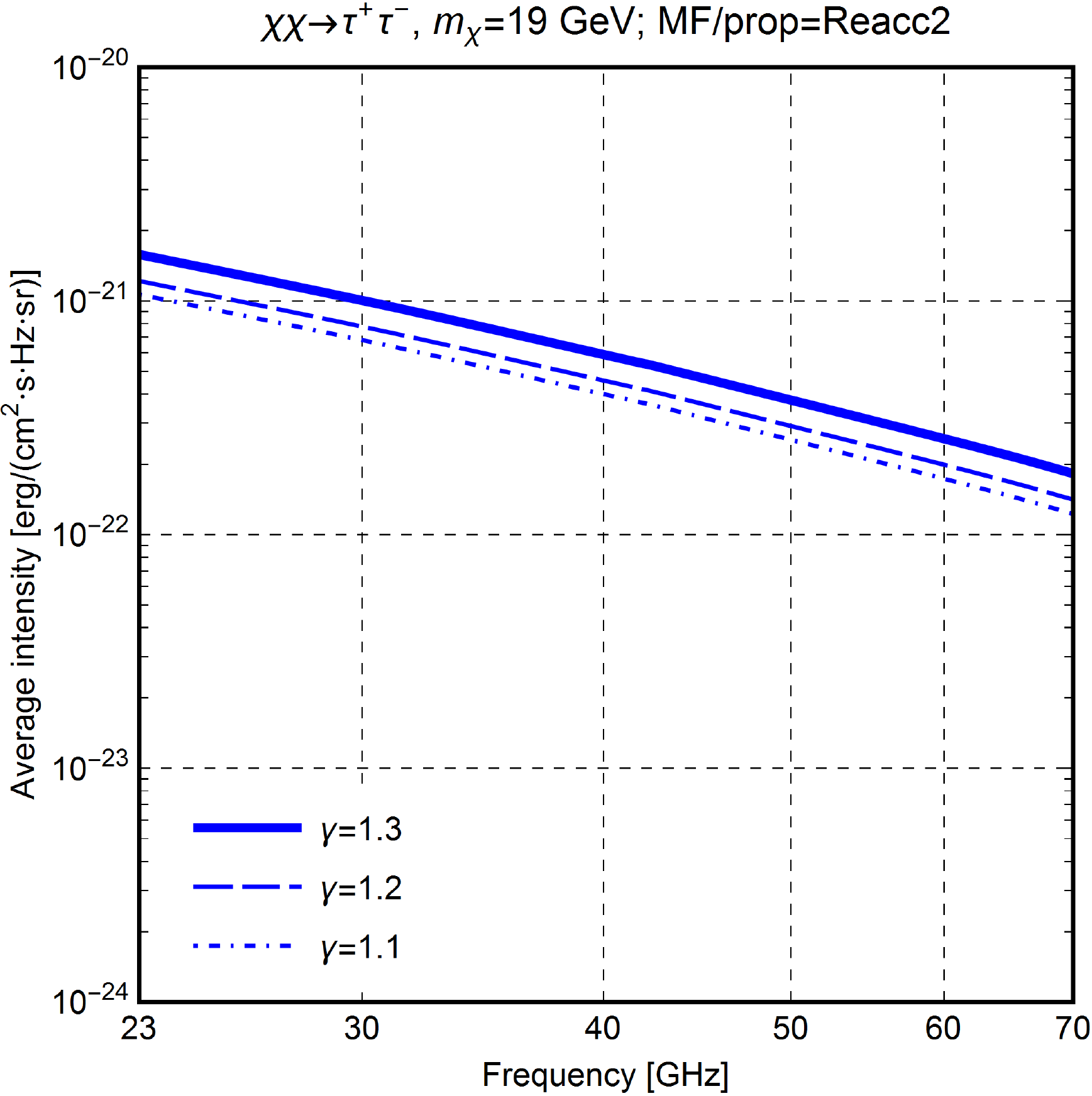}

\vspace*{0.25cm}

\includegraphics[width=0.495\textwidth]{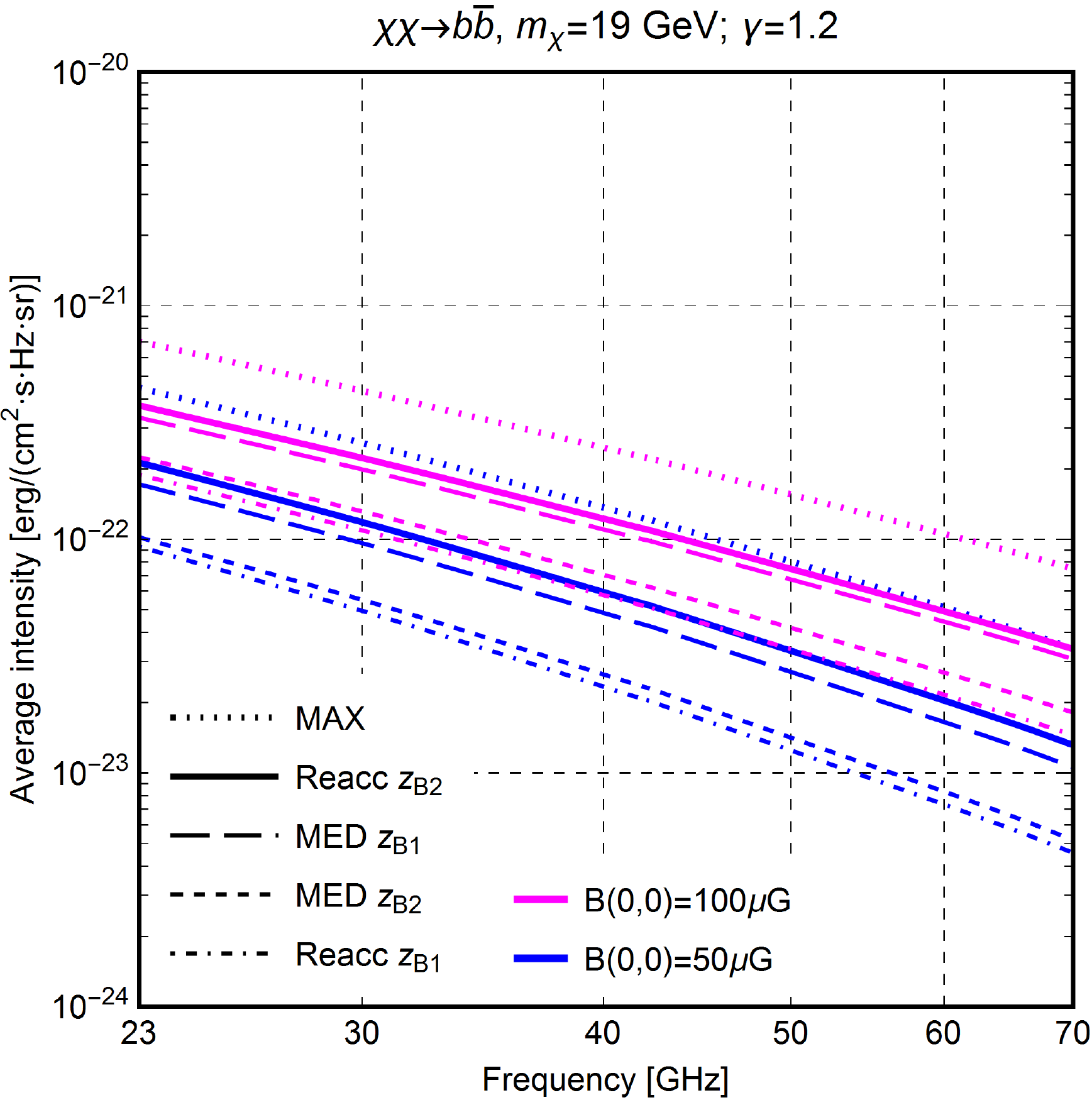}
\includegraphics[width=0.495\textwidth]{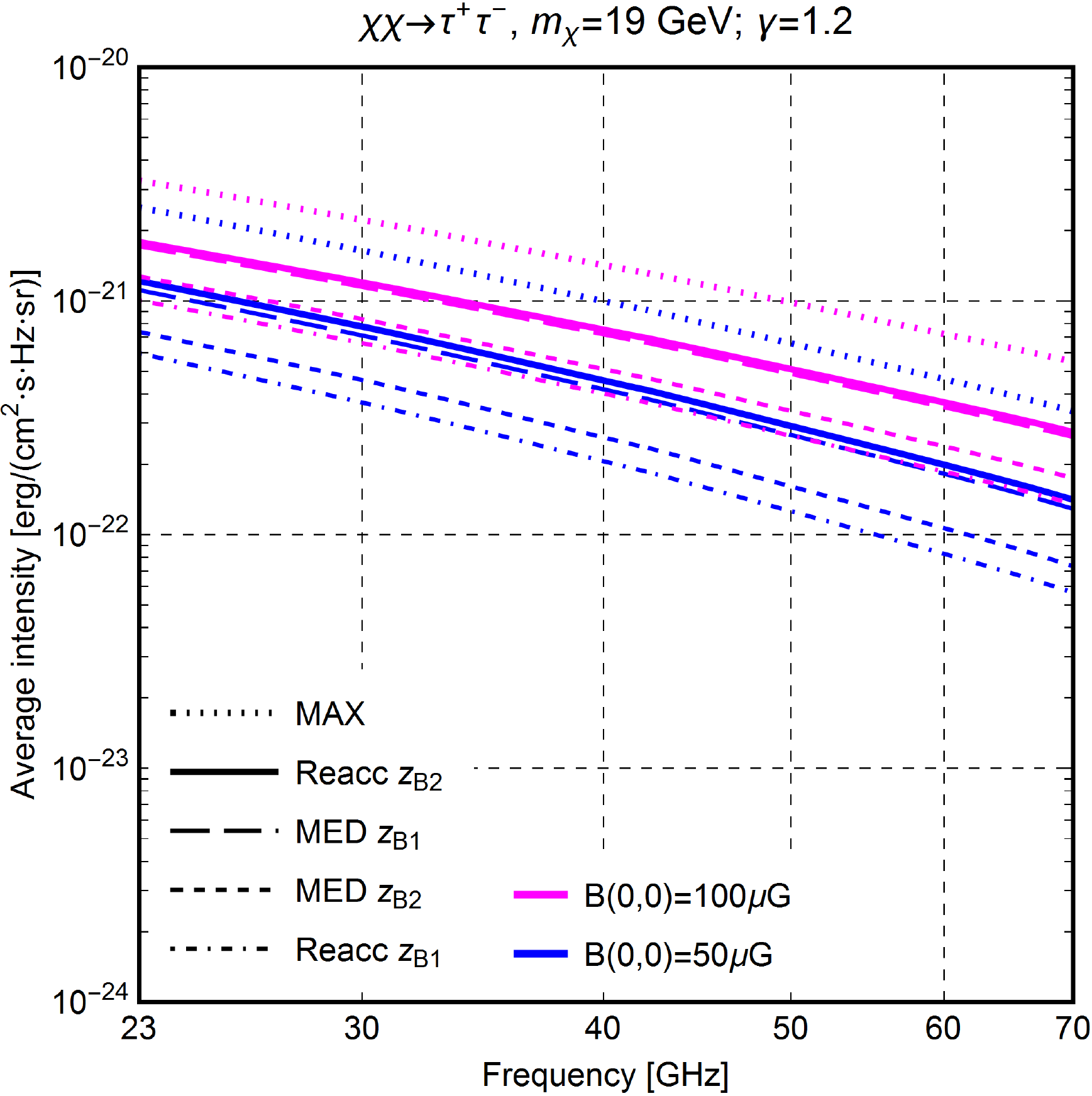}
\end{center}
\end{figure}
\begin{figure}[H]
\caption{\label{fig:dm_sp}  \textit{(preceding page)} Dependence of the DM emission spectrum on various model parameters. The ROI over which the DM intensity was averaged is shown in the right panel of Fig.~\ref{fig:masks}.  Left column in this Figure represents the $b\bar{b}$ annihilation channel, while the right column shows results for the $\tau^+ \tau^-$ channel. The annihilation cross section $\langle \sigma v \rangle = 3 \cdot 10^{-26}$ cm$^3$/s in all panels.  Points were calculated from the DM maps at each frequency and then interpolated.  The first row shows the spectral dependence on the DM mass, the second row shows the dependence on the inner profile slope $\gamma$, and the third row shows the dependence on the MF and propagation model.}
\end{figure} 

As for the MF and propagation model, higher MFs produce larger intensities. The spectral slope seems to be largely independent of the MF and propagation models, however the overall amplitude strongly depends on the propagation model, varying by up to one order of magnitude.
As expected, configurations with higher vertical MF extent $z_B$ produce larger emission. For the same $B(0,0)$ and $z_B$, models with reacceleration generate slightly larger intensities than those without reacceleration. Such inferences about DM spectra have interesting implications for a comparison with real sky data. In particular, the Planck team claimed in \cite{:2012rta} that the average spectral index of synchrotron emission from electrons from Galactic astrophysical sources is $\approx -1.1$. The anomalous microwave excess around GC, on the other hand, has a very different index $\approx -0.6$. This is one of the main motivations for the suggestion that the Haze comes from a distinct population of electrons, such as from DM annihilation. Comparing the DM indexes estimated above with the ones derived for the Haze and the synchrotron component of the WMAP and Planck data, we notice that while DM is able to generate an arbitrary  amplitude for the intensity,  the spectrum of the emission has a limited range of slopes (at least for DM masses of interest). Therefore, the DM component is only marginally able to mimic / contribute to regular synchrotron (for the largest WIMP masses considered), and at the same time its spectrum is too steep  to  fit the excess, which has $\beta \approx -0.6$. These simple considerations suggest that, even in the case that a DM signal is present, it is likely that another mechanism contributes the majority of the Haze signal. This was one of the motivations for us to include a possible counterpart of the Fermi Bubbles in the component separation procedure presented in \S\ref{sec:Fitting} too. Indeed, higher WIMP masses $\sim 100$ GeV may potentially provide the spectral slopes needed for the Haze. However, in this work we intentionally restricted our analysis to only those DM models which fit the gamma-ray excess.

\section{Conservative constraints on DM annihilation}
\label{sec:cmb_constraints}

In this section we present conservative DM constraints derived from the Planck sky maps when the emission is assumed to be composed only of  CMB emission and DM emission, with no other foregrounds  considered. Since the intensity of the DM emission is linearly proportional to the annihilation cross section $\langle \sigma v \rangle$, an upper limit on the total observed non-CMB emission at a given frequency can be directly converted into an upper limit on $\langle \sigma v \rangle$ for all WIMP masses of interest.
In performing this estimate, we consider only the CMB emission as it is very bright and well characterized and thus very simple to  model extremely accurately.
 
For this analysis, we chose to use Planck Low Frequency Instrument (LFI)  channels only (28, 44 and 70 GHz) as the Planck noise level is far superior to that of WMAP while the frequency range covered is comparable.

The Planck 2013 maps that have been released have the monopole and dipole largely subtracted, but, after careful component separation, an estimate of the residual monopole and dipole in the maps was made and published by the Planck team. These residual values with their errors are shown in Table~2 of \cite{Ade:2013hta}. We performed subtraction of the quoted residual monopole and dipole values from all three Planck LFI channels. We then computed the average intensities in each frequency map. The impact of subtracting residual monopole and dipole on the average intensity is shown in Table~\ref{tab:cmb-noise}.  The first line of this table shows the average intensity inside our ROI from Fig.~\ref{fig:masks} (left panel)  without any map processing. The second and third lines of Table~\ref{tab:cmb-noise} show the ROI intensities after subtraction of the residual monopole and  monopole+dipole respectively. These have a significant impact on the 70 GHz channel in particular, where the intensity decreases by about a factor of two.

\begin{table}[tbp]
\caption{Average intensities in our ROI from the Planck LFI temperature maps and their associated uncertainties (all in units of erg/(cm$^2$ s sr Hz), converted from CMB temperature units using: $\Delta I (\nu) = \left. \frac{\partial B_{bb}(\nu,T)}{\partial T} \right|_{T=2.726K} \Delta T (\nu)$).  These were used to derive the intensity upper limits from the data for the subsequent derivation of DM constraints. See \S\ref{sec:cmb_constraints} for more details.}
\label{tab:cmb-noise}
\centering
\begin{tabulary}{1\textwidth}{|C|C|C|C|}
\hline
Map & 28 GHz & 44 GHz & 70 GHz \\
\hline
Nominal Planck LFI maps & $3.69 \cdot 10^{-20}$ &	$2.38 \cdot 10^{-20}$ &	$3.79 \cdot 10^{-20}$ \\
Maps after residual monopole subtraction & $3.50 \cdot 10^{-20}$ &	$2.27 \cdot 10^{-20}$ &	$1.79 \cdot 10^{-20}$ \\
Maps after residual mono/dipole subtraction & $3.64 \cdot 10^{-20}$ &	$2.27 \cdot 10^{-20}$ &	$1.79 \cdot 10^{-20}$ \\
Standard deviation of residual monopole & $4.72 \cdot 10^{-22}$ &	$5.66 \cdot 10^{-22}$ &	$1.33 \cdot 10^{-21}$ \\
Standard deviation of residual dipole & $6.17 \cdot 10^{-22}$ &	0 &	0 \\
Standard deviation of CMB fluctuations & $6.00 \cdot 10^{-21}$ &	$1.44 \cdot 10^{-20}$ &	$3.37 \cdot 10^{-20}$ \\
Total standard deviation & $6.05 \cdot 10^{-21}$ &	$1.44 \cdot 10^{-20}$ &	$3.38 \cdot 10^{-20}$ \\
\hline
\end{tabulary}
\end{table}

Once the monopole and dipole were subtracted and the average intensities in the ROI were computed, we proceeded to estimate uncertainties due to Planck noise, errors in the residual monopole and dipole, and CMB fluctuations. The instrumental noise level was estimated using the maps provided as a part of the Planck data products, and it was found to be largely subdominant compared to the other uncertainties.
The monopole and dipole uncertainties were taken from Table~2 of \cite{Ade:2013hta} and assumed to be the standard deviations of respective Gaussian distributions.  Numerical values for the monopole and dipole ``noise'' (denoted  $\sigma_{\nu}^{\rm mon}, \;\sigma_{\nu}^{\rm dip}$) are shown in the fourth and fifth line respectively of Table~\ref{tab:cmb-noise}. The dipole-induced uncertainty slightly exceeds the monopole uncertainty  for 28 GHz and does not exist for other channels by construction as it was not modeled in the Planck team fitting.
 
The most important contribution to the intensity uncertainty comes from the CMB temperature fluctuations, which we take to be an unknown realization of the underlying cosmological model, in our specific ROI\@. 
Here we aim at deriving the standard deviation $\sigma_{\nu}^{\rm CMB}$ of the mean ROI intensity $I = 1/n \sum\limits_{i=1}^n T_i$, where $n = 6427$ denotes the number of map pixels in our ROI at our resolution and $T_i$ is the intensity (or, equivalently, thermodynamic temperature) value in each pixel $i$. 

The variance of $I$ was calculated as follows:
\begin{equation}\label{eq:var(T)}
	\langle (I-\langle I\rangle )^2 \rangle = \langle I^2 \rangle = \langle \frac{1}{n^2}\sum\limits_{i=1}^n T_i \sum\limits_{j=1}^n T_j \rangle = \frac{1}{n^2}\sum\limits_{i,j=1}^n \langle T_i T_j \rangle = \frac{1}{n^2} \sum\limits_{i,j=1}^n \sum\limits_{l} \frac{2l+1}{4\pi} C_l B_l^2 W_l^2 P_l(\cos \theta_{ij}).
\end{equation}
Here the brackets $\langle \rangle$ mean averaging effectively over all possible realizations of the underlying cosmological model in an area that matches our ROI\@. When considering CMB fluctuations,  $\langle I \rangle = 0$. The well-known expression for $\langle T_i T_j \rangle$ in terms of  the CMB power spectrum $C_l$  can be found in, e.g., Eq.~(18.29) in \cite{Peacock}.

The beam  $B_l$ and pixel window function  $W_l$ in Eq.~\ref{eq:var(T)} account for
the smoothing procedure we applied to the data (we smoothed the Planck maps down to $1\degree$ convolving them with a 2D Gaussian function)
and map pixel size ($\sim 0.5\degree$ in our case).
$P_l(x)$ denotes the Legendre polynomials of order $l$ and $\theta_{ij}$ the angular distance between the $i$'th and $j$'th pixels.  The standard deviation given by Eq.~\ref{eq:var(T)} was computed using the HEALPix tools \cite{Healpix} and summing from $l=2$ to $l_{max} = 60$; it was found to be $25.4 ~\mu K_{\rm CMB}$ (identical for all frequencies). 
As input $C_l$'s we used the power spectrum theoretically derived from the conventional best-fit $\Lambda$CDM model \cite{Ade:2013zuv}.
Specifically, we took numerical values of $C_l$'s from the ``Ancillary Data'' section of  \cite{maps-url} and the Planck+WP+highL+lensing parameter set there.
The $\sigma_{\nu}^{CMB}$ values are shown in the sixth line of Table \ref{tab:cmb-noise}.

The three sources of noise described above are uncorrelated, so the total noise is the sum in quadrature of $\sigma_{\nu}^{\rm mon}$, $\sigma_{\nu}^{\rm dip}$, and $\sigma_{\nu}^{\rm CMB}$.
 Numerical values are shown in the last line of Table~\ref{tab:cmb-noise}, where we see that the CMB fluctuations contribution completely dominates over the others. 
 Regarding the signal to noise ratio,  the  CMB noise is at the level $\approx 20\%$ with respect to the mean intensity at 28 GHz, but it increases fast with frequency, and at 70 GHz it already exceeds the mean signal.  We compute the 95\% confidence level (CL) limit by setting $I_{\nu}+1.64\sigma_{\nu}$ as the upper limit at each frequency. This chosen cut corresponds to the 95\% upper limit  for a one-dimensional Gaussian distribution. Our choice is justified by the fact that the main source of noise is CMB fluctuations, which are correlated between frequency channels.  

After setting $I_{\nu}+1.64\sigma_{\nu}$ as the intensity upper limit, we required that the DM emission intensity in our ROI should not exceed these limits in {\it any}  of the frequency channels. 
Since the  DM intensity is proportional to $\langle \sigma v \rangle$, this requirement translates to an upper limit for the cross section, which we display in Fig.~\ref{fig:dm_c_cmb}. 
In general, our exclusion curves are quite far from the canonical thermal relic cross section $\langle \sigma v \rangle = 3 \cdot 10^{-26}$ cm$^3$/s (red dashed line) for all considered configurations. The best constraint among all is $\langle \sigma v \rangle \lesssim 2 \cdot 10^{-25}$ cm$^3$/s. 
 No clear dependence of the exclusions on the profile slope $\gamma$ is apparent. 
 However, the dependence on the annihilation channel is quite strong: constraints for $\tau^+ \tau^-$ are $\gtrsim 3$ times stronger than for $b \bar{b}$ for the same MF/propagation models. Also the $\tau^+ \tau^-$ channel forms  a slightly narrower band of exclusion lines. 
 Increases in both $z_B$ and $B(0,0)$ values strengthen the constraints. 
 All these parameter dependencies are indeed in agreement with those for DM intensities, which were described in \S\ref{sec:dmspectra}. 
The most pessimistic constraints are typically realized by the Reacc $z_{B1}~ B(0,0) = 50~\mu$G configuration, while the most optimistic ones by MAX $B(0,0) = 100~\mu$G. The more moderate configuration is Reacc $z_{B2}~ B(0,0) = 50~\mu$G, which we set as our representative constraint (shown by the blue solid line in Figs.~\ref{fig:dm_sp}, \ref{fig:dm_c_cmb} and \ref{fig:dm_c}). 
For this we chose  the $B(0,0) = 50~\mu$G case rather than $B(0,0) = 100~\mu$G due to the MF considerations explained in \S\ref{sec:MF}.
Note also that the fiducial configuration corresponds to the Reacc diffusion model, which is expected to be newest and most accurate among all three considered.

Along with our exclusion lines,  Fig.~\ref{fig:dm_c_cmb} shows the parameter regions at 95\% CL (unless stated otherwise), which are required to fit the GC gamma-ray excess by WIMP annihilation; see figure caption for details. In general, there is a good agreement between the parameter regions derived in independent studies for the same channel and profile slope pairs.
\begin{figure}[H]
\begin{center}
\includegraphics[width=0.495\textwidth]{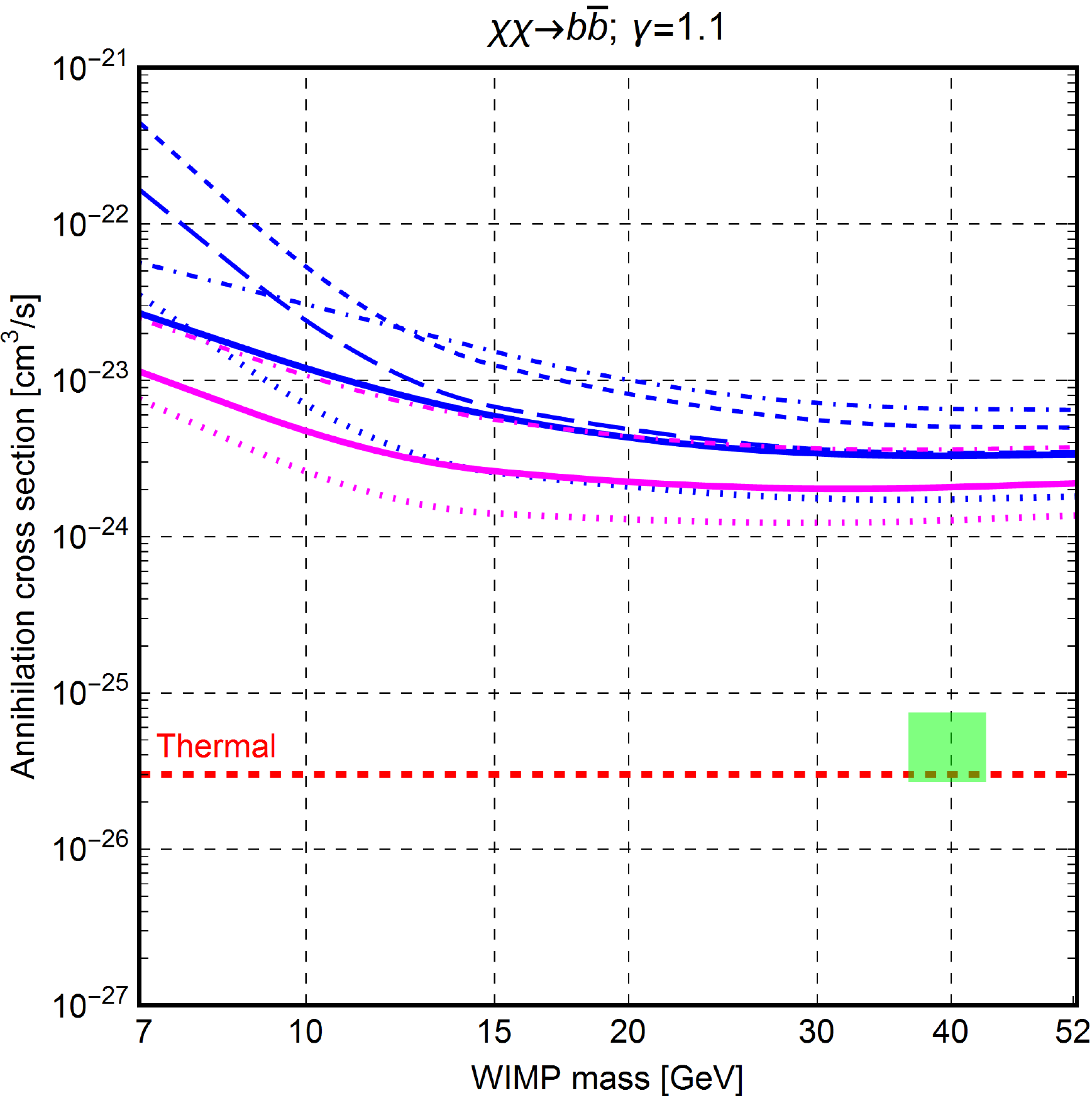}
\includegraphics[width=0.495\textwidth]{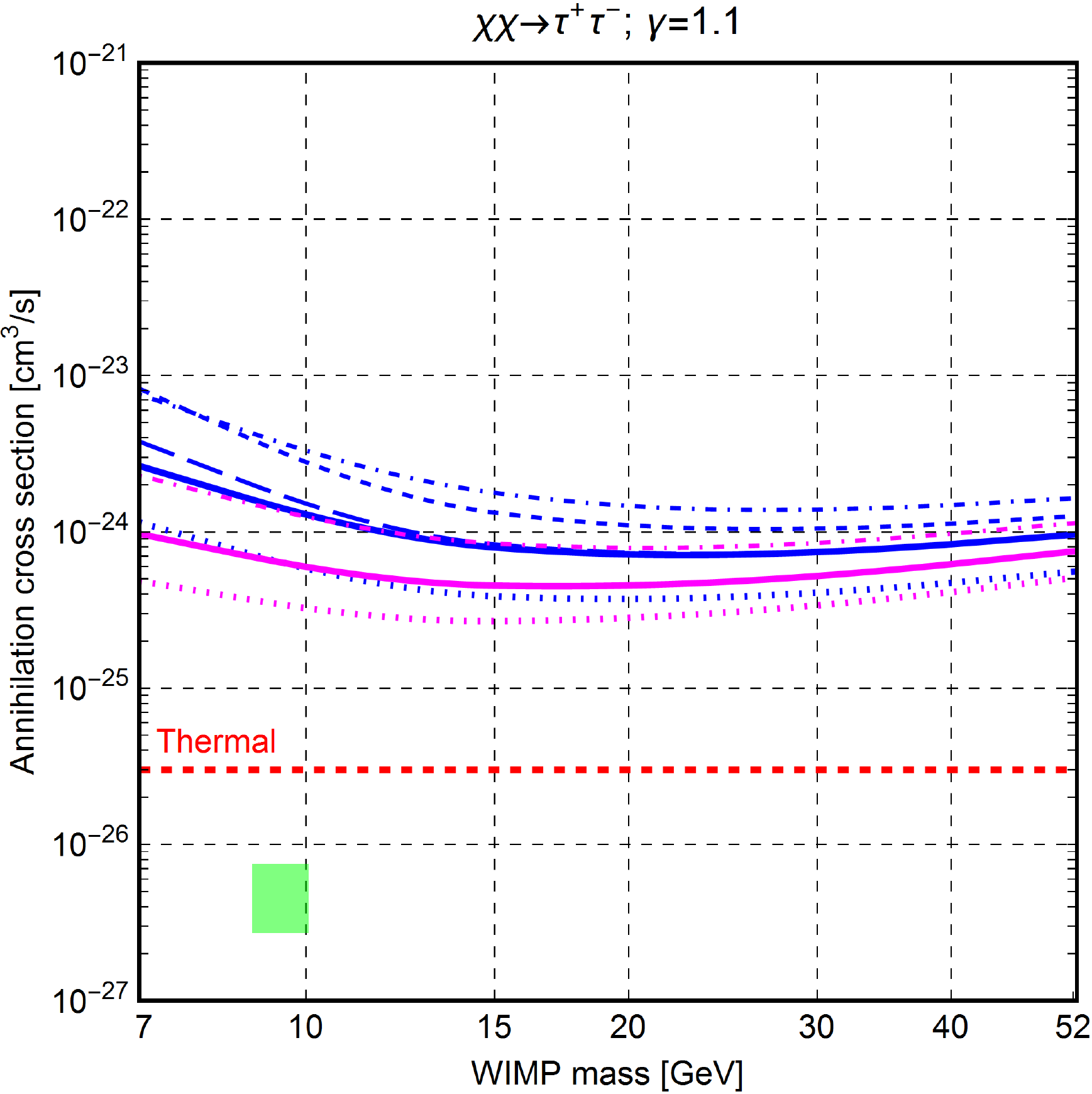}

\vspace*{0.25cm}

\includegraphics[width=0.495\textwidth]{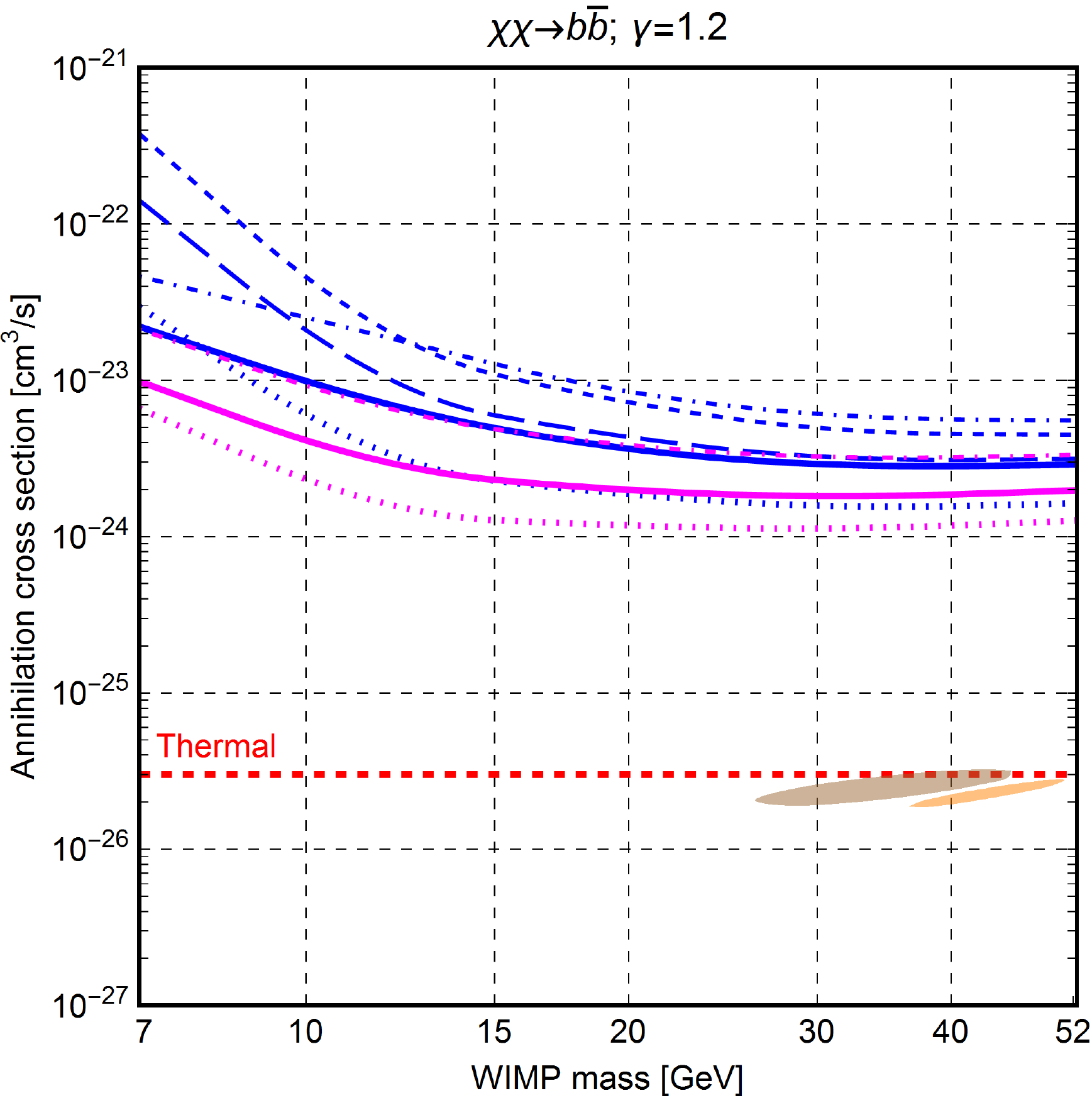}
\includegraphics[width=0.495\textwidth]{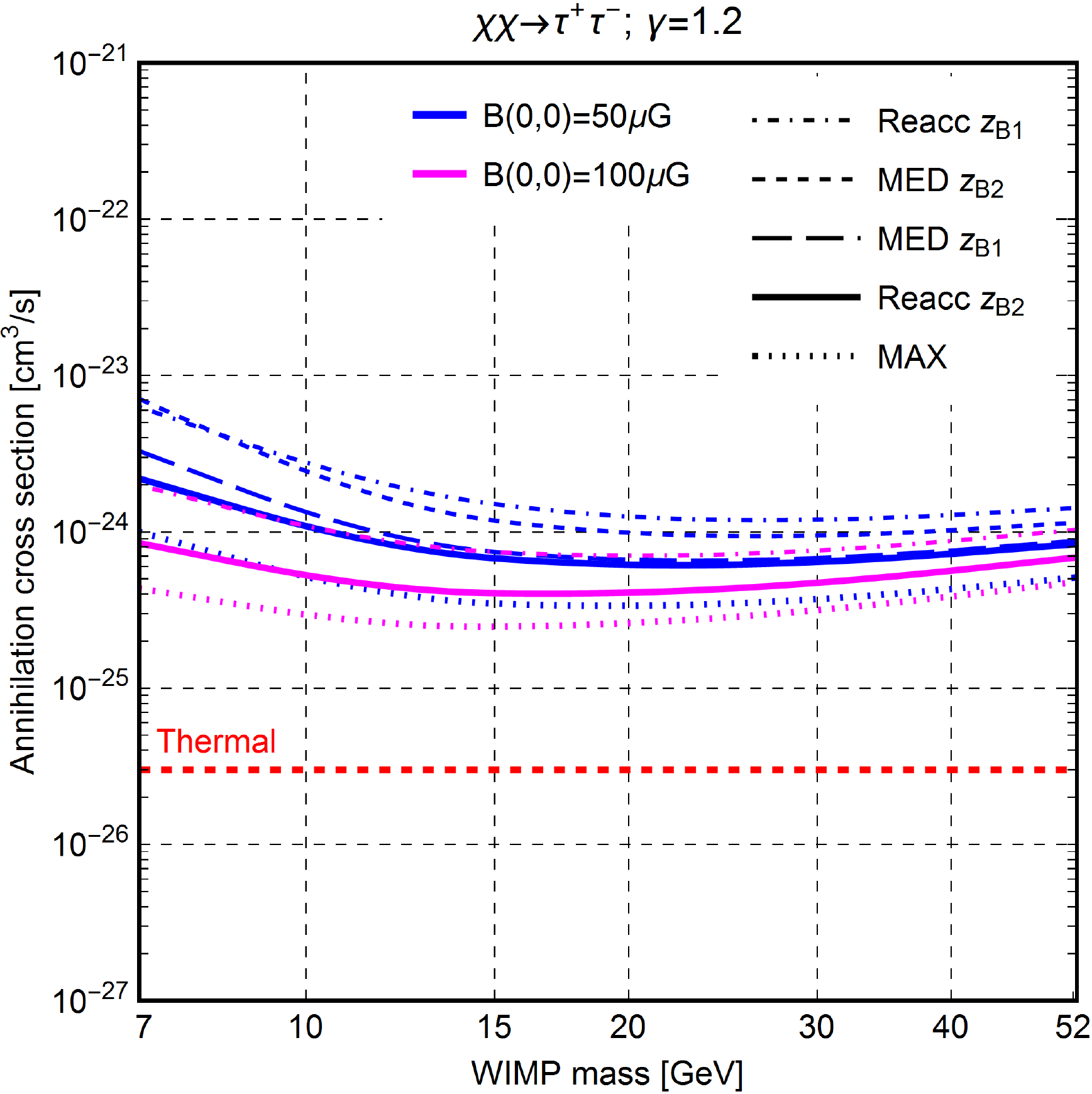}

\vspace*{0.25cm}

\includegraphics[width=0.495\textwidth]{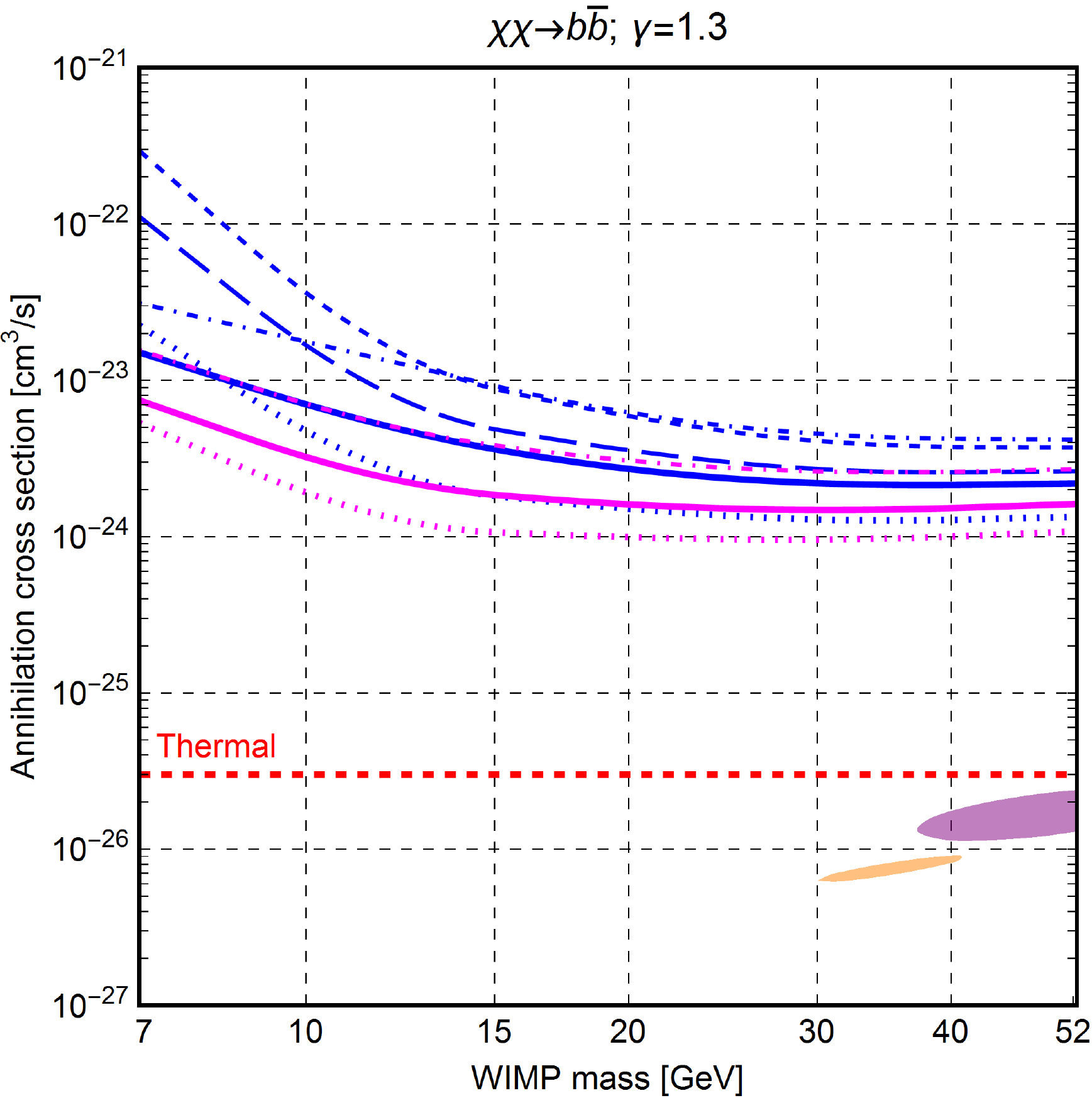}
\includegraphics[width=0.495\textwidth]{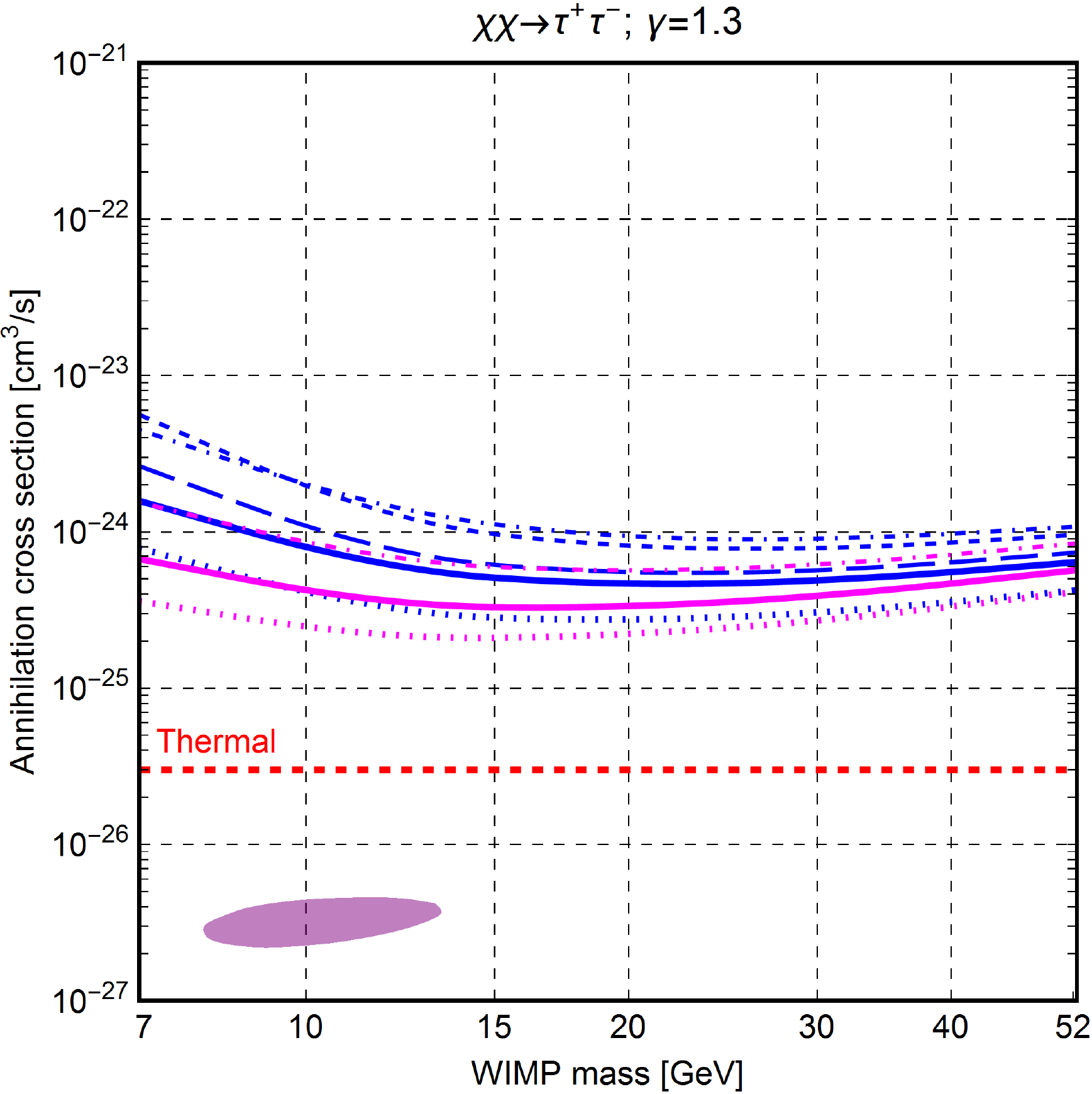}
\end{center}
\end{figure}
\begin{figure}[H]
\caption{\label{fig:dm_c_cmb} \textit{(preceding page)}  DM constraints derived at 95\% CL resulting from the Planck temperature maps with subtraction of the CMB emission only for various DM/MF/propagation models ($\langle \sigma v \rangle$ above the lines are excluded). The red dashed horizontal line indicates the thermal relic cross section $\langle \sigma v \rangle = 3 \cdot 10^{-26}$ cm$^3$/s. Besides exclusion curves we also show here WIMP parameter regions (at 95\% CL unless stated otherwise) needed to fit the GC gamma-ray excess. The brown region for the case $\chi \chi \rightarrow b\bar{b}; \gamma = 1.2$ was derived in the work \cite{Gordon:2013vta}, the orange ones for $\chi \chi \rightarrow b\bar{b}$ in \cite{Daylan:2014rsa}, the green ones for $\gamma = 1.1$ in \cite{Abazajian:2014fta} (68\% CL regions), and the purple ones for $\gamma = 1.3$ in \cite{Calore:2014nla}.  For the optimistic high MF models with $B(0,0) = 100~\mu$G, only three exclusion curves out of five computed in total are shown. The two curves not shown lie inside the other curves. More details are in \S\ref{sec:cmb_constraints}.}
\end{figure}
While all these works only include DM prompt gamma rays in the fit,  \cite{Lacroix:2014eea} showed that including secondary gammas from ICS and bremsstrahlung from DM leptons marginally changes the result (their signal region, derived for the configuration $\chi \chi \rightarrow b\bar{b}; \gamma = 1.2$,  is almost completely inside the brown region shown). Therefore, overall our constraints  on the cross section show no tension with the signal regions derived from gamma-ray data analyses of the excess.

An alternative way to account for CMB fluctuations in the emission map is to subtract at every frequency  the reconstructed CMB fluctuation map as provided by the Planck team (see ``all-sky-maps'' folder at \cite{maps-url}). This map is the result of the emission component separation procedure performed by the team and reflects the most up-to-date knowledge of the actual realization of  CMB fluctuation  field. We attempted this strategy and found a relatively minor improvement in our derived constraints of  $\approx 16\%$. Although more stringent, these constraints depend on the specific component separation procedure adopted to derive the map, and therefore we opted instead to show the results from the procedure described earlier in this section.

The DM limits described in this section are relatively weak, however we expect that foregrounds other than the CMB play a potentially important role in this region of the sky: dust, free-free from ionized hydrogen, synchrotron from regular CRs and the Bubbles.  In the following section we perform a full component separation analysis to more accurately characterize a possible DM contribution.

\section{DM constraints from a component separation approach}
\label{sec:Fitting}

\subsection{The WMAP-Planck Haze: history }
\label{sec:Fitting.review}

A mysterious microwave excess above the canonical astrophysical foregrounds was first reported  with the first  WMAP data release \cite{Finkbeiner:2003im}, and was later  interpreted as a possible  signal of DM annihilation \cite{Hooper:2007kb}. Further WMAP-based studies confirmed the existence of the  Haze  and characterized it in detail \citep{Dobler:2007wv, Pietrobon:2011hh, Dobler:2012ef}. The most recent and comprehensive Haze study was performed after the Planck data releases in \cite{:2012rta}. This work combined WMAP-9 and Planck data to separate out the Haze emission at 6 frequencies via frequency-by-frequency template fitting. The results showed the Haze properties to be highly consistent between the two data sets. This can be seen in Fig.~8 in \cite{:2012rta}, where the derived Haze spectrum is presented. As we already mentioned, those authors claimed that the Haze spectrum $I \sim \nu^{-0.56}$ is indeed significantly harder than the usual synchrotron $I \sim \nu^{-1.1}$, which suggests the origin to be from a distinct population of electrons. The overall amplitude of the Haze emission was found to constitute $\approx 33\%$ of the total synchrotron and $\approx 23\%$ of the total Galactic emission at 23 GHz in the relevant region (Fig.~\ref{fig:masks}, left). Figs.~9-10 in \cite{:2012rta} show the Haze morphology, which appears to be consistent with that of the Fermi Bubbles. This fact and other considerations (see, e.g., \cite{Crocker:2014fla}), led the community to favor a scenario where the Haze is simply the microwave counterpart of the Fermi Bubbles.

The DM and the Bubbles interpretation of the Haze are in fact not mutually exclusive, and thus it is necessary to consider both in the analysis.  In what follows, we will investigate whether the data are best fit with one of these components at a time, or both together.  In this part of the analysis, we will adopt the ROI shown in the right panel of Fig.~\ref{fig:masks} for the component separation analysis.

\subsection{Methodology}
\label{sec:Fitting.procedure}

Our procedure  relies on template fitting, which is  described by the following equation:
\begin{equation} \label{eq:fit}
	d(\nu,b,l) = \sum_i a_i(\nu) t_i(\nu,b,l) + r(\nu,b,l), 
\end{equation}
where $d(\nu,b,l)$ is the data map  at frequency $\nu$ in a pixel with coordinates $(b,l)$; $t_i(\nu,b,l)$ is an assumed template intensity of the $i$-th component; $r(\nu,b,l)$ are residuals; $a_i(\nu)$ is a dimensionless amplitude of the $i$-th component, which is determined  by the fitting. 
As for  the sky maps $d(\nu,b,l)$, we used four WMAP channels: 23, 33, 41, 61 GHz (K, Ka, Q, and V bands from \cite{w-maps-url}) and three Planck LFI channels: 28, 44, 70 GHz from \cite{maps-url}. 
In contrast to \cite{:2012rta},  we also include in the fit  the 70 GHz  frequency channel, which provides further leverage for DM fitting and allows a better investigation of the frequency dependence of foregrounds.

In addition to DM, we included in the fit as foregrounds: dust, free-free, synchrotron and the Bubbles. A monopole and dipole component, as well as the CMB, are initially fit to the whole sky map, and removed from the data, so there is no monopole, dipole, or CMB fitting performed in the chosen ROI\@.
As described in \S\ref{sec:Computation}, the DM emission modeling  predicts a specific emission at each frequency,   therefore  the DM  has only one free parameter, namely an overall amplitude $a_{DM}$, which  sets the annihilation cross section value required by the fit. We ran the fitting procedure  separately for each DM/MF/propagation model. We assume spatially-constant power laws for all Galactic foregrounds. This means that the templates are rescaled by a single number, $a_{i}(\nu)$, in the ROI at every frequency independently. Therefore, unlike for the DM component, we do not impose a frequency dependence to foregrounds in the fit. We compute it, however, a posteriori.

\begin{figure}[H]
\begin{center}
\includegraphics[width=0.495\textwidth]{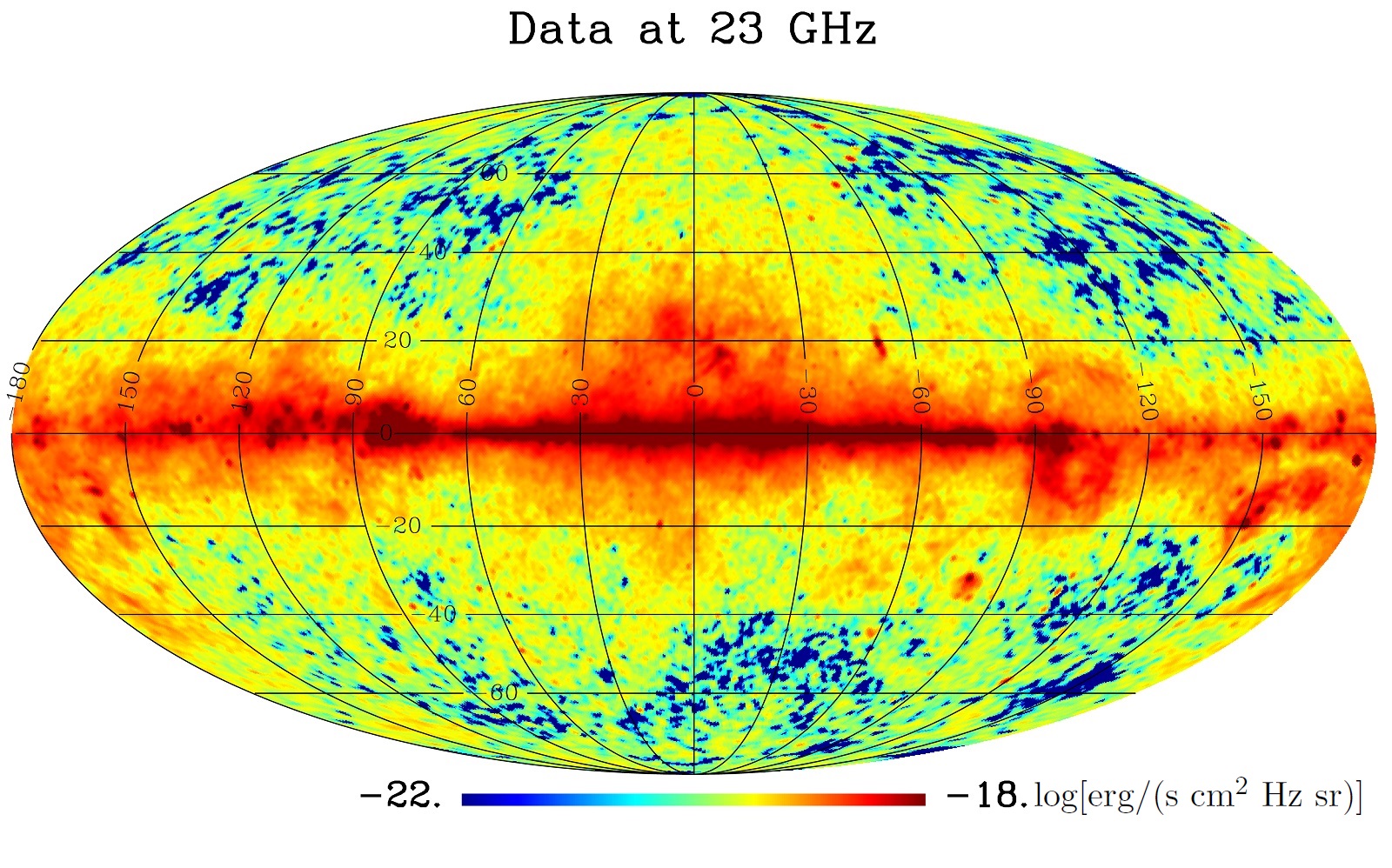}
\includegraphics[width=0.495\textwidth]{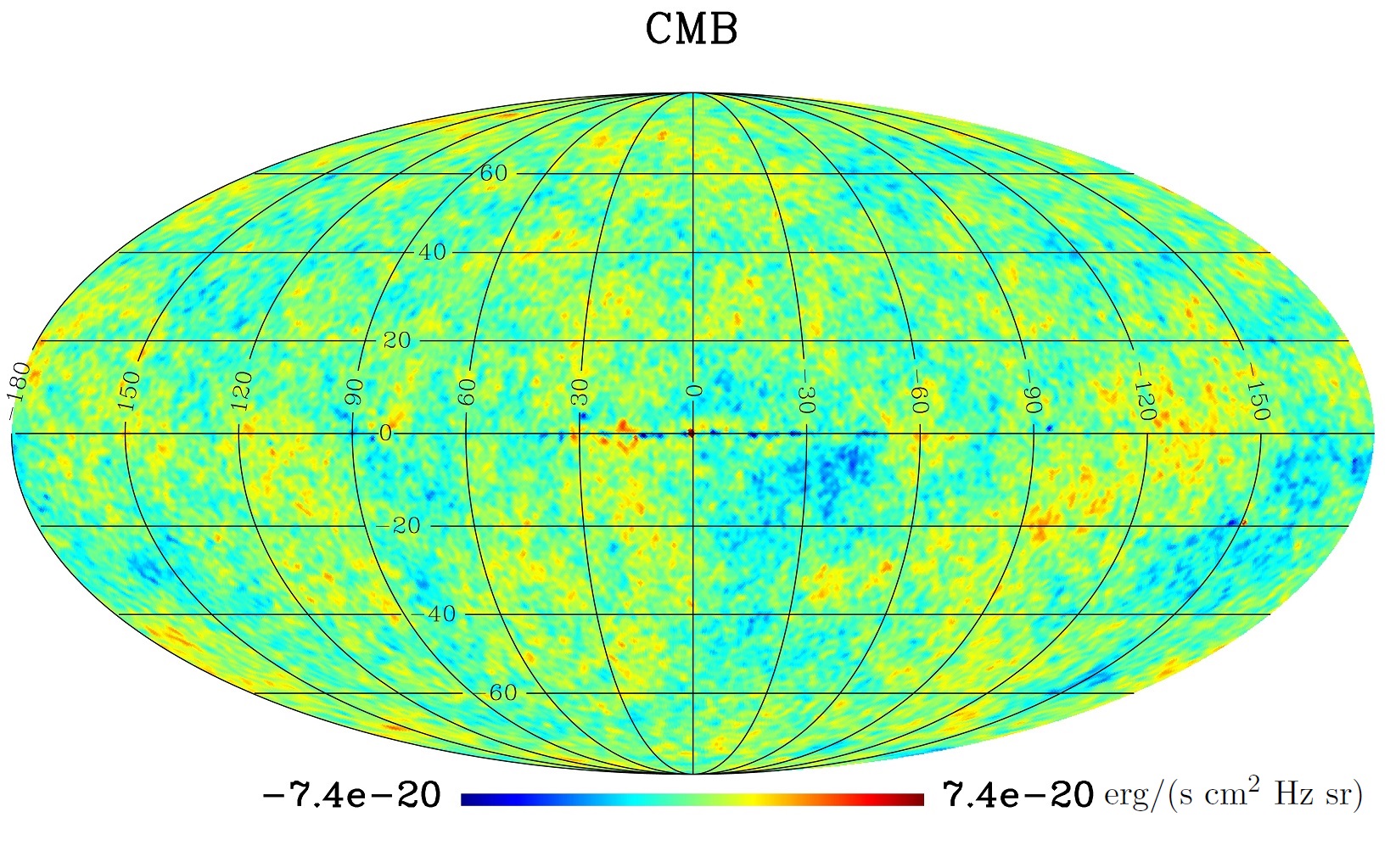}
\includegraphics[width=0.495\textwidth]{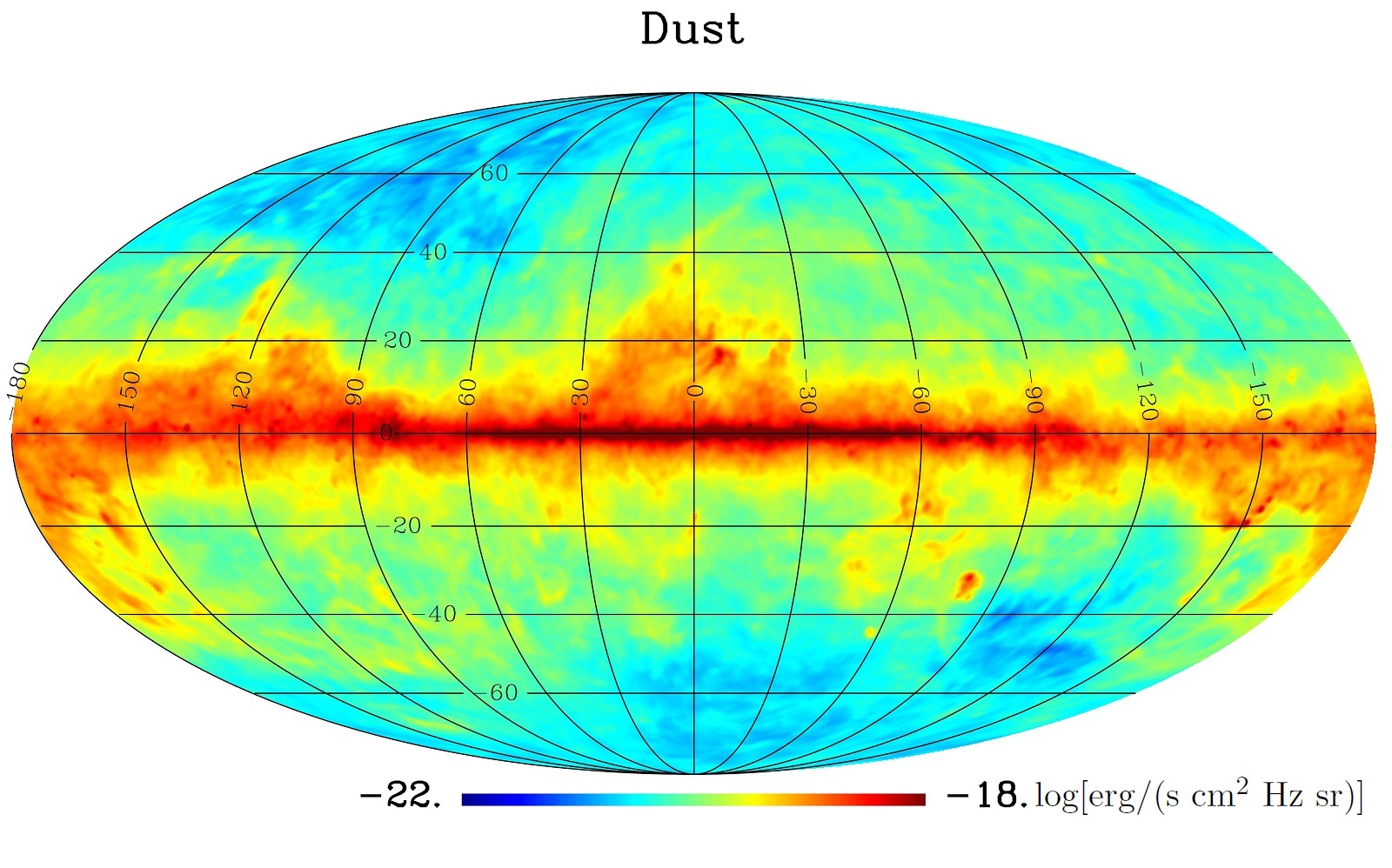}
\includegraphics[width=0.495\textwidth]{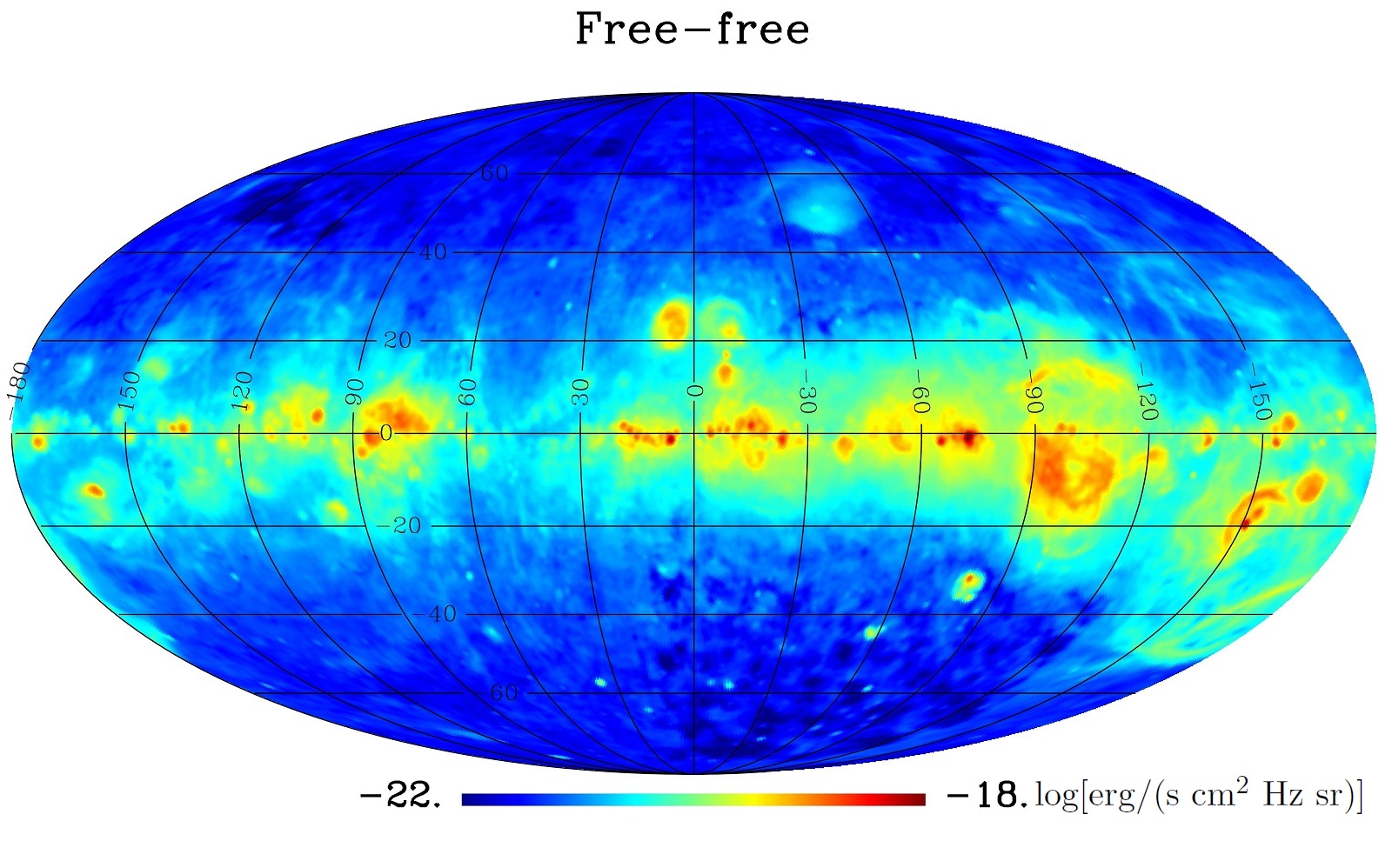}
\includegraphics[width=0.495\textwidth]{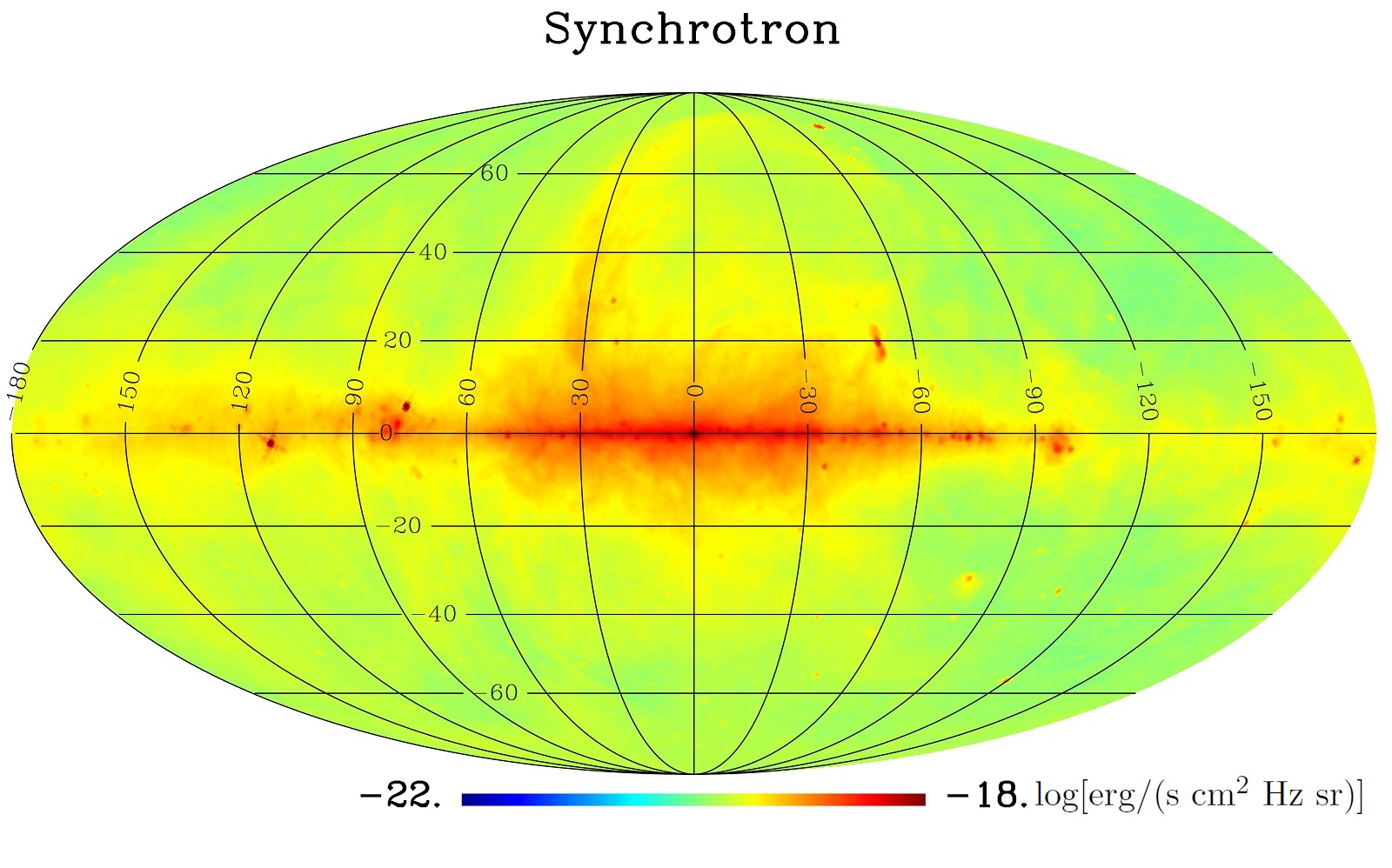}
\includegraphics[width=0.495\textwidth]{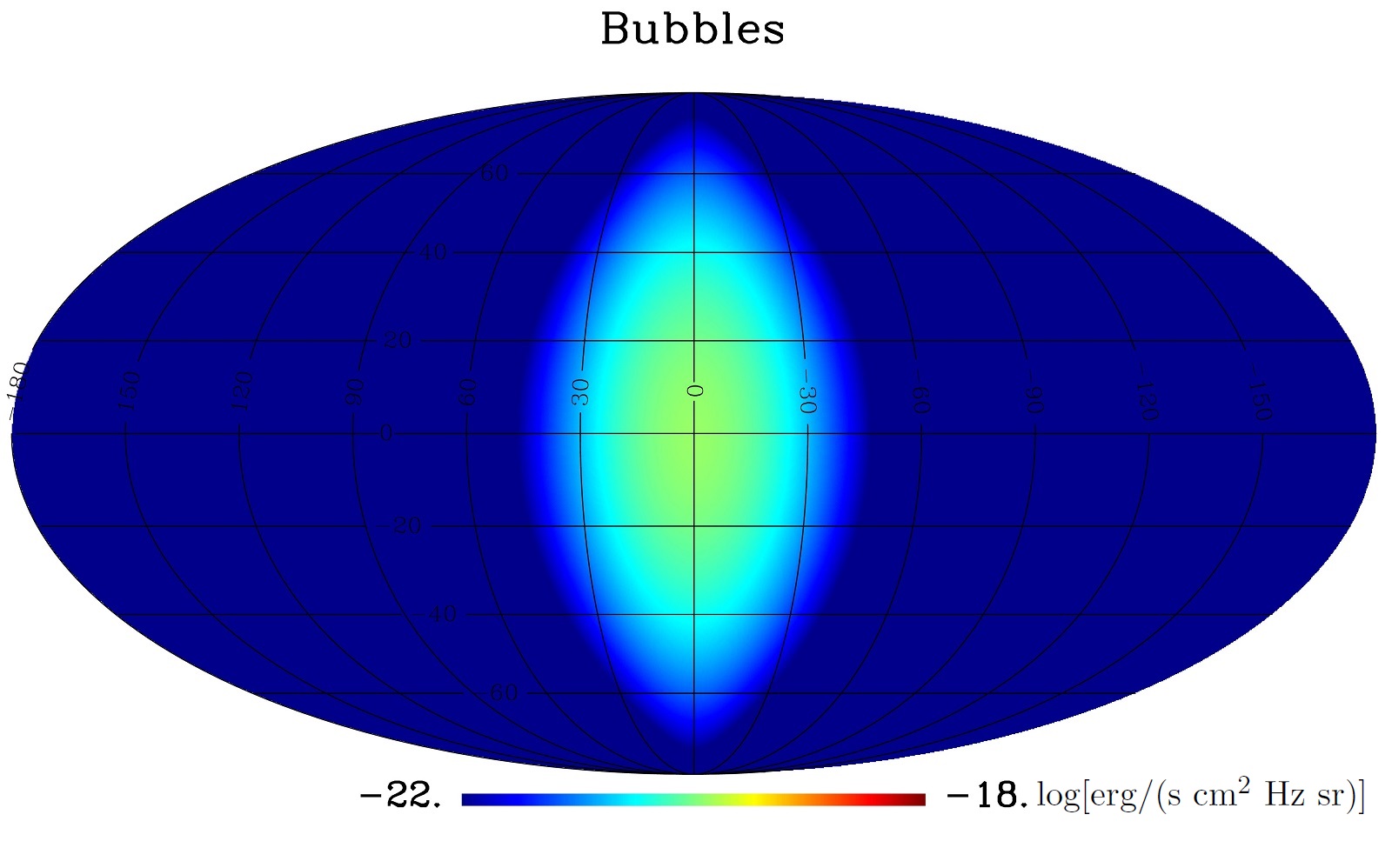}
\includegraphics[width=0.495\textwidth]{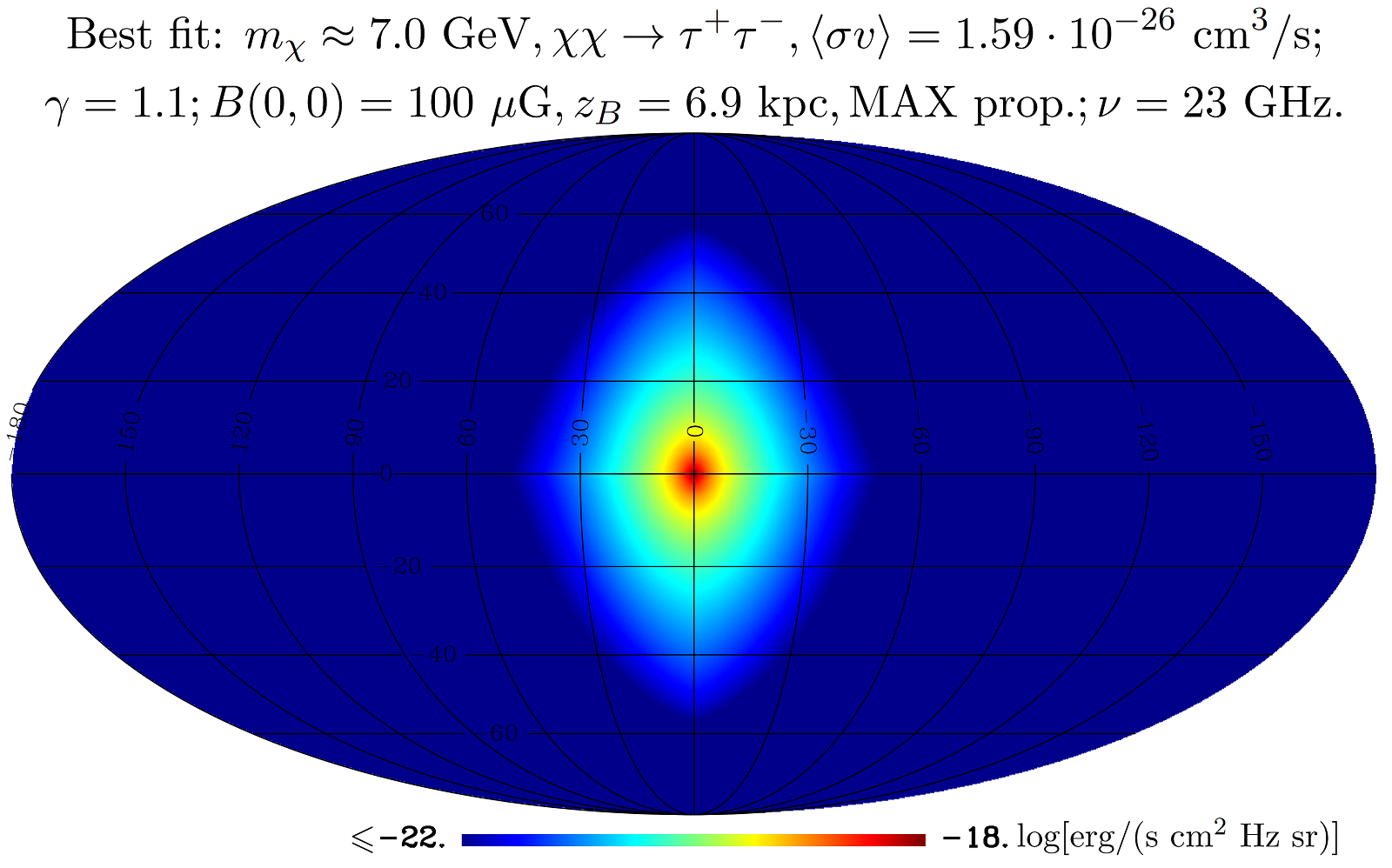}
\includegraphics[width=0.495\textwidth]{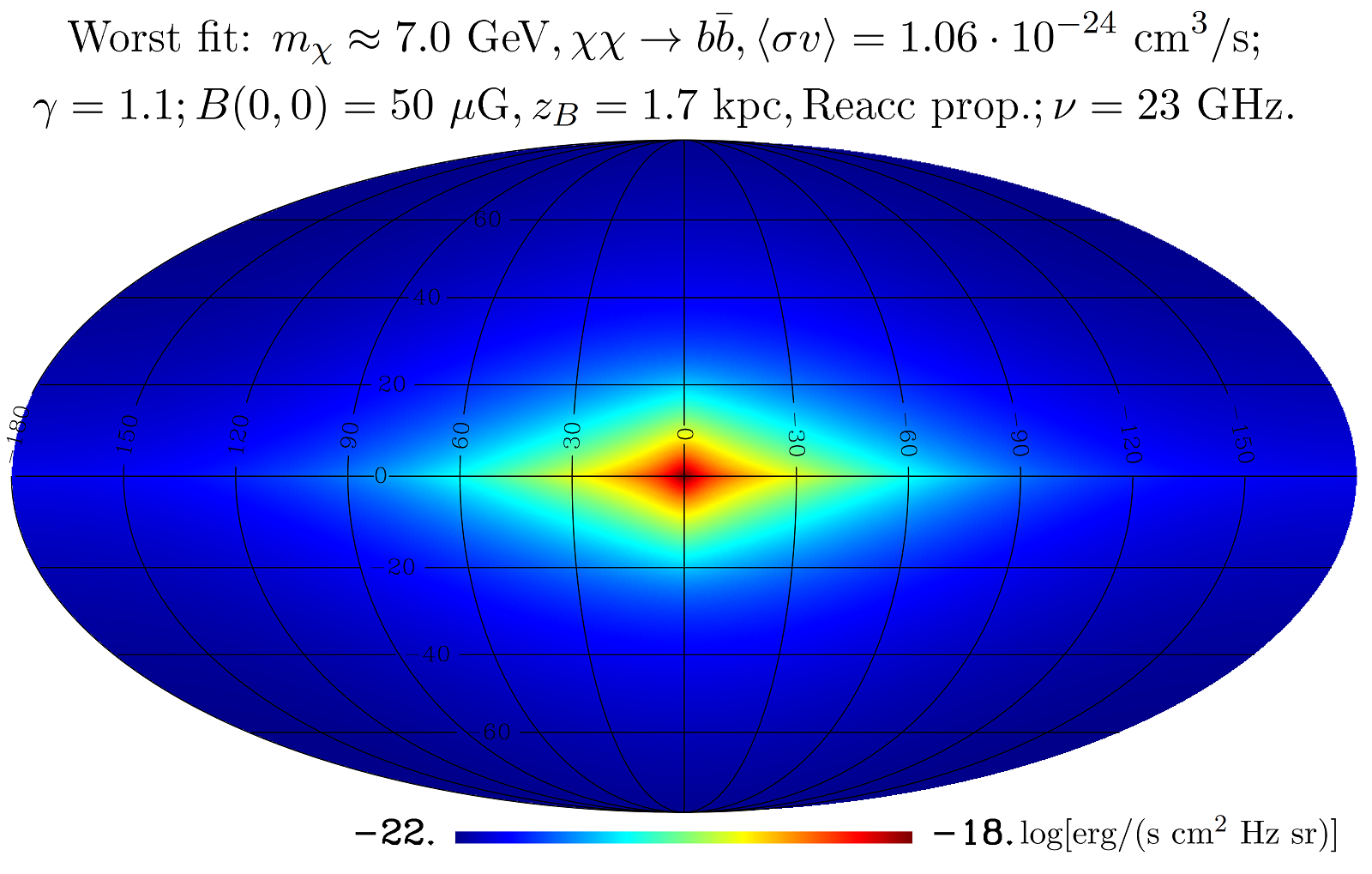}
\caption{Data (total observed intensity at 23 GHz, top left panel) and templates (all other panels) used in the analysis. Morphology of all components except DM was assumed to be frequency-independent. Normalization of each template corresponds to that of the best-fit model at 23 GHz. The bottom panels show the best-fit and worst-fit DM sky maps among all the models considered. See more details in \S\ref{sec:Fitting.procedure}. }
\label{fig:t}
\end{center}
\end{figure}

The data at 23 GHz and the foreground templates are shown in Fig.~\ref{fig:t}. Note that our units can be converted into thermodynamic units: $10^{-20}$ erg/(s cm$^2$ Hz sr) $\approx 62~\mu K_{\rm CMB}$ at 23 GHz at temperatures of interest. For the CMB template we employed the Internal Linear Combination from Planck HFI channels (weights were derived from one single region, outside of the type-III Planck 2013 CO-based mask \cite{PlanckCO}). The reason for this choice rather than the Planck official map (SMICA) is that we wanted to avoid any of the LFI channels contributing to the CMB solution in order to avoid any circularity in the fit.  The Galactic dust emission map is described in \cite{ReddeningSFD, DustFDS}.
Free-free emission naturally correlates with $H_{\alpha}$, which provides an estimate of its morphology \cite{HalphaF}.
The synchrotron from conventional astrophysical CRs is assumed to have a similar morphology at radio and microwave frequencies (which is essentially equivalent to an assumption of constant spectral index across the sky). This allows us to use the canonical Haslam 408 MHz map as the synchrotron template \cite{Haslam}. Finally for the Bubbles we adopted the elliptical Gaussian template (with $\sigma_l = 15\degree$ and $\sigma_b = 25\degree$) proposed by \cite{:2012rta} for the Haze emission, which was suggested to be suitable for the Fermi Bubbles counterpart. 
Indeed an exact Bubbles morphology is not known and may be quite complicated with both vertical and horizontal asymmetries (see, e.g., \cite{Carretti:2013sc} for the morphology in radio). Our choice has the advantage of allowing comparison of our results with existing results in the literature (see \cite{:2012rta} and references therein).

Additional foreground components could also be considered. For example, the authors of \cite{:2012rta} also introduced a ``disc'' component  (an elliptical Gaussian with $\sigma_l = 20\degree$ and $\sigma_b = 5\degree$)  to account for a potential difference in the synchrotron morphology between 408 MHz and the typical CMB frequency range. This was intended to account for the energy dependence of the electron propagation length. 
Our attempt to add such component to the fitting  provided unphysical results. Also, considering the relatively minor importance of the disc reported in \cite{:2012rta},  we decided not to  include any further correction to the synchrotron emission in our final procedure.

The fitting procedure determines the amplitudes $a_i(\nu)$ by minimizing the residuals $r(\nu,b,l)$ in Eq.~\ref{eq:fit}.  They are
the solution of the following set of equations:
\begin{equation}
\frac{\partial \chi^2}{\partial a_i(\nu)} = 0.
\label{eq:differential}
\end{equation}
If the sky model only included free amplitudes for each component at each frequency,
the solution of the linear system Eq.~\ref{eq:differential} would be:
\begin{eqnarray}
&& \mathbf{a}(\nu) = \mathbf{T}^{-1} \, B_\nu \\
&& \Delta a(\nu)^2 = {\rm diag} \frac{1}{2} \; \mathbf{T}^{-1}
\end{eqnarray}
where $B_{\nu} = (d(\nu,b,l)/\sigma(\nu,b,l)) \cdot (t(b,l)/\sigma(\nu,b,l))$ and $T_{ij}=t_i(b,l)/\sigma(\nu,b,l) \cdot t_j(b,l)/\sigma(\nu,b,l)$.
However, our DM modeling couples all frequencies so that we have to resort to a Markov-Chain procedure to find the values and errors for $a_i(\nu)$.

Modeling foregrounds by means of templates and assuming spatially-constant spectral indexes across the whole sky is a very crude approximation. Moreover, we expect the DM signal to be prominent near the GC but not everywhere. On the other hand, a residual monopole and dipole might be present in the WMAP and Planck maps, as well as in the templates, as result of the map-making and calibration procedures, and on small areas of the sky these components are not properly fit.

For these reasons, we decided to perform the fit in two steps. First, we fit the data masking the Galactic plane and bright sources with a mask very similar to the one used by \cite{:2012rta}, and obtained the template amplitudes at all frequencies; we did not include the DM in this nearly full-sky fit. We then subtracted the CMB and monopole and dipole from the data and solved for the dust, free-free, synchrotron, Bubbles, and DM amplitudes in the larger ROI (right panel of Fig.~\ref{fig:masks}). 

In order to determine the best-fitting DM/foreground amplitudes in our chosen ROI,  we ran our own Markov Chain Monte Carlo (MCMC) sampler to study the multi-parameter likelihood, which is assumed to be:  
\begin{equation} \label{chisq}
	L \sim \exp(-0.5\chi^2), ~ \chi^2 = \sum_{j=1}^7 \sum_{i=1}^{n_2} \frac{r_{ij}^2}{\sigma_{ij}^2},
\end{equation}
where $r_{ij}$ is the residual in pixel $i$ at frequency channel $j$ according to Eq.~\ref{eq:fit}, $\sigma_{ij}$ are map noise, and the summation runs over all seven frequencies and all $n_2=15481$ pixels in our ROI\@. The reduced $\chi^2$ ($\chi^2_{\rm r}$), is computed dividing the full $\chi^2$ in Eq.~\ref{chisq} by the number of degrees of freedom $N_{\rm dof} = 7 \cdot n_2 - 7 \cdot 4 - 1 = 108338$.

Eq.~\ref{chisq} implicitly assumes that the noise in each pixel is  uncorrelated. In fact, noise correlations are expected because of the spacecraft scanning strategy and the  adopted  smoothing procedure.  Despite being sub-optimal, this procedure is commonly used (e.g., \cite{:2012rta})  because of the difficulty to treat the full complexity of the noise at intermediate and high resolution.
The noise maps were computed as the standard deviation of full resolution Gaussian noise realizations, smoothed to one degree and downgraded to $N_{side}=128$ (to match the DM template). The RMS maps derived were further rescaled to match the noise power spectrum level of the full-resolution smoothed maps. We did not put any  prior on the amplitudes, aside from imposing that they should be positive.

\subsection{Fitting results}
\label{sec:Fitting.results}

We ran the MCMC fit for our grid of DM/MF/propagation models and obtained $\chi^2_{\rm r}$ for each of them. Overall the quality of the fit is quite poor, with a $\chi_{\rm r}^2$ ranging between 6.87 and 7.03. This is simply a confirmation of the challenge that the component separation poses, the limitations we face with the current, non-Planck-based foreground templates, and the simplicity of template fitting. As a result, the $\chi_{\rm r}^2$ difference cannot be considered as a powerful discriminant between models. Nonetheless, we can rank the models based on their likelihood and draw conclusions. The parameter values of the best-fitting (BF) and worst-fitting (WF) configurations are presented in Table~\ref{tab:bfwf}. This table also shows for an illustration the fit quality of the other representative DM/MF/propagation models, which differ from the BF model by one or two parameters.

\begin{table}[tbp]
\caption{All parameter values for the best-fitting (BF, the first row) and worst-fitting (WF, the second row) DM/MF/propagation models. Lower rows illustrate the fit quality for other models, which differ from the BF model by one or two parameters indicated in the respective columns. }
\label{tab:bfwf}
\begin{center}
\begin{tabulary}{1\textwidth}{|C|C|C|C|C|C|C|C|}
\hline
Mo-del/ $\chi^2_{\rm r}$ & Cross section $\langle \sigma v \rangle$ [cm$^3$/s] & Mass $m_{\chi}$ [GeV] & Channel & Profile slope $\gamma$ & Central MF $B(0,0)$~[$\mu$G] & Vertical MF scale $z_B$ [kpc] & Diffusion model \\
\hline
BF/ 6.87 & $1.6 \cdot 10^{-26}$ & 7.0 & $\tau^+\tau^-$~~~~~~ & 1.1 & 100 & 6.9 & MAX \\
WF/ 7.03 & $1.1 \cdot 10^{-24}$ & 7.0 & $b\bar{b}$ & 1.1 & 50 & 1.7 & Reacc \\
\hline
6.92 & $1.9 \cdot 10^{-26}$ & 19 &  &  &  &  &  \\
6.99 & $4.7 \cdot 10^{-26}$ & 52 &  &  &  &  &  \\
6.91 & $1.3 \cdot 10^{-25}$ &  & $b\bar{b}$ &  &  &  &  \\
6.89 & $9.2 \cdot 10^{-27}$ &  &  & 1.3 &  &  &  \\
6.89 & $2.9 \cdot 10^{-26}$ &  &  &  & 50 &  &  \\
6.91 & $1.8 \cdot 10^{-26}$ &  &  &  &  & 2.7 & Reacc \\
6.89 & $6.3 \cdot 10^{-26}$ & 19 & $b\bar{b}$ &  &  &  &  \\
6.92 & $8.8 \cdot 10^{-26}$ & 52 & $b\bar{b}$ &  &  &  &  \\
\hline
\end{tabulary}
\end{center}
\end{table}

While overall the poor $\chi^2_r$ values indicate a poor fit, it is sensible to ask whether the Bubbles and/or DM substantially improve the fit, with respect to the case in which only standard foregrounds are included. The respective fit qualities for various setups are shown in Table~\ref{tab:chisq}, with fits to each  individual frequency and  all frequencies together reported.  If only standard foregrounds are considered, the fit is very poor. The lowest frequencies show the  largest residuals, hence they likely contain the largest contribution from the Haze. Above 40 GHz,  residuals are less pronounced and quite acceptable, with a possible overfitting in the 60 GHz channel where the foregrounds are typically at their minimum. When adding the Bubbles template, we see a very significant improvement at low frequencies. Similarly, when adding only DM, there is an improvement is the fit.  The value of $|\Delta \chi^2_{\rm r}|$ is slightly less than in the Bubbles-only case, but quite comparable. When allowing for both Bubbles and DM in the fit, $\chi^2_{\rm r}$ reduces by almost a factor of two; however remaining quite large (since naively we expect it to take the value of one in the case of a good fit).

\begin{table}[tbp]
\caption{Effects of addition of the Bubbles and DM on the quality of the fit. The BF DM model shown is described in Table~\ref{tab:bfwf}. The second column shows results with ``standard'' foregrounds only. The third column  shows the $\chi^2_{\rm r}$ improvement after adding the Bubbles.  The fourth column shows improvements after adding the BF DM model only (no Bubbles). The last column shows improvements in the most general case of the Bubbles and DM considered together.}
\label{tab:chisq}
\begin{center}
\begin{tabulary}{1\textwidth}{|C|C|C|C|C|}
\hline
$\nu$ & $\chi^2_{\rm r}$ with CMB, mono/dipole, f-f, synchr., dust & $\Delta \chi^2_{\rm r}$ after addition of the Bubbles & $\Delta \chi^2_{\rm r}$ after addition of the BF DM model & $\Delta \chi^2_{\rm r}$ after addition of the Bubbles and BF DM model\\
\hline
23 & 37.4 & -15.9 & -15.4 & -17.8 \\
28 & 21.4 & -6.35 & -5.42 & -6.86 \\
33 & 10.0 & -3.13 & -2.67 & -3.47 \\
41 & 2.23 & -0.475 & -0.363 & -0.533 \\
44 & 3.31 & -0.343 & -0.219 & -0.379 \\
61 & 0.563 & -0.023 & -0.013 & -0.023 \\
70 & 2.29 & -0.078 & -0.007 & -0.077 \\
\hline
Gl. & 11.0 & -3.75 & -3.44 & -4.16 \\
\hline
\end{tabulary}
\end{center}
\end{table}

\begin{figure}[t]
\begin{center}
\includegraphics[width=0.325\textwidth]{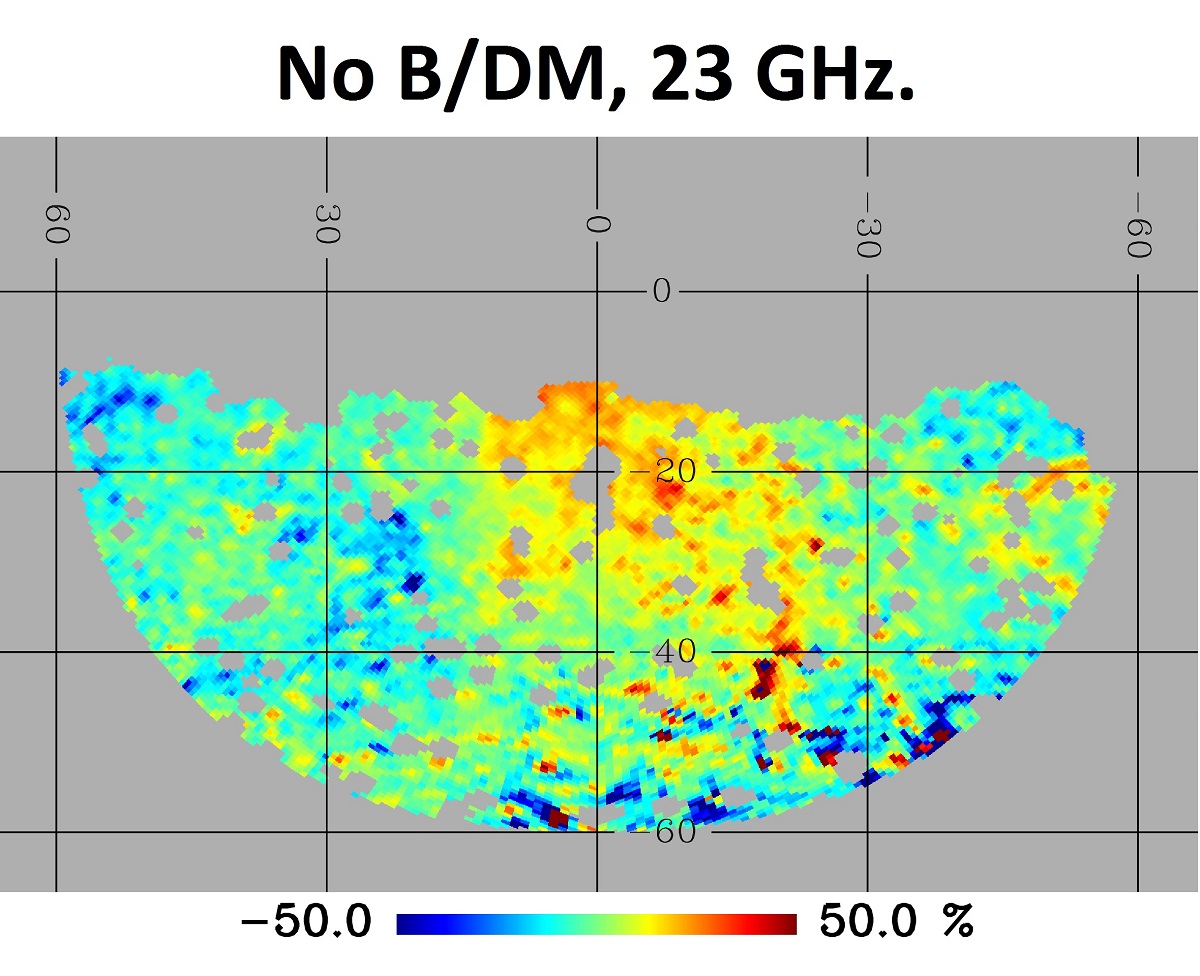}
\includegraphics[width=0.325\textwidth]{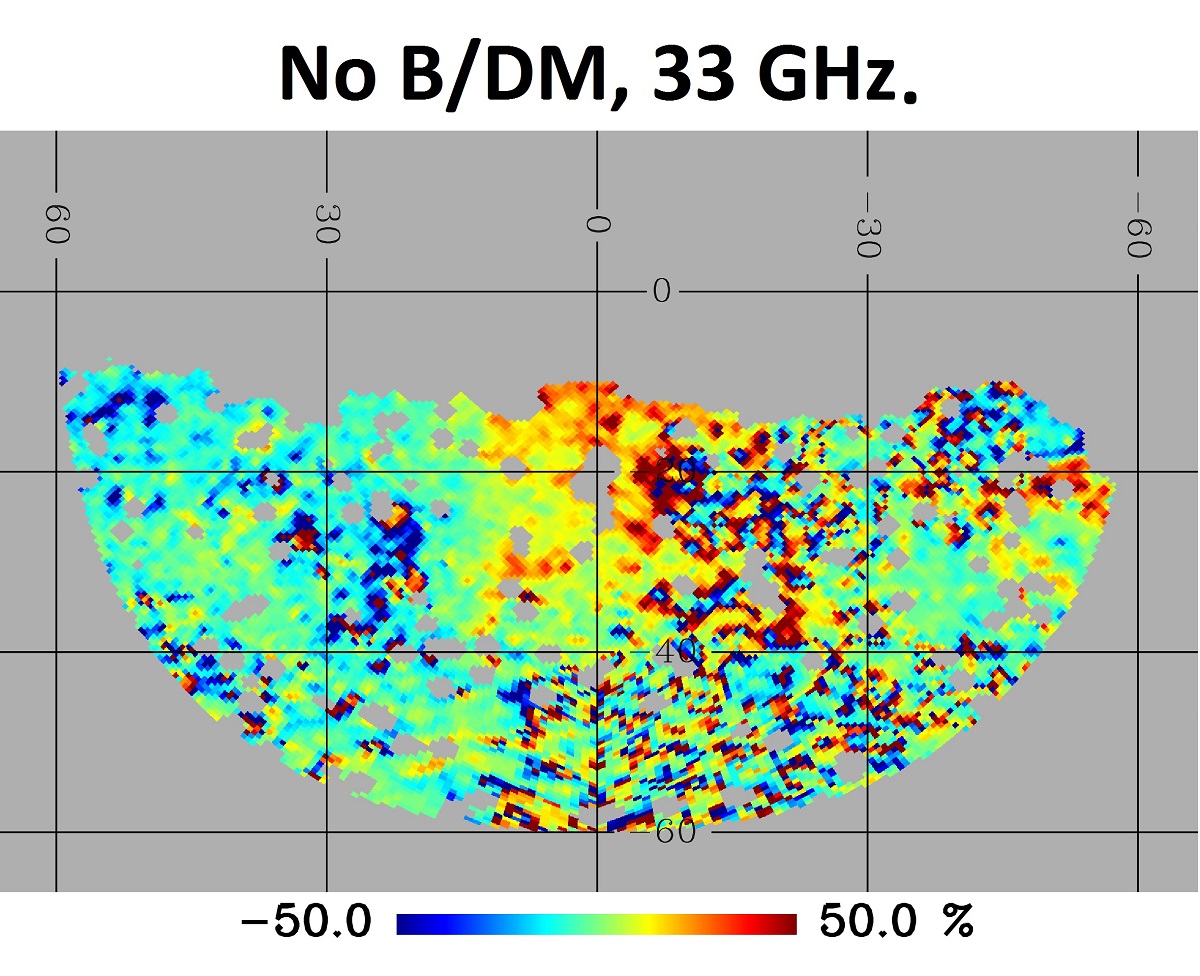}
\includegraphics[width=0.325\textwidth]{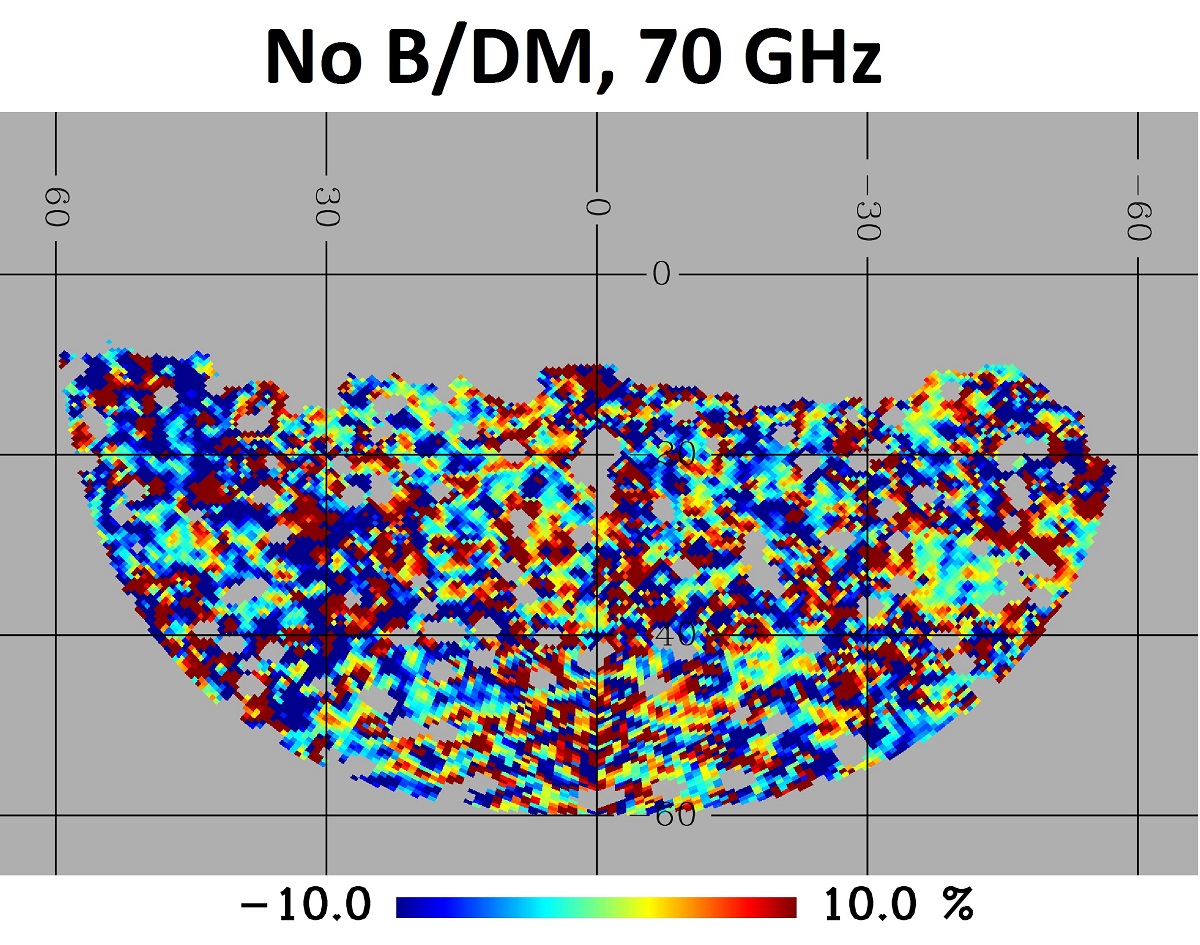}

\includegraphics[width=0.325\textwidth]{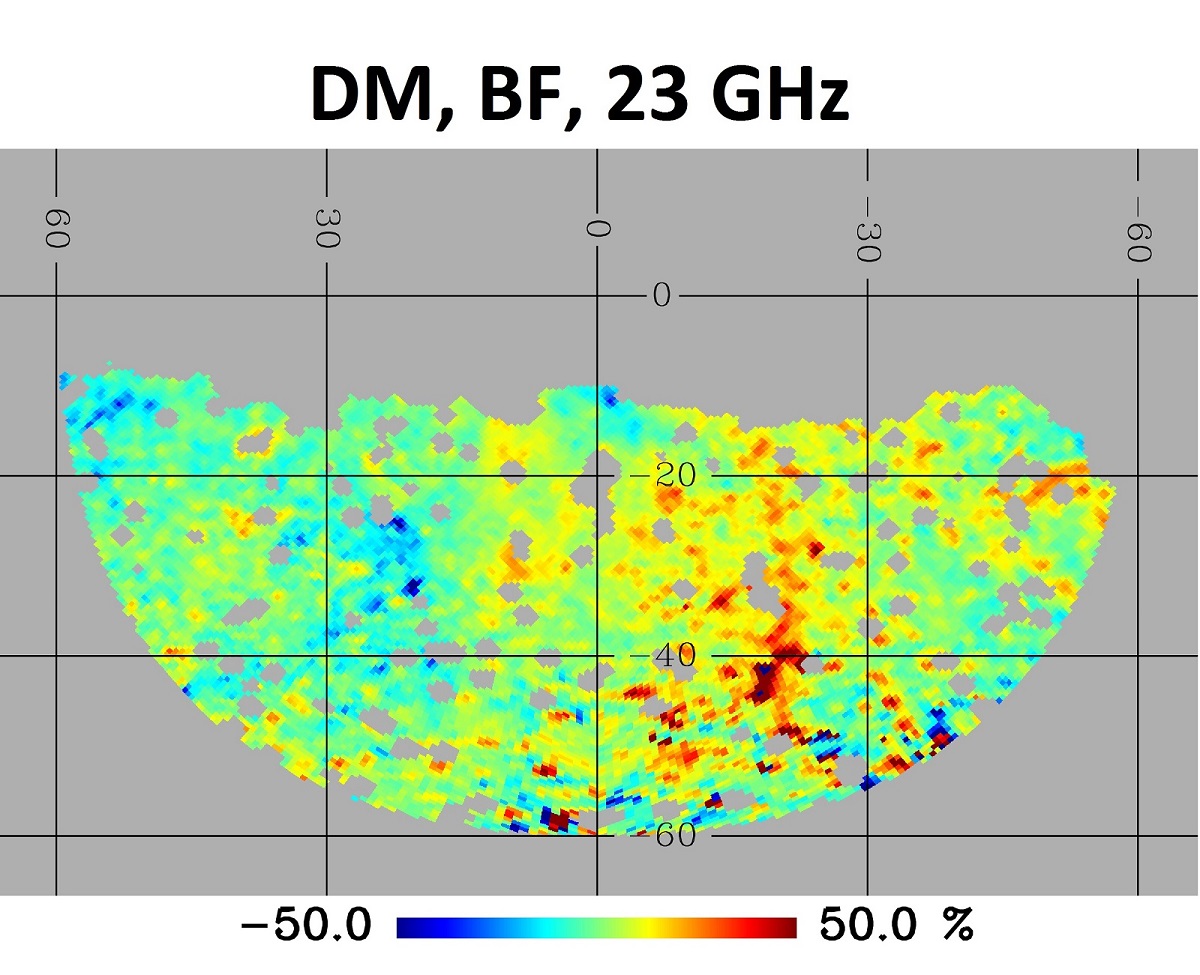}
\includegraphics[width=0.325\textwidth]{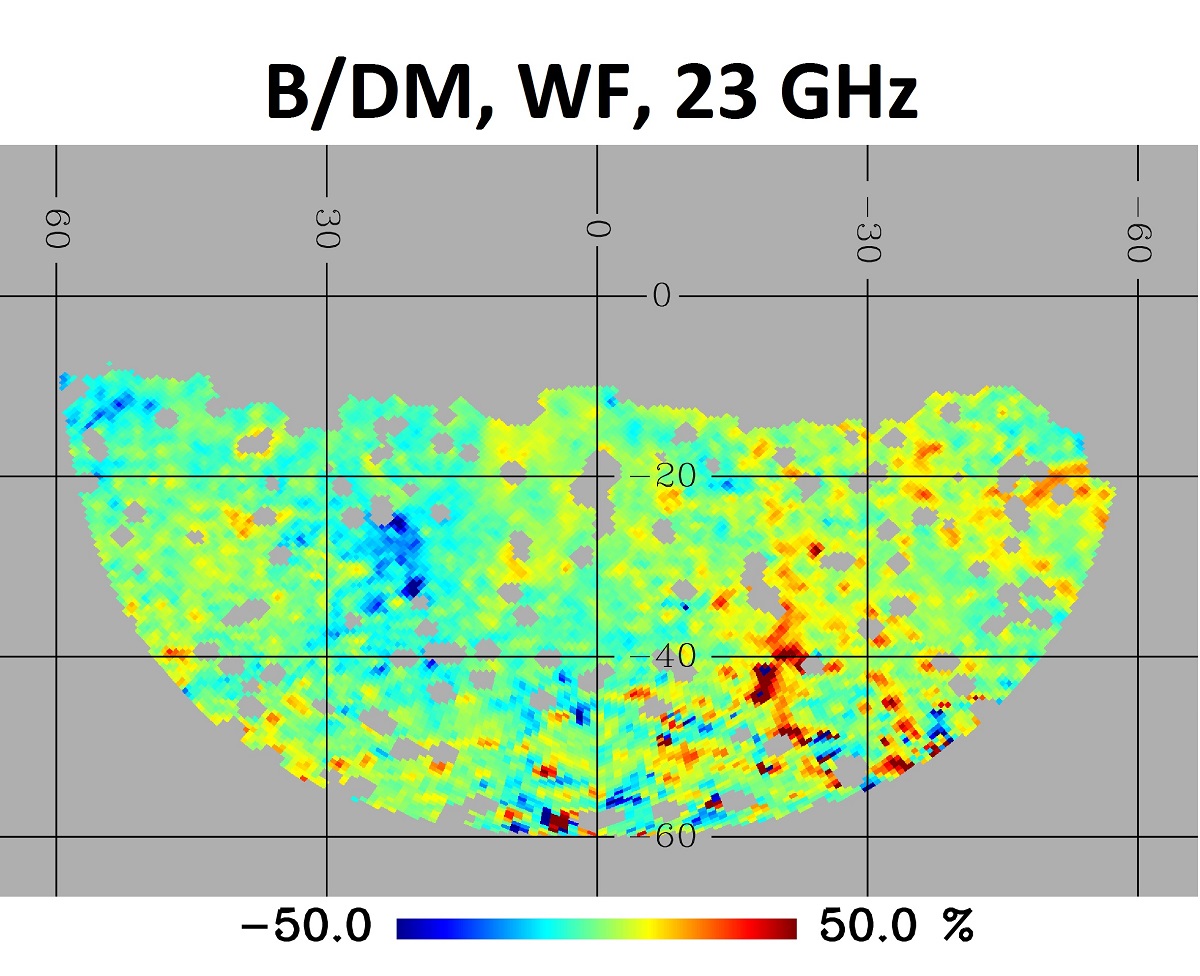}
\includegraphics[width=0.325\textwidth]{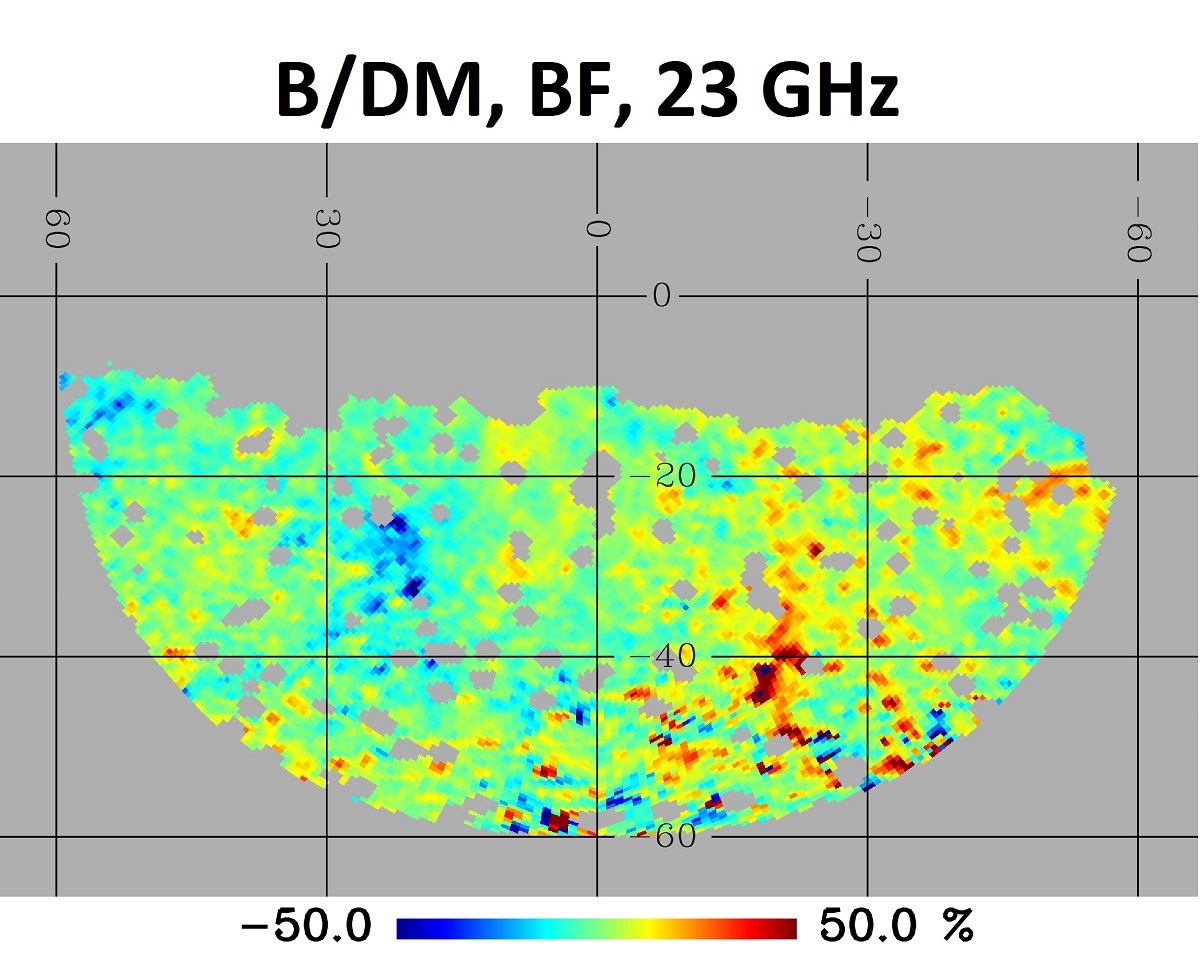}
\caption{Residuals resulting from our component separation procedure in units of the total measured sky intensity. The top row shows the residuals at different frequencies without introducing either the Bubbles or DM template. We can see a clear excess in this case at low frequencies. The bottom row shows residuals at 23 GHz with DM only and Bubbles+DM. Adding DM improves residuals slightly, while adding both the Bubbles and DM improves the residuals much more. At the same time the difference between the BF and WF configurations is almost absent. More details are in \S\ref{sec:Fitting.results}.}
\label{fig:res}
\end{center}
\end{figure}

In Fig.~\ref{fig:res} we show maps of the relative residuals $r(\nu,b,l)/[d(\nu,b,l)-md(\nu,b,l)]$ with  $r(\nu,b,l)$  and  $d(\nu,b,l)$ defined as   in Eq.~\ref{eq:fit} and $md(\nu,b,l)$ the fitted mono/dipole.
The top row of Fig.~\ref{fig:res} shows the frequency dependence of the residuals with neither  DM nor the Bubbles in the fit. At 23 GHz the Haze is very clearly  visible, reaching up to about 50\% of the total sky intensity at peak points. Note the vertically-elongated Haze morphology (which would be absorbed by the Bubbles template) slightly bent to the right relative to zero longitude (which visually matches the shape derived in \cite{Carretti:2013sc}). 
The Haze becomes less prominent with frequency, and is almost absent at 70 GHz. 
This spectral trend agrees  well with that in Table~\ref{tab:chisq}, which shows that  $\chi^2_{\rm r}$ reaches considerably lower values at higher frequencies, even if 
only standard foregrounds are considered.
Moreover,  looking at the top left panel, it is clear that the Haze tends to have rather irregular morphology with sharp edges, which seems more consistent with the Bubbles origin rather than smooth DM emission (e.g., Fig.~\ref{fig:dm_maps}).
The bottom row shows the residuals at the lowest and most relevant frequency for the cases of DM only, DM+Bubbles WF and DM+Bubbles BF\@. 
We can see that the  addition of  DM  alone reduces residuals somewhat. Adding both the Bubbles and DM further improves the results, almost entirely absorbing the Haze. 
There is, however, essentially no difference in the DM BF and WF cases, which is consistent with our previous finding that the $\chi^2_{\rm r}$ differs very little for these 
two models.

\begin{figure}[t]
\begin{center}
\includegraphics[width=1\textwidth]{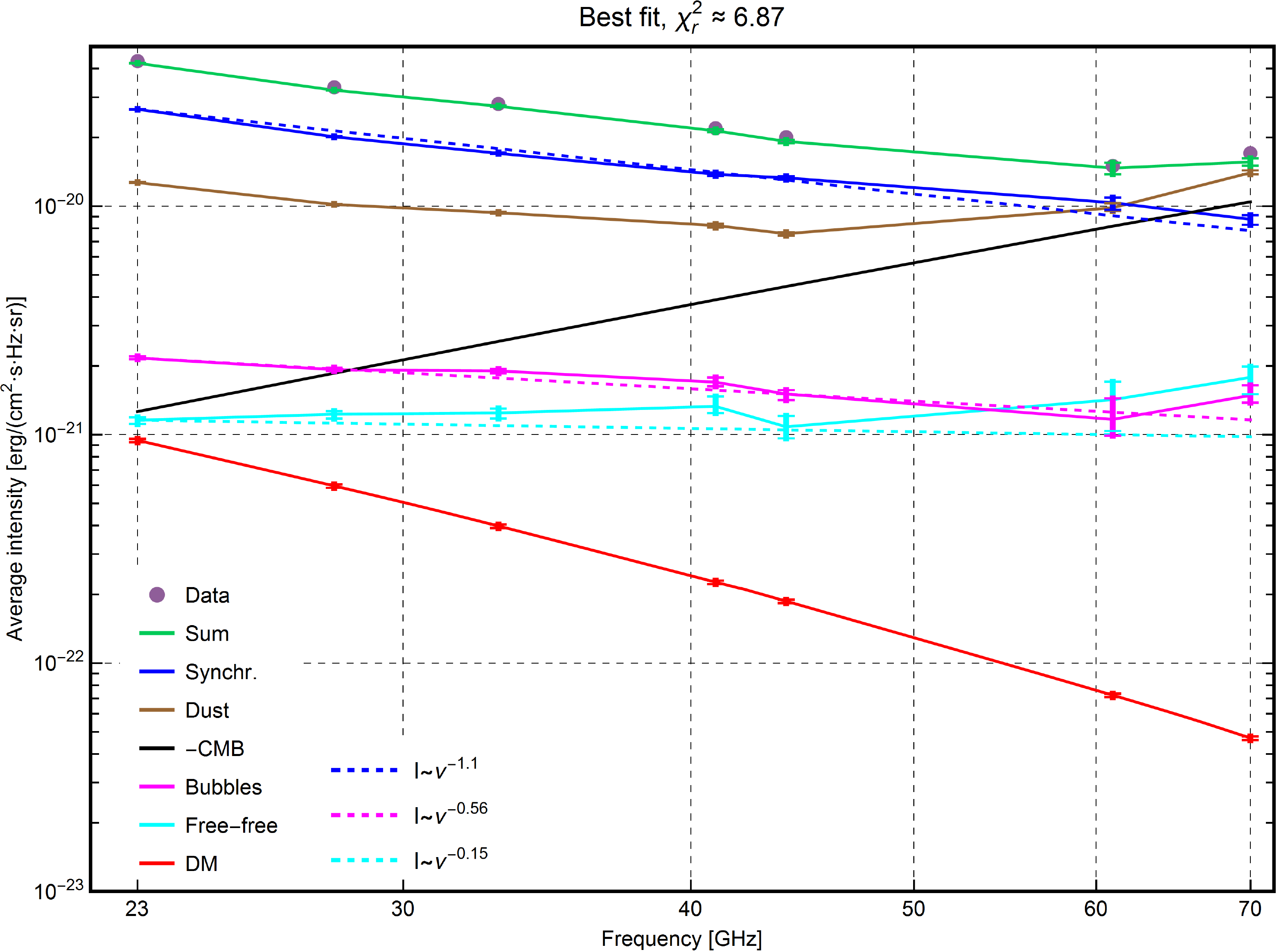}
\caption{\label{fig:sp}   Spectra of all physical components (and their sum) for the BF DM configuration. Error bars correspond to statistical 68\% CL. The data reported here have monopole and dipole subtracted, and the noise does not exceed the marker size used. 
Solid lines are results of this work, dashed lines are power-law spectra typically expected for the components considered. More details are in \S\ref{sec:Fitting.results}. }
\end{center}
\end{figure}

The computation of the fitting coefficients $a_i(\nu)$ also allows us to evaluate the average intensity of each component within the ROI and its frequency dependence, as well as the  components' relative amplitudes. This is reported in Fig.~\ref{fig:sp} for the BF DM model. Other models did not show significant qualitative differences. Hence, for simplicity, in what follows we focus on the BF case. At essentially all frequencies, there is a hierarchy in intensity (from higher to lower): synchrotron, dust, Bubbles, free-free and DM. The CMB fluctuations average is negative in our ROI, and its absolute value grows from a subdominant 2.9\% at 23 GHz up to a significant 61\% at 70 GHz, which reflects the black-body spectrum and the overall foreground frequency dependence. DM has the smallest contribution at all frequencies,  ranging from 2.2\% to 0.27\%. The synchrotron dominates at almost all frequencies, decreasing from 61\% to 51\%. Also we see that we reproduced well the canonically-expected (e.g., \cite{:2012rta}) synchrotron spectral index $I \sim \nu^{-1.1}$. The dust spectrum is not monotonic; it contributes 29\% at 23 GHz. The free-free contributes 2.7\% and matches quite well the expected slope $I \sim \nu^{-0.15}$ (e.g., \cite{Adam:2015wua}). The Bubbles contribution is relatively flat in frequency and equal to 5-9\%. Our Bubbles spectrum agrees very well with $I \sim \nu^{-0.56}$ derived in \cite{:2012rta}. 
 
We also compare our  foreground component intensities with those derived in \cite{:2012rta}, for the frequency where the Haze is most pronounced.  For this exercise, we considered the ROI used in  \cite{:2012rta}  and shown in Fig. \ref{fig:masks} (left panel). This comparison is shown in Table~\ref{tab:for}.
While all standard astrophysical components have similar derived properties in both that work and ours, the Haze component in this work is about a factor of two smaller.
This might be due to the mono/dipole removal strategy described above and/or to the modeling of the Haze which receives contributions from both Bubbles and DM in our work. 
Lastly, note that the sum of all components is  well within the 3$\sigma$ error bars on the data in all channels. 

\begin{table}[tbp]
\caption{The breakdowns of the Galactic components (in \% of the total non-CMB emission) at 23 GHz in the ROI (Fig. \ref{fig:masks}, left) obtained in our work (BF case) and by the Planck team in \cite{:2012rta}. More details are in \S\ref{sec:Fitting.results}.}
\label{tab:for}
\begin{center}
\begin{tabulary}{1\textwidth}{|C|C|C|C|}
\hline
Component & Planck team & Our work \\
\hline
Synchrotron & 43 & 56 \\
Dust & 30 & 26 \\
Free-free & 4 & 3.1 \\
Haze (DM+Bubbles) & 23 & 12 \\
\hline
\end{tabulary}
\end{center}
\end{table} 

We also investigate the model dependence of the Bubbles and DM contributions.
In Table~\ref{tab:dmb} we report the mean and variation ranges of the contributions of all 
DM/MF/propagation models at 23 and 70 GHz. The Haze contribution does not  significantly fluctuate from model to model.  At 23 GHz its amplitude is well above systematics due to model variations, whereas at higher frequencies it is of the same order of magnitude as the residuals. 
The mean flux of the Haze represents 7.8\% of the total sky emission compared to the best-fit amplitude of 7.2\%.
DM contributes a  significant part of the Haze at 23 GHz: 21\% on average (30\% in the BF case). Because DM spectra usually fall off with frequency faster than the Bubbles spectrum,  the DM contribution to  the Haze decreases with frequency.  

Note that in Table~\ref{tab:dmb}  we also quote the range and mean of the residuals normalized to the data, which can give an indication of the systematic uncertainties (error bars in Fig.~\ref{fig:sp} reflect statistical errors only). Systematic uncertainties may be due to inaccuracy in  the templates, the foreground modeling error introduced by our component separation setup, as well as residual systematics in the WMAP and Planck data (e.g., calibration and map-making). Studying these in detail is beyond the scope of our work, however note that they are below 10\% of the sky signal at all frequencies.

\begin{table}[tbp]
\caption{Variation ranges and mean of the contributions for all our DM/MF/propagation models. The last row shows the residuals normalized to data. More details are in \S\ref{sec:Fitting.results} and \S\ref{sec:Fitting.DM}.
}
\label{tab:dmb}
\begin{center}
\begin{tabulary}{1\textwidth}{|C|C|C|}
\hline
Parameter & 23 GHz & 70 GHz \\
\hline
Range/mean of Haze contribution to the data, \% & [7.2-10]/7.8 & [8.7-10]/9.2 \\
Range/mean of DM contribution to the data, \% & [0.43-6.3]/1.7 & [0.0043-4.7]/0.59 \\
Range/mean of DM contribution to the Haze, \% & [5.5-62]/21 & [0.047-45]/6.5 \\
Range/mean of the residuals, \% & [1.7-2.1]/2.0 & [7.7-9.1]/8.6 \\
\hline
\end{tabulary}
\end{center}
\end{table} 

\begin{figure}[t]
\begin{center}
\includegraphics[width=0.325\textwidth]{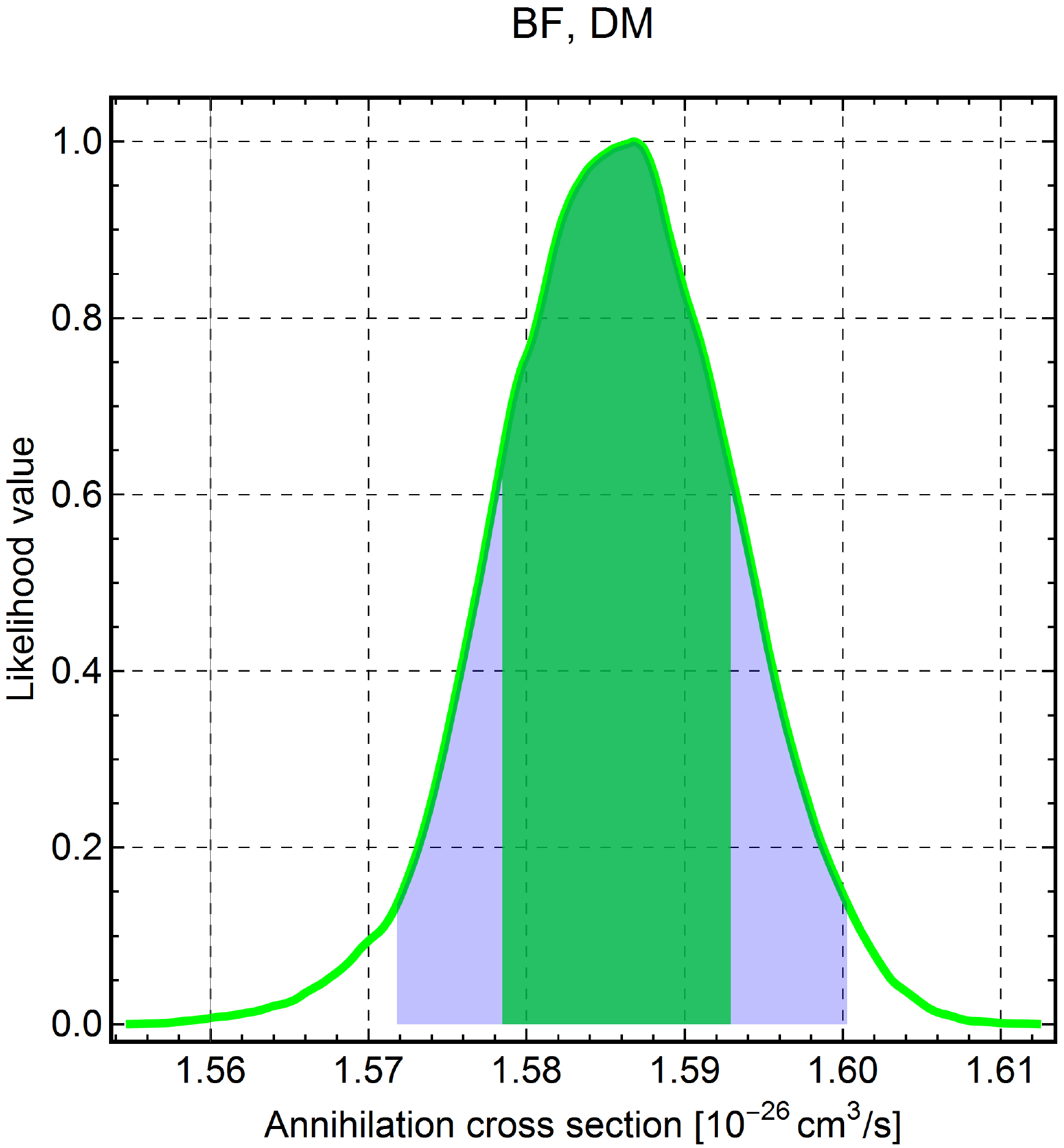}
\includegraphics[width=0.325\textwidth]{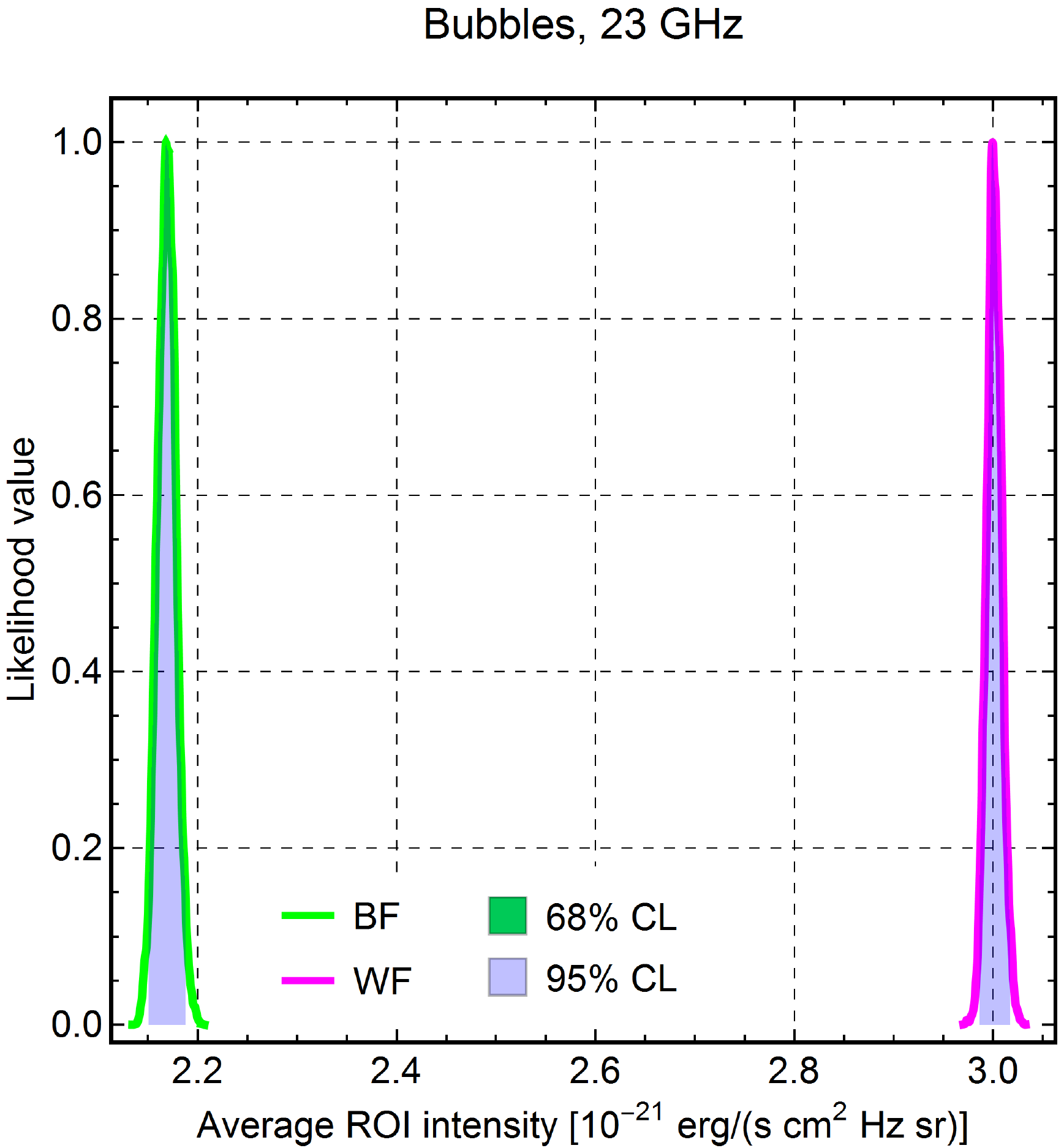}
\includegraphics[width=0.325\textwidth]{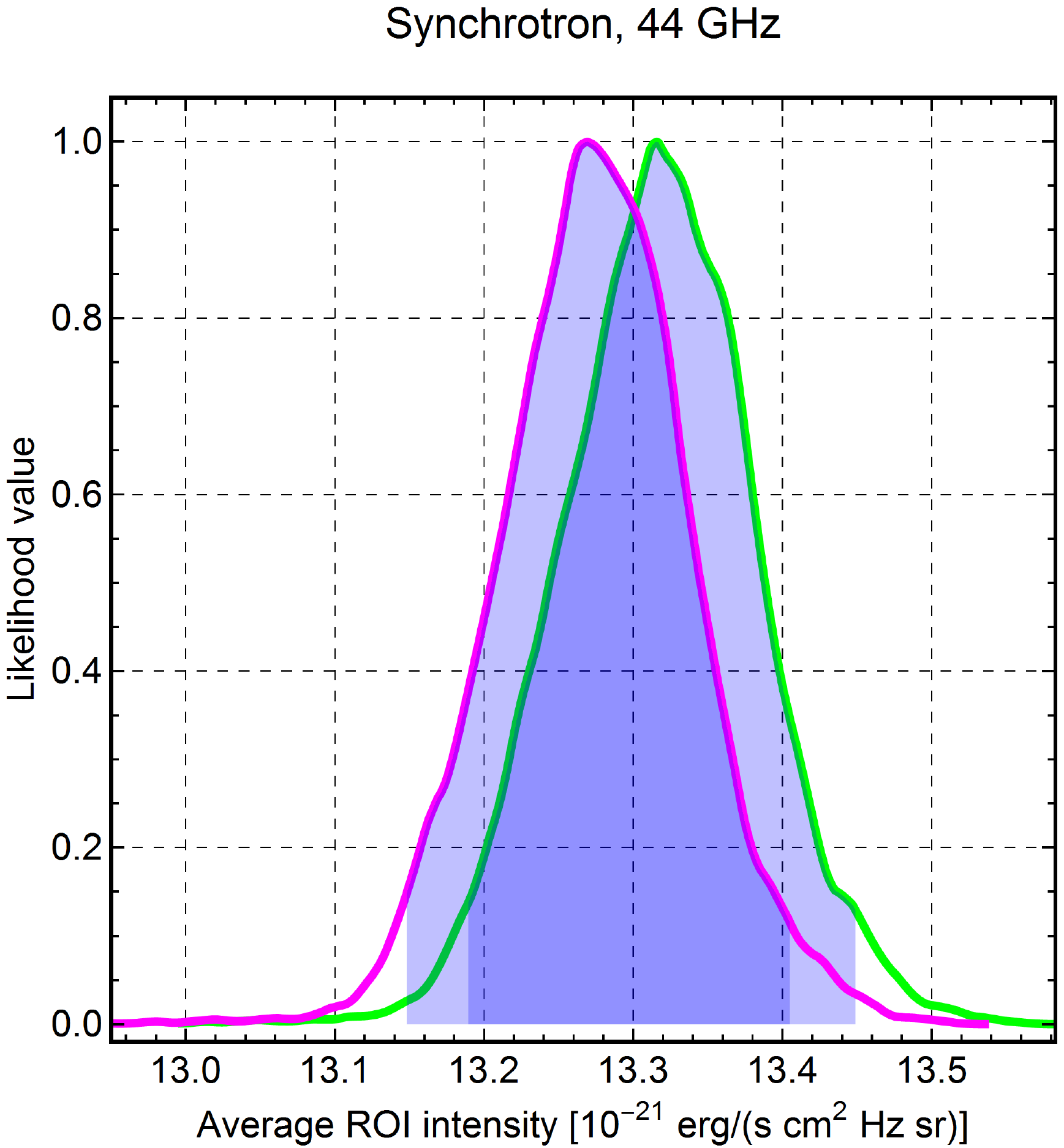}

\vspace*{0.25cm}

\includegraphics[width=0.325\textwidth]{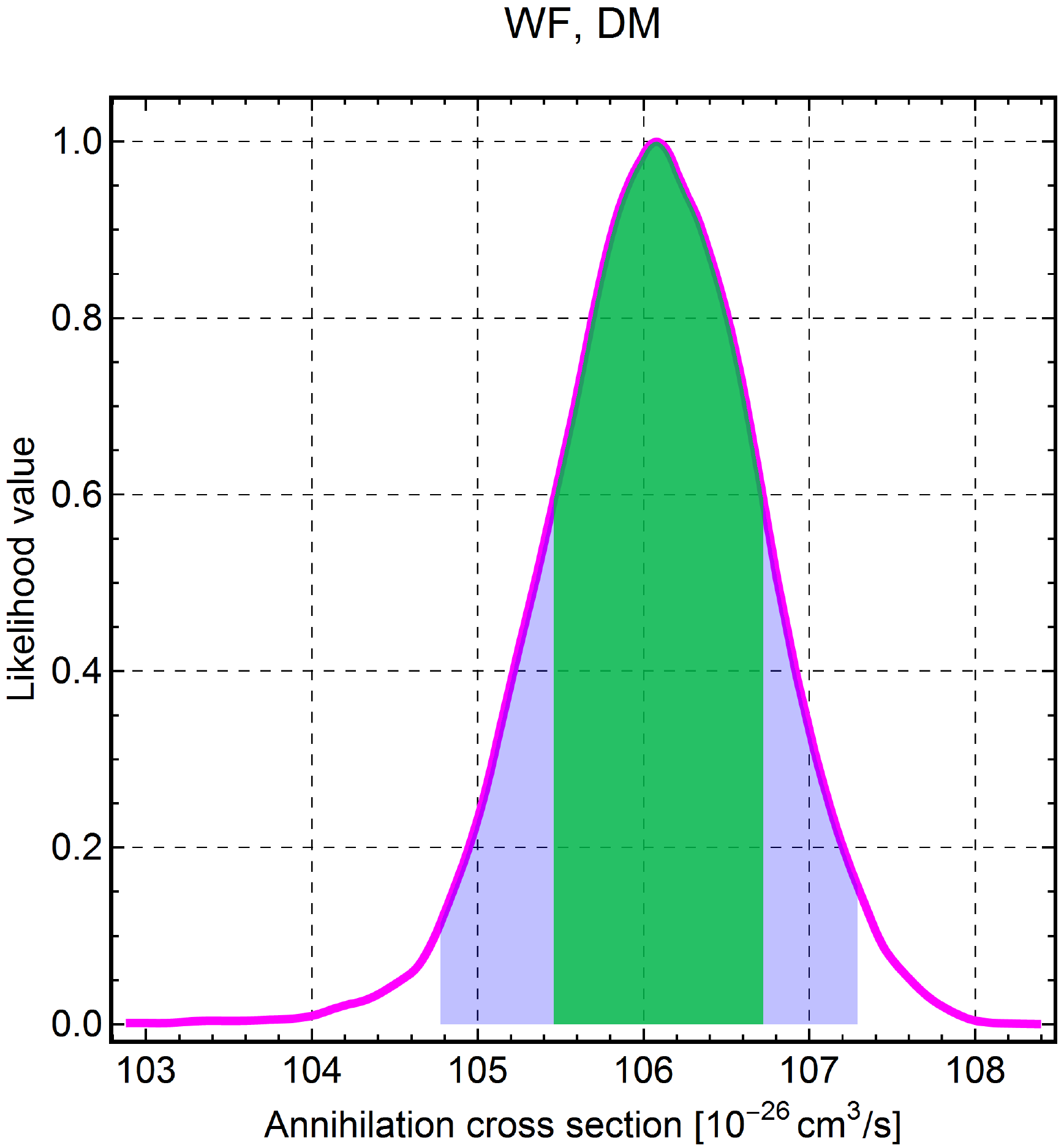}
\includegraphics[width=0.325\textwidth]{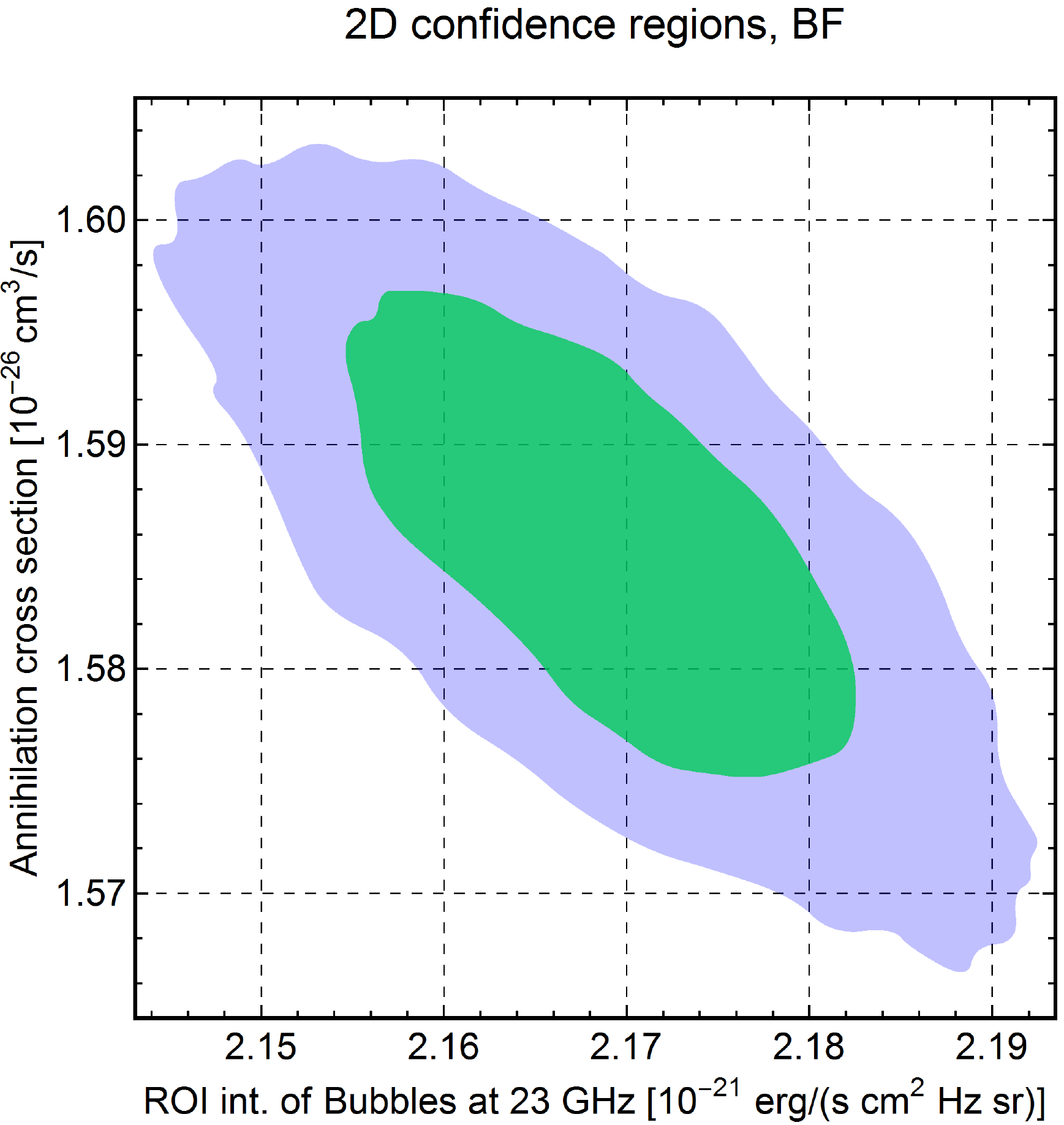}
\includegraphics[width=0.325\textwidth]{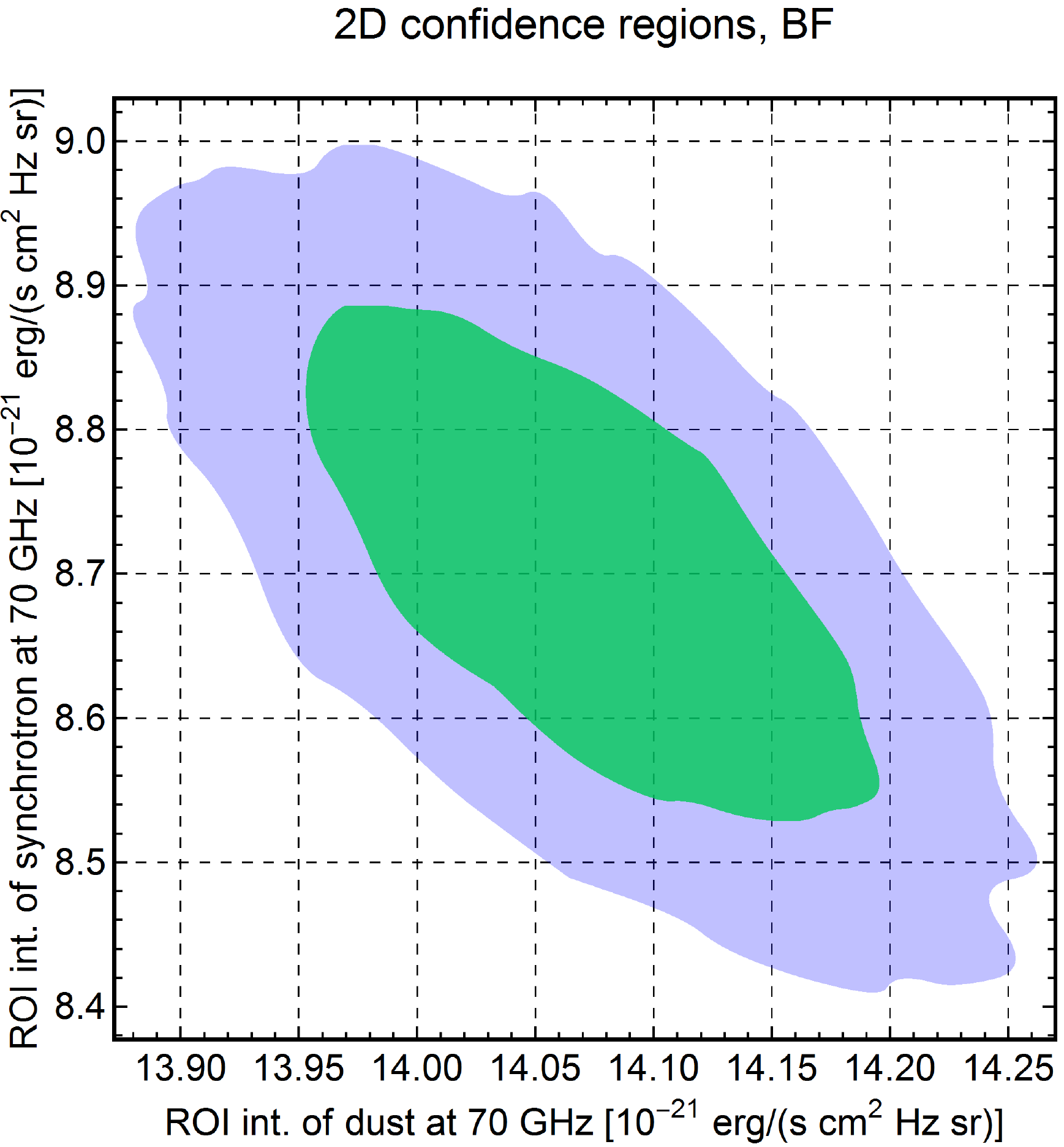}
\caption{Examples of the marginalized 1D and 2D likelihoods for the components mentioned on each plot. The green regions show 68\% confidence intervals while the blue regions 95\%. 1D likelihoods are shown for both the BF and WF cases.  2D confidence regions are shown for the pairs of most correlated component intensities in the BF case. More details are in \S\ref{sec:Fitting.results}.
}
\label{fig:likes}
\end{center}
\end{figure}

Statistical uncertainties in  Fig.~\ref{fig:sp}  rarely exceed  a few percent. 
The largest uncertainties are at high frequencies for the Bubbles and free-free component. 
Given the low map noise, our component separation  strategy yields quite narrow N-dimensional  likelihoods (N=29) for all parameters.
Fig.~\ref{fig:likes} shows some examples of the marginalized likelihoods, for some components' amplitudes.
The distributions are typically symmetric but not necessarily Gaussian, and the dispersion is below a few percent. Two of the bottom panels also show the 2D marginalized confidence regions for the most correlated pairs of components in the BF case. Thus, we see that the largest correlation is observed between the DM and Bubbles amplitudes at 23 GHz with the correlation coefficient $R=0.70$ (defined in the standard way as the covariance divided by the square root of the product of the variances). This degeneracy between DM and the Bubbles is likely caused by similarity in the templates. The second largest correlation, with $R=0.65$, is seen between the dust and synchrotron at 70 GHz.

\subsection{DM implications}
\label{sec:Fitting.DM}

Table~\ref{tab:dmb} shows that the derived DM contribution to the data, considering all 
DM/MF/propagation models, is small:  at 23 GHz the mean contribution is 1.7\% and the BF model 2.2\%.
This is at the level of the residuals at 23 GHz, implying no robust detection of DM\@.
The signal-to-noise ratio does not improve at higher frequencies because the DM emission is less pronounced and the noise increases.
Some models show a 6\% contribution of DM to the data at 23GHz, which is well above the residuals. 
However, this is also not a highly significant detection, and the global fit is, in general, poor with $\chi^2_{\rm r} \simeq 7$ for all models. 

We now examine correlations between the fit quality and DM parameters. 
For example, one may naively expect a preference for a specific DM spectral slope by the data. The DM spectral slope $\beta$ (defined as $I \sim \nu^{\beta}$) varies over a wide range $-5.0 < \beta < -1.1$ for all models.  The best fitting models have $\beta \approx -2.7$, which is much smaller than that of both the Bubbles and synchrotron (for which we find fitted values of $-0.44 \pm 0.05$ and $-1.17 \pm 0.01$ respectively).
However, $\beta$ is not linked unambiguously to any single DM/MF/propagation parameter. 
Looking at  more physical parameters, one can see the following correlations. The fit quality mainly  depends on the MF/propagation model, it depends significantly on WIMP mass, and is almost insensitive to the annihilation channel and profile slope $\gamma$. 
The smallest $\chi^2_{\rm r}$ typically corresponds to the strongest MF ($B(0,0) = 100~\mu$G) with medium to high vertical extent ($z_B$ = 2.7, 6.9 kpc), strong diffusion models MAX and Reacc (see Table~\ref{tab:MF/prop}), and intermediate WIMP masses around 19 GeV. Instead, the worst fits are observed for low $z_B$ = 1.7-2.0 kpc, MED and Reacc diffusion, and low WIMP masses around 7 GeV. 
These general  trends should not be considered in a strict sense. Particularly, one may notice that the same WIMP mass 7 GeV serves for both BF and WF models (Table~\ref{tab:bfwf}), which suggests only a weak correlation with mass. The correlation with MF/propagation configuration is stronger, and this has a quite straightforward interpretation. Comparing the BF and WF DM maps in Fig.~\ref{fig:t} two main differences are apparent: the BF map has a steeper brightness profile and a vertically elongated morphology instead of the horizontal morphology observed in the  WF\@. The vertical elongation can be naturally explained by a strong and highly extended MF with the parameters mentioned above. The reason for such a preference by the fit could be related to the degeneracy between DM and the Bubbles: the BF DM template looks much more similar to the Bubbles template than the WF template. 

Fig.~\ref{fig:dm_c} presents our main result: the best-fitting WIMP annihilation cross sections and masses for various DM/MF/propagation models. 
As for the more conservative constraints (Fig.~\ref{fig:dm_c_cmb}), $\langle \sigma v \rangle$ values vary over an order of magnitude for different MF/propagation models,  with  the $\tau^+ \tau^-$ channel showing less of a spread than $b\bar{b}$ does. Indeed, the MF/propagation model uncertainties significantly exceed the statistical ones. The BF $\langle \sigma v \rangle$ values reside close to the canonical cross section $3 \cdot 10^{-26}$ cm$^3$/s, especially for $\tau^+ \tau^-$ channel. It is interesting to compare the constraints in the $\langle \sigma v \rangle ~ - ~ m_{\chi}$ plane with those required to fit the GC gamma-ray excess, also reported in Fig.~\ref{fig:dm_c}. In most cases we see no intersections between the allowed regions determined by gamma-ray data analysis and our BF values. A small intersection is visible only in the case $\chi \chi \rightarrow b\bar{b}, \gamma = 1.1$. On one hand, very optimistic MF/propagation models with $B(0,0) = 100~\mu$G would be required in this case. On the other hand, the green rectangles reflect just 68\% CL regions, which strengthen the potential overlap. Thus, we can conclude that for realistic parameter configurations our fit of the microwave data with DM does not prefer similar DM parameters as the gamma-ray data fits \cite{Gordon:2013vta, Daylan:2014rsa, Abazajian:2014fta, Calore:2014nla}, but rather favors higher cross section values (by a factor $\sim (2-7)$) for similar masses and annihilation channels. However, in case of a push of both the gamma-ray fit and MF/propagation parameters towards an optimistic side, one can reconcile the DM parameters in the case $\chi \chi \rightarrow b\bar{b}, \gamma = 1.1$. 

Finally, comparing  DM best-fit values resulting from the component separation and the more conservative constraints  derived in \S\ref{sec:cmb_constraints} (Fig.~\ref{fig:dm_c} vs. Fig.~\ref{fig:dm_c_cmb}), it is apparent that foreground cleaning may improve the sensitivity of a DM search by a factor of 10-100. However, as was already mentioned in \S\ref{sec:cmb_constraints}, any component separation procedure yields results whose validity depends on the foreground and signal modeling.

\section{Discussion and conclusions}
\label{sec:Discussion}

In this paper, we addressed the DM implications of the WMAP-Planck Haze in the context of models that could explain the GC gamma-ray excess. We investigated whether the microwave observations are compatible with a DM contribution to the Haze emission, derived constraints on DM particle parameters, and compared them to those needed to fit the GC gamma-ray excess.

We modeled the synchrotron emission in the microwave portion of the electromagnetic spectrum due to DM annihilation, making various assumptions about the DM density distribution, the DM particle model, as well as the MF and CR propagation models.  DM emission intensity sky maps for various DM/MF/propagation models were created and compared with the data at seven WMAP-Planck frequencies. To test the consistency of the DM models with the data, we adopted two approaches.  First we adopted a conservative approach which only considered CMB fluctuations and DM as contributing to the total sky intensity, while second we attempted to separate all the emission components including a potential DM signal using template fitting. The first approach yields relatively weak, but very robust DM constraints (see Fig.~\ref{fig:dm_c_cmb}). 
\begin{figure}[H]
\begin{center}
\includegraphics[width=0.495\textwidth]{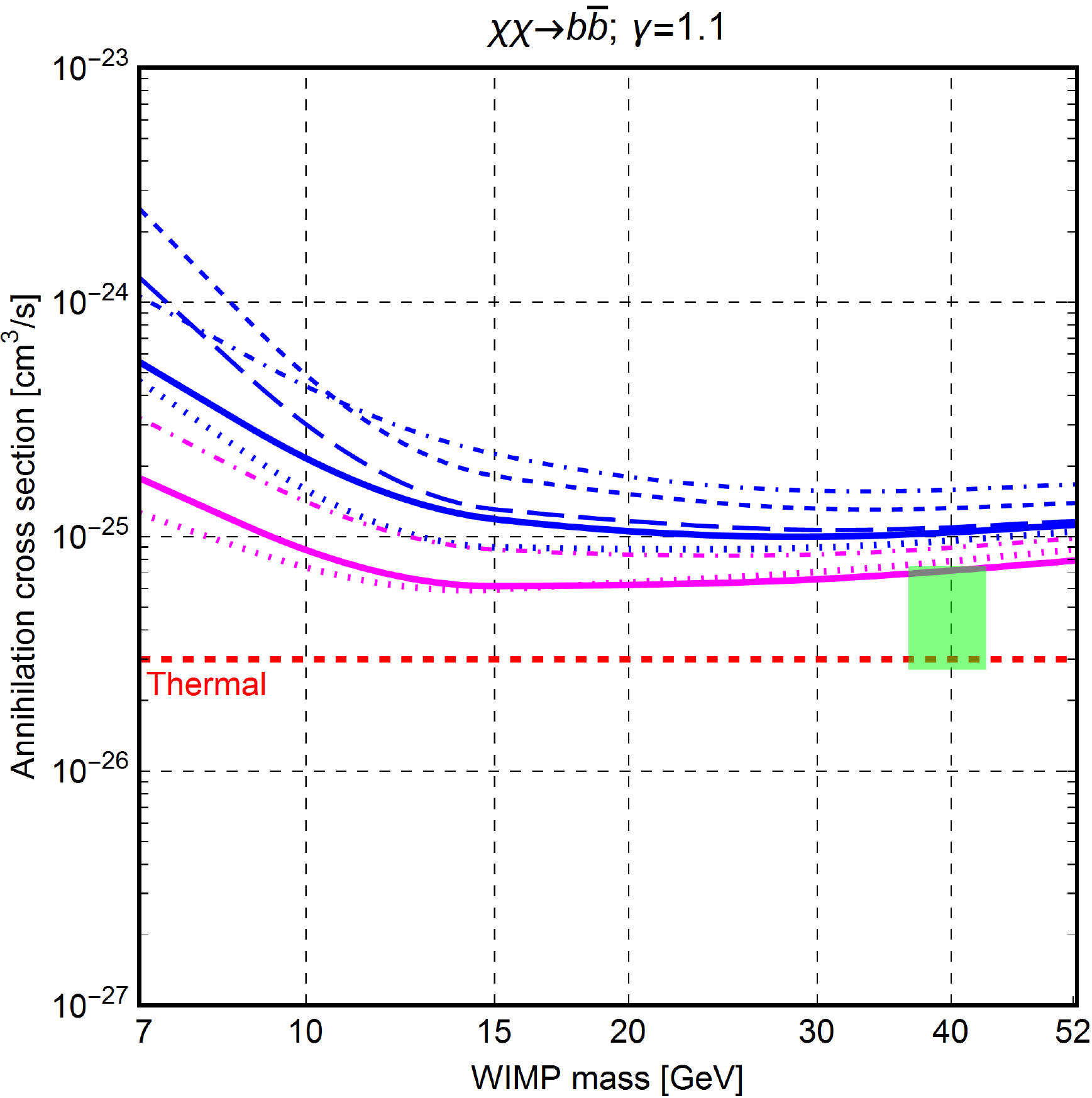}
\includegraphics[width=0.495\textwidth]{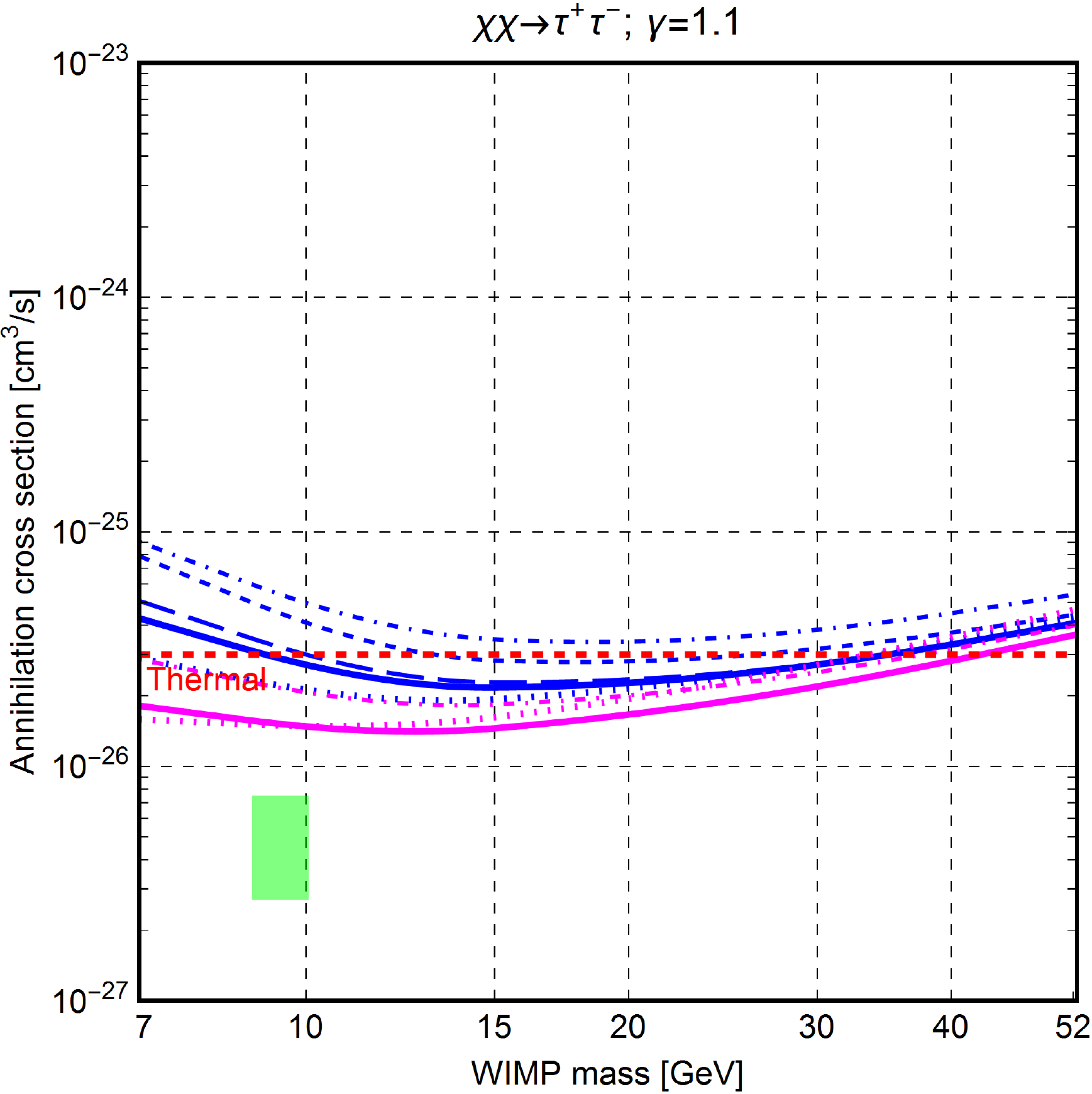}

\vspace*{0.25cm}

\includegraphics[width=0.495\textwidth]{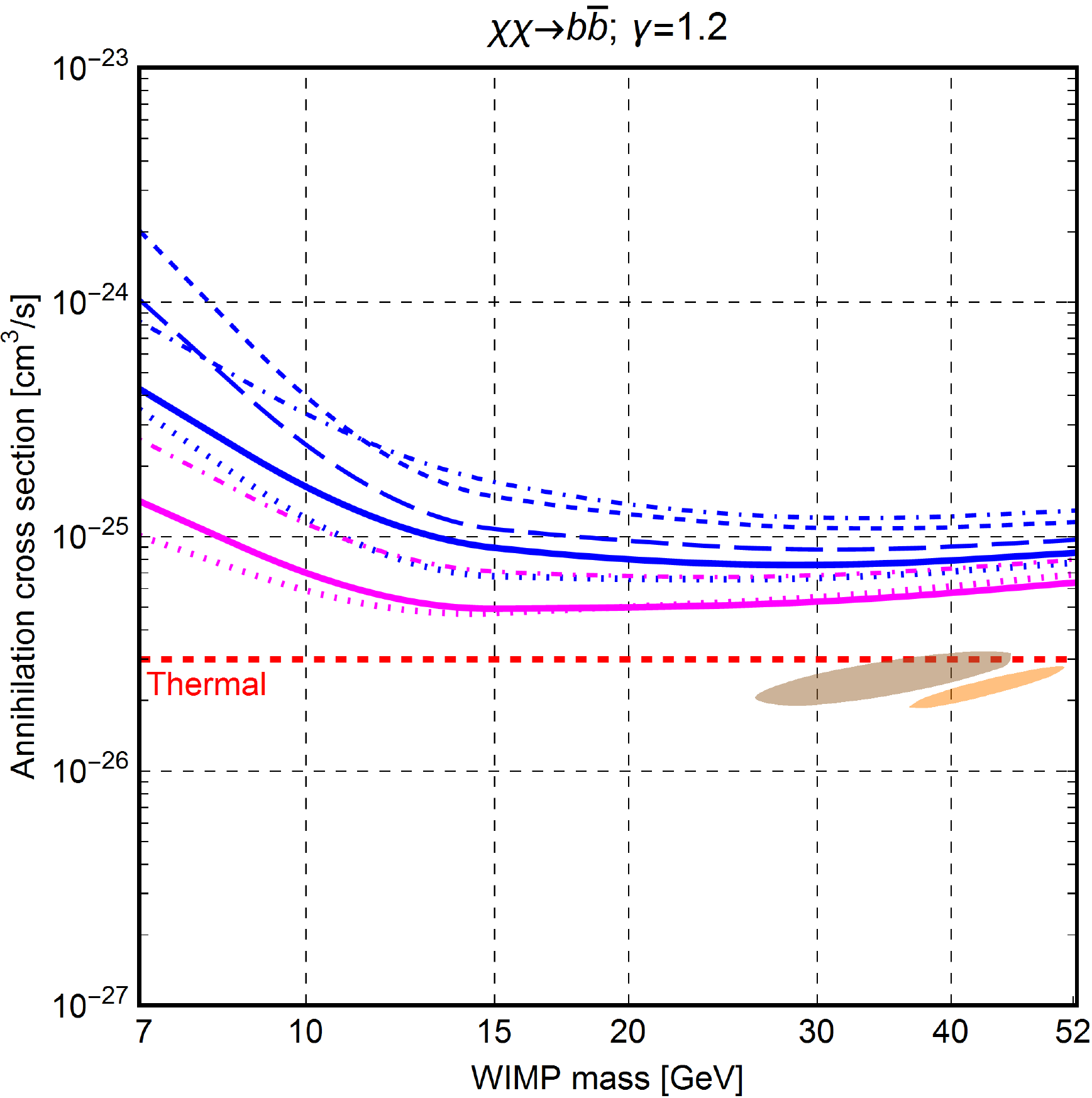}
\includegraphics[width=0.495\textwidth]{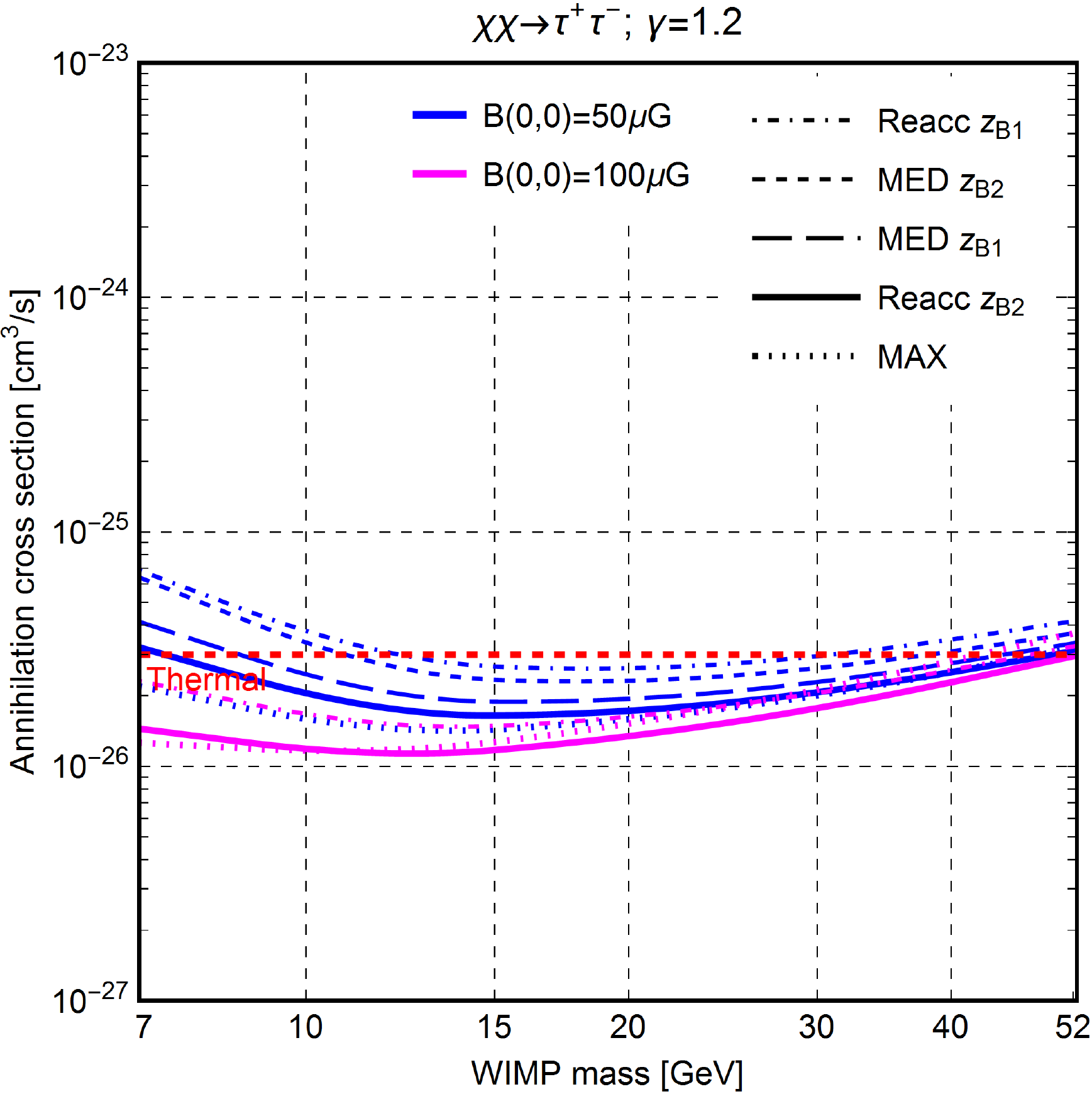}

\vspace*{0.25cm}

\includegraphics[width=0.495\textwidth]{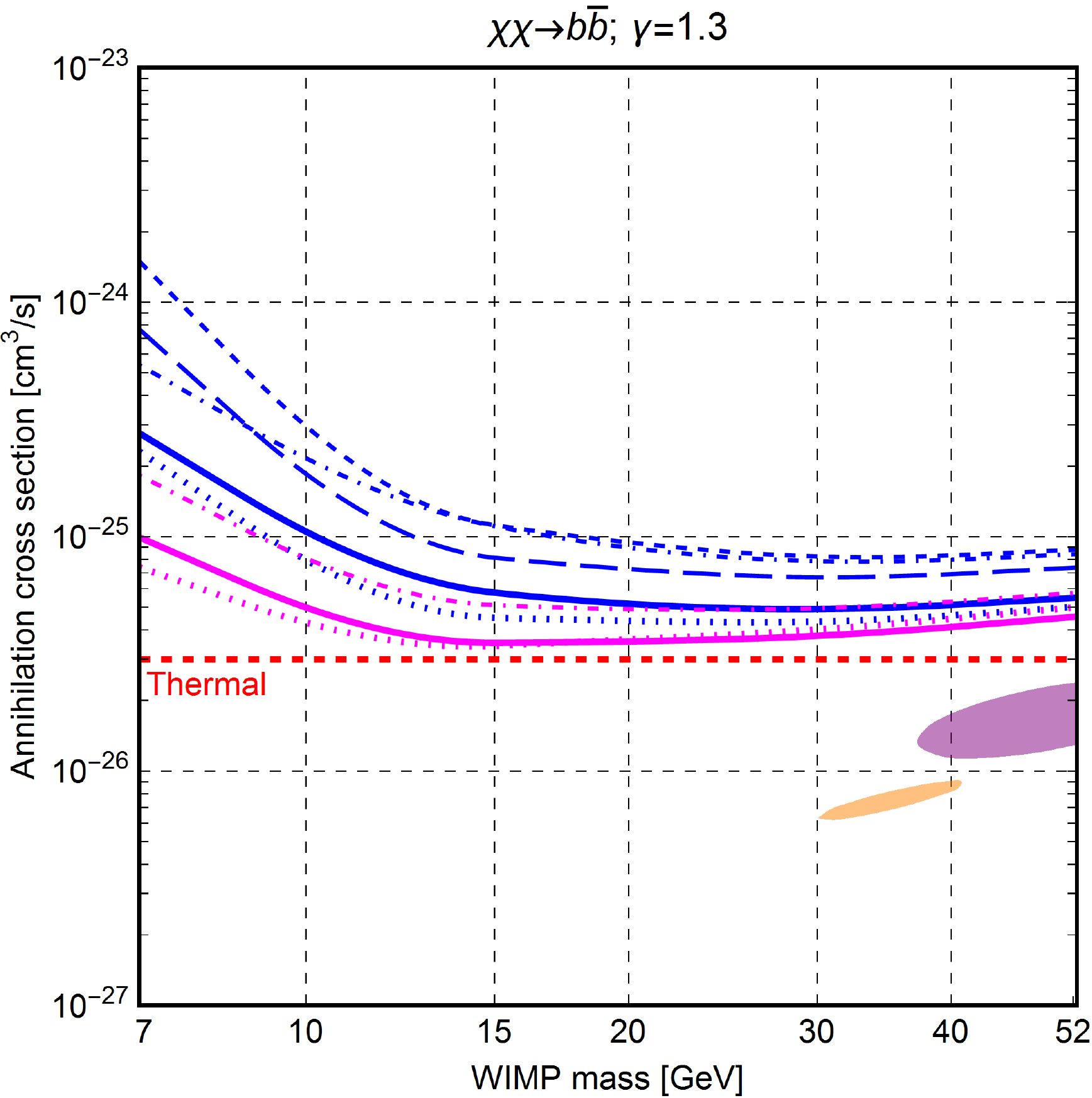}
\includegraphics[width=0.495\textwidth]{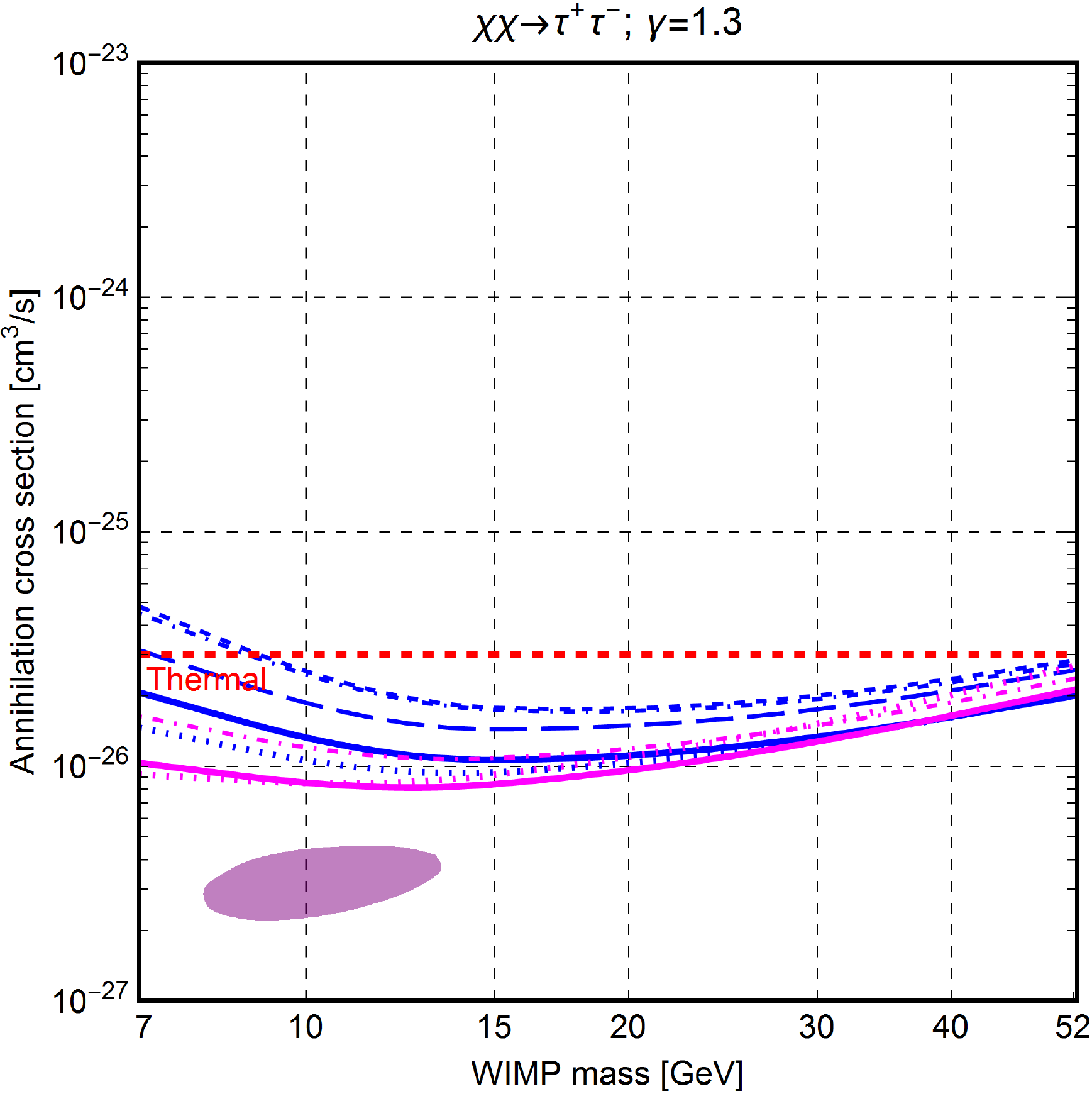}
\end{center}
\end{figure}
\begin{figure}[H]
\caption{\label{fig:dm_c} \textit{(preceding page)} DM constraints derived at 95\% CL resulting from our component separation procedure for various DM/MF/propagation models. Limiting cross section values are indistinguishable from the BF ones at this line resolution. The red dashed horizontal line indicates the thermal relic cross section $\langle \sigma v \rangle = 3 \cdot 10^{-26}$ cm$^3$/s. In addition to the best-fit curves we also present here DM parameter regions (same as in Fig.~\ref{fig:dm_c_cmb}) needed to fit the GC gamma-ray excess. We see again that our constraints mainly do not contradict them. See more details in \S\ref{sec:Fitting.DM}.  }
\end{figure}
The second one probes much lower cross sections, with the best-fit cross-sections reaching the thermal relic expectation $\langle \sigma v \rangle = 3 \cdot 10^{-26}$ cm$^3$/s (see Fig.~\ref{fig:dm_c}), however this approach is more dependent on the details of the modeling and the results are less straightforward to interpret due to degeneracies between DM and other model components.

We summarize our results as follows:
\begin{itemize}

	\item  The (weak) upper limits obtained with the conservative approach are compatible with both the thermal cross section and the signal regions of DM models that fit the gamma-ray excess that have been determined in other works.
	
	\item In the component separation analysis, the overall quality of our fits is poor, indicating that we lack sufficient understanding of foregrounds, or that template fitting is a too simplistic procedure for explaining the observed signal in the sky area under consideration.
	
	\item Adding additional components that produce emission at tens of degrees around the GC to the standard astrophysical foregrounds and CMB considerably improves the fit. Thus we confirm the clear presence of the Haze: a microwave excess not explained by the CMB, (spinning) dust, free-free, and synchrotron emission, which contributes (7.2-10)\% of the intensity at 23 GHz and (8.7-10)\% at 70 GHz in our selected sky region near the GC\@. We also derive the Haze spectrum (here represented by the sum of the DM and Bubbles components), obtaining $I \sim \nu^{-0.84 \pm 0.04}$. This is somewhat different from the slope of -0.56 reported in \cite{:2012rta}. However, this is not a completely equivalent comparison due to the presence of the DM template in our Haze fit. Our derived Haze spectrum significantly differs from that of the usual synchrotron ($I \sim \nu^{-1.17 \pm 0.01}$), which suggests the Haze originates from a population of electrons distinct from typical Galactic CR electrons.
	
	\item We considered two possible origins of the Haze in constructing our templates: i) DM annihilation and ii) a counterpart of the Fermi Bubbles.  We showed that the data do not have a strong preference for either of these two components when considered separately; they provide comparable fit qualities with a mild preference towards the Bubbles. However, inclusion of both templates together provides a better fit than adding either template individually. Thus, both emission mechanisms can be contributing to the Haze. In this scenario, DM constitutes (5.5-62)\% (on average 21\%) of the Haze emission and (0.43-6.3)\% (on average 1.7\%) of the total emission at 23 GHz. The DM spectrum, as derived from our modeling of the DM emission, is generally steeper than the Bubbles fit-inferred spectrum. The amplitude of the latter exceeds that of the DM, especially  at high frequencies.
	
	\item Given the overall poor quality of the fits,  it is not possible to confidently attribute any of the emission to a DM model, nor to single out a particular DM model as the one preferred by the data. In general the data show a mild preference for a strong MF that is extended above the plane.  This kind of MF generates DM emission with a vertically elongated morphology that is aligned with the Bubbles observed in gamma rays.
	 
	\item For realistic MF/propagation parameter values, the best-fit cross sections as determined by the component separation procedure are a factor $\approx$(2-7) higher than those favored to explain the GC gamma-ray excess, and are in tension with other constraints from indirect DM searches (e.g., \cite{Ackermann:2015zua}). Note, however, that in the case of $\chi \chi \rightarrow b\bar{b}, \gamma = 1.1$ an agreement between microwave and gamma-ray fits exists for optimistic MF/propagation configurations.
	 	
\end{itemize}

Looking forward, although multi-frequency studies clearly offer complementary information to that from indirect searches in gamma rays, to make these searches competitive it will be necessary to better understand the various non-DM contributions to the total microwave emission. This could be pursued by considering other component separation strategies and an optimization of the ROI\@.

\acknowledgments
It is a pleasure to thank Loris Colombo for extensive help with the CMB emission modeling and Planck data interpretation, Andrew Strong for valuable consultations on GALPROP, and Antony Lewis for assistance with GetDist. AE and EP acknowledge the support of NASA  grant  NNX13AP25G and the usage of the computers at the USC High Performance Computing Center.  JG acknowledges support from NASA through Einstein Postdoctoral Fellowship grant PF1-120089 awarded by the Chandra X-ray Center, which is operated by the Smithsonian Astrophysical Observatory for NASA under contract NAS8-03060, and from a Marie Curie International Incoming Fellowship in the project ``IGMultiWaveÕÕ (PIIF-GA-2013-628997) -- ADD''. We also acknowledge the use of the LevelScheme package for Mathematica developed by Mark Caprio \cite{LS}, it greatly helped in preparation of high-quality plots. In addition, this research used resources of the National Energy Research Scientific Computing Center, which is supported by the Office of Science of the U.S. Department of Energy under Contract No.~DE-AC02-05CH11231. Some of the results in this paper have been derived using the HEALPix \cite{Healpix} package.

\bibliography{PlanckHazeDM_bib}

\end{document}